\documentclass[12pt,a4paper]{report}
\usepackage{charter}
\usepackage{graphicx}

\usepackage{vmargin}
\setmarginsrb{2.5cm}{2cm}{2.5cm}{2cm}{0mm}{10mm}{0mm}{10mm}
\usepackage{amsmath}
\usepackage{amssymb}
\usepackage{enumitem}
\usepackage{calligra}
\usepackage{tikz}
\usetikzlibrary{shapes,arrows}
\usepackage{fancybox}

\begin{document}
\newpage
\renewcommand{\thepage}{\roman{page}}
\setcounter{page}{0}
\begin{center}
\LARGE \textbf{A THEORETICAL STUDY ON A TWO-DIMENSIONAL FLAP-TYPE WAVEMAKER}\\
\vspace{2cm} 
\large \textbf{Final Project}\\ 
\textbf{Proposed as a requirement for an undergraduate degree}\\
\vspace{5cm} 
\large \textbf{by}\\ 
\Large \textbf{Natanael Karjanto}\\ 
\textbf{10197045}\\ 
\vspace{5cm}
\begin{figure}[h]
\begin{center}
\includegraphics[scale = 0.8]{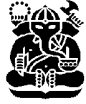}
\end{center}
\end{figure}
\normalsize 
\textbf{DEPARTMENT OF MATHEMATICS} \\ 
\textbf{FACULTY OF MATHEMATICS AND NATURAL SCIENCES} \\ 
\textbf{BANDUNG INSTITUTE OF TECHNOLOGY}\\ 
\textbf{2001}
\end{center}

\newpage
\setcounter{page}{0}
\begin{center}
\LARGE \textbf{A THEORETICAL STUDY ON A TWO-DIMENSIONAL FLAP-TYPE WAVEMAKER}\\
\vspace{0.5 in} \large \textbf{Approval Page}\\ \vspace{1 in}
\Large \textbf{\large It has been examined and approved by the Supervisor:}\\
\vspace{1.25 in} \textbf{\underline{Dr. Andonowati}}\\
\textbf{NIP. 131803263}\\ \vspace{1.5 in} \Large 
\textbf{and Examiners:}
\Large \vspace*{3.2cm}\\
\hspace*{0.5cm}\textbf{\underline{Prof. Dr. M. Ansjar}} \hfill \textbf{\underline{Dr. Wono Setya Budhi}\hspace*{0.5cm}} \\
\hspace*{0.8cm}\textbf{NIP. 130143972} \hfill \textbf{NIP. 131284801} \hspace*{1cm}
\end{center}

\newpage
\begin{verse}
\setcounter{page}{0}
\vspace*{2cm}
\begin{center}
\emph{For my beloved father, mother, and sister.}
\end{center}
\end{verse}

\begin{abstract}
\setcounter{page}{4}

{\normalsize A mathematical model for unidirectional wave generation explained in this study. The model consists of Laplace's equation in the semi-infinite two-dimensional water interior, dynamic and kinematic boundary conditions at the free surface, lateral boundary condition on the wavemaker, and fixed wall at the bottom. The model is for a flap-type wavemaker that is commonly used in a towing tank of a hydrodynamic laboratory. To simplify the problem, some assumptions are made, namely that water is an ideal fluid. Linear wavemaker theory is used and a generation of monochromatic wave (single frequency) is considered. The relation between the wavenumber, wave height, and wavemaker stroke is further derived. \par} \vspace*{1cm}

\noindent 
{\normalsize \textbf{Keywords:} ideal fluid, wavemaker, Laplace equation, kinematic free surface, dynamic free surface, linear wavemaker theory, monochromatic wave, wavenumber, wave height, and wavemaker stroke. \addcontentsline{toc}{chapter}{Abstract} \par}
\end{abstract}

\begin{center}
\huge {\textbf{Foreword}}
\end{center}
\setcounter{page}{6}

\normalsize {\Huge \textbf{T}}hank you to the readers who take the time to open this final project report. First of all, allow me to thank God Almighty for blessing me to complete this project and to finish my undergraduate degree. I am also grateful to the following individuals who have supported me during my study.
\addcontentsline{toc}{chapter}{Foreword}
\begin{enumerate}[leftmargin=1.2em]
\item Dr. Andonowati who was willing and patiently supervised this project.

\item Professor Dr. M. Ansjar and Dr. Wono Setya Budhi who have become the examiners for the project presentation on 15 December 2000. 
I thank Professor Ansjar for trusting me as his teaching assistant in Mathematical Methods during Fall/Autumn 2000.
I am grateful to Dr. Wono for patiently teaching me \emph{Maple} and \LaTeX \ so that this report can be completed nicely.

\item Dr. Nana Nawawi Gaos for being my academic adviser during Common First Year 1997/1998 and Intensive Semester of 1998.

\item Dr. Ahmad Muchlis for being my academic advisor during my sophomore until senior years. Thank you for motivating me to complete my undergraduate study in three and half a year.

\item Warsoma Djohan M.Si. for delegating me as his teaching assistant for Calculus 1 during Fall 1999 and Fall 2000.
I thank Dr. Jalina Widjaja for having me as her teaching assistant in Engineering Mathematics, Matrix and Vector Spaces and Calculus 1 as well as Dr. Nuning Nuraini whom I assist in her Calculus 2 course. I would not forget Koko Martono M.Si. for delegating me as his teaching assistant in Multivariable Calculus and Complex Function as well as training me to write scientific articles. I acknowledge other instructors and professors who have contributed me in developing my mathematical maturity.

\item My parents who have supported me financially and their incessant prayer during my study. My beloved sister who has supported and motivated me to study diligently.

\item Hadi Susanto for being my friend, both in good and bad times, particularly before he departed to the Netherlands. \textsl{Ik wil U bedanken omdat U mij heel goed heeft gemotiveend om extra hard te studeren. Bedankt, Hadi!}

\item Maykel, Luis, Dina, Ety, Sica, Anna, Sondang, Rilyovira and other friends from the Class of 1997 whom I could not mention individually. Thanks for our friendship. I also thank both my seniors and juniors who have assisted me during my study.

\item Toto Nusantara (MA-S3), Lylye Sulaeman (P4M) and Surya (MA96) who have helped me in \TeX, \LaTeX, and \emph{Scientific Work Place} to type this report.

\item Wili (MA97), Henry (FI97), Albert(TF97), Wila (TI97), Aan (TK97), Faiq (TG97), Krshna (EL97), Fitra (TA98), Dindin (TL98), Hidayat (FA98), Dwi Susanti (FA98), Mia (KI99), Erika (FA99), Dwi Hesti (TG2000) and other friends as well as juniors from SMA Negeri 4 Bandung who have motivated and supported me to complete my study in less than eight semesters. I am grateful to you all!
\end{enumerate}

\emph{Experientia est optima rerum magistra} is a well-known Latin expression for ``Experience is the best teacher''. Like other events in life, completing the undergraduate study as well as finishing this report is a unique and interesting experience for me, become a `teacher' in happiness and sorrow. I have learned many things when writing this report, not only academically but also growing mature thinking.

With a humble heart, I am fully aware that as a person, I am merely an individual from a crowd who comes and goes and attempts to imprint an academic achievement at this department and university. Although I might not be able to present my best for the progress of the department, I have striven to do my best in completing this project. I also hope that this `tiny' writing could become \emph{cr\'eme de la cr\'eme} from all the work that I have made albeit there are many limitations in various aspects.

Therefore, I would like to take this opportunity to apologize to the readers of this final project report. I hope this piece of work could be beneficial not only for interested readers but also for the progress of Mathematics.
\begin{flushright}
Bandung, mid-January 2001 \vspace*{0.5cm}\\
{\Large \calligra Natanael Karjanto \ }
\end{flushright}

\newpage
\pagestyle{headings} \setcounter{page}{9}
\addcontentsline{toc}{chapter}{Table of Contents} \tableofcontents

\newpage
\addcontentsline{toc}{chapter}{List of Figures} \listoffigures

\chapter{Introduction}

\pagenumbering{arabic}

\section{Background and Problem Formulation}

\subsection{Background}

As a nation progresses, the need for research in science and technology is increasing as well. More than two-thirds of the Indonesian territory consists of seas with abundant natural resources. Indonesia has been known as a maritime nation since ancient times. As science and technology progress, there is a need for research in the maritime area. 

In Ocean Engineering, the need to model phenomena in the open ocean mathematically motivates the research in this area. This leads to interdisciplinary collaboration among various fields of science and engineering. Mathematics as the queen and servant of science has the powerful ability to model many natural phenomena. In this final project report, we will discuss a theoretical study for a wave generation in a towing tank.\footnote[1]{A towing tank is a facility in a hydrodynamic laboratory, the long pond contains water with a wavemaker on one side and a wave absorber on the other side. Some examples are listed in Appendix~C.}

Modeling the wave generation process in a hydrodynamic laboratory is useful for ship testing as if the ship is sailing in the open ocean.
For example, a ship that plans to sail in the Javanese Sea should be tested with the ocean waves characterizing the ones in the Javanese Sea.
Keeping this in mind, we could obtain information on how strong the ocean waves are and this translates to building a strong ship that could handle the pressure from those ocean waves.

\subsection{Problem Formulation}

Based on the above background, we propose a mathematical model for ocean waves using the wavemaker theory utilized in a laboratory.
We would like to investigate how large we should deviate the wavemaker if we wish a wave profile with particular wave height. 
Additionally, we would like to examine the resulting wave profile if we move the wavemaker with a particular frequency.
Hence, we could formulate a mathematical model for water wave evolution in a towing tank.

\section{Study Coverage}

Based on the above problem formulation, the study coverage of this report is proposing a simple mathematical model of a two-dimensional flap-type wavemaker in a towing tank.
We will adopt relevant assumptions to simplify the problems. We only consider the output of monochromatic waves. Using computer software, a simple simulation describing the wave evolution will be presented.

\section{Purpose}

A subjective purpose of writing this final project report is fulfilling the requirement for completing an undergraduate degree at the Department of Mathematics, Faculty of Mathematics and Natural Sciences, Bandung Institute of Technology. The meeting for the degree conferring was held on 24 January 2001.

An objective purpose of this writing is to deepen the study of Mathematics, particularly its applications in physical problems.
Additionally, by modeling problem from outside Mathematics, the insight regarding the interconnection of science and technology will also be expanded.

\section{Basic Assumptions}

The proposed model depends on the following factors:
\begin{itemize}[leftmargin=1.2em]
\item The assumption of fluid characteristics leads to mathematical formulation.
\item There exists a governing differential equation with several boundary conditions.
\item An assumption regarding the type and the characteristics of the output waves.
\end{itemize}

\section{Hypothesis}

If we adopt an assumption that water is an ideal fluid, then we could obtain a partial differential equation with several boundary conditions as a governing differential equation. If we also adopt another assumption that the output wave is monochromatic and periodic, then we could propose a linear, two-dimensional wave generation theory.

\section{Research Methodology}

\subsection{Methodology}

We implement the theoretical and analytical methodology. This means analyzing theoretically a model related to the wave generation and deriving equations related to this model. We also implement computational analysis methodology to analyze and solve problems using a computer.

\subsection{Data Collection}

Data collection includes literature study and regular meetings with the supervisor. A literature study covers independent learning references related to Fluid Mechanics and Mathematical Modeling, particularly related to wave generation theory. The list of references can be found in the References for this report. The supervising activity includes reading assignments, discussion and problem-solving. Presentations were also conducted regularly. 

\section{Discussion Flow}

After this introduction, we will discuss the characteristics of ideal fluid flow. Chapter~\ref{ideal} also explains the physical characteristics of the fluid, the derivation of the continuity equation, Euler's equation, and Bernoulli's equation using the Laws of Mass and Momentum Conservation, as well as using the Transport Theorem. Several assumptions and equations covered in this chapter will be utilized in the subsequent chapters.

Chapter~\ref{wavemaker}, the main part of this report, covers the problem formulation and the study coverage. We will not only discuss the simplified wave generation theory but also more realistic linear theory by solving the Laplace equation as the governing equation with a lateral boundary condition at the wavemaker, a boundary condition at the bottom of the towing tank, and boundary conditions at the surface of the water. The latter includes kinematics and dynamic boundary conditions.

While the theoretical approach and derivation dominate the preceding chapters, Chapter~\ref{numeric} explains computer simulation using {\slshape Maple V Release 5}. We calculate the values of progressive and traveling wavenumbers. We will display some figures related to the movement of the wavemaker and the produced monochromatic wave profiles.

The final chapter concludes this final project report. We also provide practical suggestions for friends and anyone interested to continue research in this area.

\chapter{Ideal Fluid Flow}    \label{ideal}

\section{Fluid Physical Characteristics}

Due to its ability to move, it is interesting to observe fluids.
A fluid is a substance that does not experience resistance when it experiences deformation and will continue to change when it is given pressures or forces.
It does not have a regular shape and always follows the space it places.
Under the influence of a force, a fluid will deform continually to form a flow.

Fluids can be categorized as liquid and gas. A liquid has a characteristic of relatively incompressible and possesses a free surface.
A gas has a characteristic of readily compressible and does not possess a free surface.
In this chapter, we will discuss liquid fluid, particularly water.
The property of water in our context belongs to the ideal fluid, i.e., incompressible and inviscid fluid.
This choice of assumptions will be used as a fundamental flow problem that we will discuss later~\cite{rs,wlh}.

\section{The Laws of Mass and Momentum Conversation}

The laws of conservation for a system of particles in Physics can also be applied to fluids since fluids are a collection of particles.
Based on this, we focus on a collection of fluid particles or a fluid material volume so that we always examine an identical particle group.

For easier symbolic writing, we express the Cartesian coordinates as the index $(1,2,3)$ with a convention that $x = x_{1}$, $y = x_{2}$, and $z = x_{3}$.
The same convention also applies for the velocity components, i.e. $u = u_1$, $v = u_2$, and $w = u_3$.

\subsection{The Law of Conservation of Mass}

The law of conservation of mass states that the total mass of a system composed of a collection of particles is always constant, it neither increases nor decreases.

In other words, there exists no fluid mass that can be created not annihilated.
The change of mass within a domain is due to the mass flow passing through a boundary.
Based on this restriction, we define fluid volume $V(t)$ bounded by a surface $S$.
If the fluid has a density $\rho$, then the total fluid mass in that volume is given by the triple integral
$\iiint \rho \, \mathrm{d}V$. In our formulation, the law of mass conservation states that the total fluid mass with density $\rho$ and volume $V$ is constant.
Stated differently, this law gives a requirement that the triple integral is constant:
\begin{equation}
\iiint\limits_{V} \rho \, \mathrm{d}V = \textmd{constant}
\end{equation}
or
\begin{equation}
\frac{d}{dt} \left(\iiint\limits_{V} \rho \, \mathrm{d}V \right) = 0. \label{ms}
\end{equation}

\subsection{The Law of Conservation of Momentum}

The law of conservation of momentum states that the total momentum of a system composed by a collection of interacting particles is constant, as long as there is no external force acting to the system.

Similarly, the density of fluid-particle momentum is $\rho\,\mathbf{u}$ with components $\rho\,u_{i}$.
In our context, the law of conservation of momentum requires that the total forces acting on a fluid volume equals the rate of change of the fluid momentum.
According to the Newtonian frame of reference, we can express the law of conservation of momentum as follows:
\begin{equation}
\frac{d}{dt}\left(\iiint\limits_{V} \rho \, u_i \, \mathrm{d}V \right) = \iint\limits_{S} \tau_{ij} \; n_{j} \; \mathrm{d}S + \iiint\limits_{V} F_i \, \mathrm{d}V, \label{mt}
\end{equation}
where $\rho\,u_{i}$ denotes the component of fluid particles momentum density, $\tau_{ij}$ denotes the pressure tensor acting on the fluid volume, $n_{j}$ denotes the unit normal vector component on the fluid surface, and $F_{i}$ denotes an external force working on the fluid particles. The surface integral is the $i$-th component from the surface forces acting on the surface $S$ and the final triple integral is the sum from external body force, including the gravitational force.

Using the Divergence Theorem,\footnote[1]{In Western countries, this theorem is known as Gauss Theorem, while in the Eastern Block territory, the theorem is known as Ostrogradsky Theorem, based on the name of one Russian mathematician~\cite{tgb}.} the first term on the right-hand side of~\eqref{mt} can be expressed as a triple integral of a divergence of a vector field covering the corresponding surface
\begin{equation}
\iint\limits_{S} \mathbf{Q}\cdot\mathbf{n} \, \mathrm{d}S = \iiint\limits_{V} \nabla \cdot \mathbf{Q} \, \mathrm{d}V.  \label{tdg}
\end{equation}
Written in component form, this equation can be expressed as
\begin{equation}
\iint\limits_{S} Q_{i} \, n_{i} \, \mathrm{d}S = \iiint\limits_{V} \frac{\partial Q_{i}}{\partial x{i}} \, \mathrm{d}V.  \label{td}
\end{equation}
Here, $\mathbf{Q}$ denotes an arbitrary continuous and differentiable vector field in the volume $V$ and the unit normal vector $\mathbf{n}$ denotes the normal vector going in the outer direction from $V$ on the surface $S$.

Using~\eqref{td} to transform the surface integral to~\eqref{mt}, we obtain
\begin{equation}
\frac{d}{dt}\left(\iiint\limits_{V} \rho \, u_i \, \mathrm{d}V \right) = \iiint\limits_{V} \left(\frac{\partial \tau_{ij}}{\partial x_{j}} + F_{i} \right) \, \mathrm{d}V. \label{tr}
\end{equation}
Equations~\eqref{ms} and~\eqref{tr} express the Laws of Conservation of Mass and Momentum for the fluid integrated over the arbitrary material volume $V(t)$, respectively.

\section{Transport Theorem}

Let a general form of the volume integral be expressed as follows:
\begin{equation}
I(t) = \iiint\limits_{V(t)} f(\mathbf{x}, t) \, \mathrm{d}V.
\end{equation}
Here, $f$ denotes an arbitrary differentiable scalar function that depends on the position $\mathbf{x}$ and time $t$ integrated over the volume $V(t)$, for which the latter can also change in time. Hence, the surface $S$ of the volume boundary will change in time as well and its normal velocity is denoted by $U_n$.

Using a common techniques employed in basic Calculus, consider the following difference:
\begin{equation}
\Delta I = I(t + \Delta t) - I(t) = \iiint\limits_{V(t+\Delta t)} f(\mathbf{x}, t + \Delta t) \, \mathrm{d}V - \iiint\limits_{V(t)} f(\mathbf{x}, t) \, \mathrm{d}V. \label{di}
\end{equation}
From the Taylor series of $f(\mathbf{x},t+ \Delta t)$, we have
\begin{equation}
f(\mathbf{x}, t + \Delta t) = f(\mathbf{x}, t) + \Delta t \, \frac{\partial f(\mathbf{x}, t)}{\partial t} + 
                              \frac{1}{2!} \, (\Delta t)^2 \,\frac{\partial^2 f(\mathbf{x}, t)}{\partial t^2} + \cdots.
\end{equation}
Neglecting the terms proportional to the order of $(\Delta t)^2$ and higher, we attain
\begin{equation}
f(\mathbf{x}, t + \Delta t) = f(\mathbf{x}, t) + \Delta t \, \frac{\partial f(\mathbf{x}, t)}{\partial t}.
\end{equation}
We can apply a similar analysis to the material volume $V(t)$
\begin{equation}
V(t + \Delta t) = V(t) + \Delta t \, \frac{dV}{dt} + \frac{1}{2!} \, (\Delta t)^2 \, \frac{d^2V}{dt^2} + \cdots.
\end{equation}
Neglecting the higher-order terms, we obtain
\begin{equation}
\Delta V = V(t + \Delta t) - V(t).
\end{equation}
Thus, $V(t + \Delta t)$ differs from $V(t)$ a thin volume $\Delta V$ included insides the boundary surfaces $S(t + \Delta t)$ and $S(t)$ and is proportional with $\Delta t$.
From~\eqref{di}, we obtain
\begin{align}
\Delta I &= \iiint\limits_{V + \Delta V} \left(f + \Delta t \, \frac{\partial f}{\partial t} \right) \, \mathrm{d}V - \iiint\limits_{V} f \, \mathrm{d}V \\
         &= \Delta t \iiint\limits_{V} \frac{\partial f}{\partial t} \mathrm{d}V + \iiint\limits_{\Delta V} f \, \mathrm{d} V + {\mathcal{O}} \left[(\Delta t)^2 \right],
\end{align}
where the final term denotes the second order error proportional to $(\Delta t)^2$. 

To evaluate the integral over the tiny volume $\Delta V$, we notice that this thin solid has the same thickness as the distance between $S(t)$ and $S(t + \Delta t)$.
This thickness equals to the normal component from the distance covered by $S(t)$ in time $\Delta t$, which is the product $U_n \, \Delta t$.
Thus, the integral of the second term only contributes to the first order, which is proportional to $\Delta t$.
The degree of accuracy of the integrand function $f$ can be assumed to be constant throughout the thin solid in the direction of the normal from the surface $S$.
Integrating only in this direction, we have the following:
\begin{equation}
\Delta I = \triangle t \iiint\limits_{V} \frac{\partial f}{\partial t} \, \mathrm{d}V + \iint\limits_{S} (U_{n} \, \triangle t) \, f \, \mathrm{d}S + 
		   {\mathcal{O}} \left[(\triangle t)^2 \right].
\end{equation}
Finally, we obtain the desired result by dividing both sides of the equation by $\Delta t$ and take the limit by allowing $\Delta t$ approaching zero
\begin{equation}
\frac{\mathrm{d}I}{\mathrm{d}t} = \lim_{\Delta t \rightarrow 0} \, \frac{\triangle I}{\triangle t} = \iiint\limits_{V} \frac{\partial f}{\partial t} \, \mathrm{d}V + 
				\iint\limits_{S} f \, U_{n} \, \mathrm{d}S.  \label{tt}
\end{equation}

Equation~\eqref{tt} is known as the Transport Theorem or the Transport Equation.
The surface integral in this equation states the transport quantity $f$ moving out from the volume $V$ as a result of the moving boundary.
In the special case when $S$ is fixed and $U_n = 0$, equation~\eqref{tt} reduces to a simple form where the differential operator can be pulled out from the integral sign.
For a complete explanation, please consult~\cite{njn}.

We have another interesting case. Since the material volume $V$ is always composed by identical fluid particles, then the surface $S$ moves with the same normal velocity with the one of the fluid itself and $U_{n} = \mathbf{u} \cdot \mathbf{n} = \mathbf{u}_{i} \, \mathbf{n}_{i}$.
In this case, applying the Divergence Theorem to equation~\eqref{tdg}, then we can express equation~\eqref{tt} in the following form:
\begin{align}
\frac{\mathrm{d}I}{\mathrm{d}t} 
&= \frac{\mathrm{d}}{\mathrm{d}t} \left(\iiint\limits_{V(t)} f \, \mathrm{d}V \right) \nonumber \\
&= \iiint\limits_{V(t)} \frac{\partial f}{\partial t} \, \mathrm{d}V + \iint\limits_{S} f\, u_{i} \, n_{i} \, \mathrm{d}S \\
&= \iiint\limits_{V(t)} \frac{\partial f}{\partial t} \, \mathrm{d}V + \iiint\limits_{V(t)} \frac{\partial}{\partial x_{i}} \, \left(f \, u_{i} \right) \, \mathrm{d}V \\
&= \iiint\limits_{V(t)} \left[\frac{\partial f}{\partial t} + \frac{\partial}{\partial x_{i}} \, \left(f \, u_{i} \right) \right] \, \mathrm{d}V.  \label{fdv}
\end{align}

\section{Continuity Equation}     \label{kontinuitas}

This equation is closely related to the Law of Conservation of Mass and the Transport Theorem since it can be derived from these two equations.
Look again equation~\eqref{ms} for the Law of Conservation of Mass. Using the result of the Transport Theorem~\eqref{fdv}, we obtain
\begin{equation}
\frac{\mathrm{d}}{\mathrm{d}t} \left(\iiint\limits_{V} \rho \, \mathrm{d}V \right) = \iiint\limits_{V} \left[\frac{\partial \rho}{\partial t} + 
\frac{\partial}{\partial x_{i}} \, \left(\rho \, u_{i} \right) \right] \, \mathrm{d}V = 0.
\label{ce}
\end{equation}
Since the last integral is evaluated at a fixed instantaneous time, the difference that $V$ is material volume is not necessary at this stage.
Furthermore, that particular volume can be composed of a group of arbitrary fluid particles.
Hence, the integrand above is identically zero throughout the whole fluid.
Therefore, the volume integral in~\eqref{ce} can be replaced by a partial differential equation expressing the Law of Conservation of Mass in the following form:
\begin{equation}
\frac{\partial \rho}{\partial t} + \frac{\partial}{\partial x_{i}} \, (\rho \, u_{i}) = 0.
\end{equation}
Written in a three-dimensional component form, this equation reads
\begin{equation}
\frac{\partial \rho}{\partial t} + \frac{\partial \rho u}{\partial x} + \frac{\partial \rho v}{\partial y} + \frac{\partial \rho w }{\partial z} = 0, 	\label{ko}
\end{equation}
or written in a differential form
\begin{equation}
\frac{\partial\rho}{\partial t} + \nabla \cdot (\rho \, \mathbf{u}) = 0.
\end{equation}

This important equation is known as the condition for the Law of Conservation of Mass or the continuity equation for incompressible fluid flow.
Equation~\eqref{ko} describes the mean rate of change of mass density at a fixed point as a result of the change in the mass velocity vector $\rho \, \mathbf{u}$.
By expanding the terms containing the product of density and the velocity components, we can derive a different form of the continuity equation
\begin{align}
\frac{\partial \rho}{\partial t} + u\,\frac{\partial \rho}{\partial x} + \rho\, \frac{\partial u}{\partial x} + v \, \frac{\partial \rho}{\partial y} + 
\rho\,\frac{\partial v}{\partial y} + w \, \frac{\partial \rho}{\partial z} + \rho \,\frac{\partial w}{\partial z} &=0 \\
\left(\frac{\partial \rho}{\partial t} + u \,\frac{\partial \rho}{\partial x} + v\,\frac{\partial \rho}{\partial y} + w\,\frac{\partial \rho}{\partial z} \right) + 
\rho \, \left(\frac{\partial u}{\partial x} + \frac{\partial v}{\partial y} + \frac{\partial w}{\partial z} \right) &=0.  \label{eks}
\end{align}
Using the material or the total derivative operator (See Appendix~A), equation~\eqref{eks} can be written as follows:
\begin{equation}
\frac{D\rho}{Dt} + \rho \, (\nabla \cdot \mathbf{u}) = 0.  \label{mat}
\end{equation}

For a steady flow, i.e., a time-independent flow, it follows that $\frac{\partial \rho}{\partial t} = 0$ and hence the continuity equation reduces to 
$\nabla \cdot(\rho \, \mathbf{u}) = 0$. Using the assumption that water is incompressible fluid, i.e., having a constant mass rate $\rho$, then the continuity equation~\eqref{eks} reduces to
\begin{equation}
\frac{\partial u}{\partial x} + \frac{\partial v}{\partial y} + \frac{\partial w}{\partial z} = 0.
\end{equation}
We obtain the following expression in the vector form, which can also be derived directly from equation~\eqref{mat}:
\begin{equation}
\nabla \cdot \mathbf{u} = 0.  \label{vpk}
\end{equation}
This relationship is known as the condition of incompressibility. This condition states the fact that the balance between outflow and inflow for a volume element or material volume is zero for all time. 

\section{Euler's Equation}

While the continuity equation is related to the Law of Conservation of Mass, Euler's equation is related to the Law of Conservation of Momentum and the Transport Theorem since its derivation makes use of these two equations.
Look again equation~\eqref{tr} for the Law of Conservation of Momentum.
Using the result from the Transport Theorem~\eqref{fdv}, we obtain
\begin{equation}
\iiint\limits_{V} \left[\frac{\partial \rho u_{i}}{\partial t} + \frac{\partial}{\partial x_{j}} \, \left(\rho \, u_{i} \,u_{j} \right) \right] \, \mathrm{d}V = 
\iiint\limits_{V} \left(\frac{\partial \tau_{ij}}{\partial x_{j}} + F_{i} \right) \, \mathrm{d}V.  		\label{eu1}
\end{equation}
Since these integrals are calculated over an identical volume, then the equation for the integrand parts of~\eqref{eu1} must be satisfied as well, it is given as follows:
\begin{equation}
\frac{\partial}{\partial t}(\rho \, u_{i}) + \frac{\partial}{\partial x_{j}} \, \left(\rho \, u_{i} \, u_{j} \right) = \frac{\partial \tau_{ij}}{\partial x_{j}} + F_{i}. \label{eu2}
\end{equation}
Next, by expanding the derivatives of the left-hand side of~\eqref{eu2} using the Chain Rule, we obtain
\begin{equation}
u_{i} \frac{\partial \rho}{\partial t} + \rho \frac{\partial u_{i}}{\partial t} + u_{i} u_{j} \frac{\partial \rho}{\partial x_{j}} + 
\rho u_{j} \frac{\partial u_{i}}{\partial x_{j}} + \rho u_{i} \frac{\partial u_{j}}{\partial x_{j}} = \frac{\partial \tau_{ij}}{\partial x_{j}} + F_{i}.
\end{equation}
Using the assumption that the observed fluid is incompressible and possesses constant mass density $\rho$, using the result of the continuity equation~\eqref{vpk}, we obtain Euler's equation,\footnote[1]{Euler, Leonhard (1707-83). \ Swiss mathematician with a calm mind who fundamentally and meaningfully contributed to various branches of mathematics and their applications, including but not limited to, differential equations, infinite series, complex analysis, mechanics and hydrodynamics, as well as calculus of variations. He was also influential in promoting the use and the understanding of analysis.~\cite{jrs}} also known as the momentum equation
\begin{equation}
\frac{\partial u_{i}}{\partial t} + u_{j} \, \frac{\partial u_{i}}{\partial x_{j}} = \frac{1}{\rho} \, \left(\frac{\partial \tau_{ij}}{\partial x_{j}} + F_{i} \right).  \label{tum}
\end{equation}
Using the material derivative operator (see again Appendix A), Euler's equation~\eqref{tum} can be expressed as follows:
\begin{equation}
\frac{\mathrm{D} u_{i}}{\mathrm{D}t} = \frac{1}{\rho} \, \left(\nabla \cdot \tau_{i} + F_{i} \right).
\end{equation}

\section{Bernoulli's Equation} 			\label{bernoulli}

Consider again equation~\eqref{tum}. In a frictionless flow, there exists neither shear stress nor normal stress for isotropic fluid flow.
We adopt a convention that the normal stresses $\tau_{11}, \tau_{22}$, and $\tau_{33}$ have positive orientation if they are tensions.
Since we assume that water is inviscid fluid, then the stress tensor $\mathbf{\tau}$ has only the normal components from the pressure.
For further explanation, please consult~\cite{hwf,kjp,njn}. We set as follows
\begin{equation}
\tau_{11} = \tau_{22} = \tau_{33} = -\,p,
\end{equation}
so that the momentum equation becomes
\begin{equation}
\frac{\partial u_{i} }{\partial t} + u_{j} \, \frac{\partial u_{i}}{\partial x_{j}} = \frac{1}{\rho} \, \left(\frac{\partial p}{\partial x_{j}} + F_{i} \right),
\end{equation}
or in a vector notation
\begin{equation}
\frac{\mathrm{D} \mathbf{u}}{\mathrm{D}t} = \frac{1}{\rho} \, \left(-\nabla p + \mathbf{F} \right),
\end{equation}
where $\mathbf{u} = (u,\,v,\,w)$, \ $\mathbf{F} = (0,\, 0,\, -\rho \,g)$, and $p$ is the pressure. Hence,
\begin{equation}
\frac{\partial \mathbf{u}}{\partial t} + (\mathbf{u} \cdot \nabla) \, \mathbf{u} = -\nabla \, \left(\frac{p}{\rho} + g \,z \right).
\end{equation}
From a vector identity, we have
\begin{equation}
\mathbf{u} \times (\nabla \times \mathbf{u}) = (\mathbf{u} \cdot \nabla) \, \mathbf{u} - \nabla \, \left(\frac{1}{2} \left|\mathbf{u} \right|^2 \right).
\end{equation}
If we also assume that water is irrotational fluid,\footnote[1]{Irrotational fluid in this context refers to the absence of eddy or whirlpool.}, then $\nabla \times
\mathbf{u} = \mathbf{0} \label{curl}$, and thus $(\mathbf{u} \cdot \nabla) \, \mathbf{u} = \nabla \, \left(\frac{1}{2} \left|\mathbf{u} \right|^2 \right)$.
Due to this assumption, the vector velocity $\mathbf{u}$ can be expressed as the gradient of the scalar or velocity potential, i.e., $\mathbf{u} = \nabla\,\phi$.
(For a detailed explanation, please see~\cite{drg} and~\cite{lld}.) The reason we implemented this simplification is for easier analysis since in general, scalar quantities are less complicated to investigate than vector quantities.

Therefore,
\begin{equation}
\frac{\partial \, \nabla \phi}{\partial t} + \nabla \, \left(\frac{1}{2} \left|\mathbf{u} \right|^2 \right) = -\nabla \, \left(\frac{p}{\rho} + g\,z \right),
\end{equation}
or
\begin{equation}
\nabla \, \left(\frac{\partial \phi}{\partial t} + \frac{1}{2} \left|\mathbf{u} \right|^2 + \frac{p}{\rho} + g\,z \right) = 0.  \label{ber1}
\end{equation}
If we integrate equation~\eqref{ber1} throughout the entire space, then we will obtain Bernoulli's equation\footnote[7]{Bernoulli, Daniel (1700-82). \ This Dutch-born scientist is a member of a famous Swiss family who consists of ten mathematicians (fathers, sons, uncles, cousins). He is well-known for his works on fluid flow and the kinetic theory of gases. His equation for fluid flow was first published in 1783. He also pursued the study and research of astronomy and magnetism and was the first scientist who solved the Riccati equation~\cite{jrs}.} 
\begin{equation}
\frac{\partial \phi}{\partial t} + \frac{1}{2} \left|\mathbf{u} \right|^2 + \frac{p}{\rho} + g\,z = \textmd{constant} \equiv f(t).  \label{ber2}
\end{equation}
\newpage
\tikzstyle{cloud} = [draw = red!50, ellipse, fill=red!30, text width=4.5cm, line width=1.0mm, minimum height=3em]
\tikzstyle{block} = [rectangle, draw = blue!50, fill=blue!30, text width=6cm, text centered, rounded corners, line width=1.0mm, minimum height=3em]
\tikzstyle{trap} = [trapezium, draw = green!80!blue, trapezium right angle = 117, fill=green!40, text width=6.5cm, text centered, rounded corners, line width=1.0mm, minimum height=3em]
\tikzstyle{line} = [draw, very thick, -latex']
\begin{center}
A schematic diagram for an ideal fluid interconnectivity. \vspace*{0.5cm} \\
\begin{tikzpicture}[auto, node distance=2cm,>=latex']
\hspace*{1cm}
\node [cloud](mass){\sffamily Mass Conservation Law};
\node [block, right of=mass, node distance = 8cm](momentum){\sffamily Momentum Conservation Law};
\hspace*{-4cm} \node [trap, below of=momentum, node distance = 2.5cm](transport){\sffamily Transport Equation};
\hspace*{ 4cm} \node [cloud, below of=mass, node distance = 5cm](continuity){\sffamily \quad Continuity equation};
\node [block, right of=continuity, node distance = 8cm](euler){\sffamily Euler's equation};
\node [block, below of=euler, node distance = 2.5cm](bernoulli){\sffamily Bernoulli's equation};
\path [line,dashed] (momentum) -- node {$+$} (transport);
\path [line] (transport) -- (euler);
\path [line] (euler) -- (bernoulli);
\draw [->, very thick, dashed] (0,-0.7cm) -- node {$+$} (0,-1.8cm);
\draw [->, very thick] (0,-3.15cm) -- (0,-4.3cm);
\end{tikzpicture}
\end{center}

\chapter{Two-Dimensional Wavemaker Theory}   \label{wavemaker}

In this chapter, we are going to compare two wavemaker theories in which the corresponding mathematical models have been proposed and investigated. The first part is a simplified theory. A basic idea comes from a displaced volume of water by a flap-type wavemaker in a towing tank. The second part covers a more detailed theory for the flap-type wavemaker. From these two theoretical approaches, we will observe the ratio between the wave amplitude~$H$ with the maximum stroke of the wavemaker~$S$.

In the context of this report, the more detailed theory is based on the linear wave theory developed by Airy\footnote[1]{Airy, Sir George Biddle (1801--1892). \ English mathematician and physicist, who became a Royal Astronomer for 46 years. He contributed not only to the theory of light and astronomy but also to gravity, magnetism, sound, wave propagation, and tidal wave~\cite{jrs}.} around 160 years ago, when he analyzed the behavior of ocean waves~\cite{aph}.

\section{Simplified Wavemaker Theory}

\begin{figure}[h]
\begin{center}
\includegraphics[scale = 3]{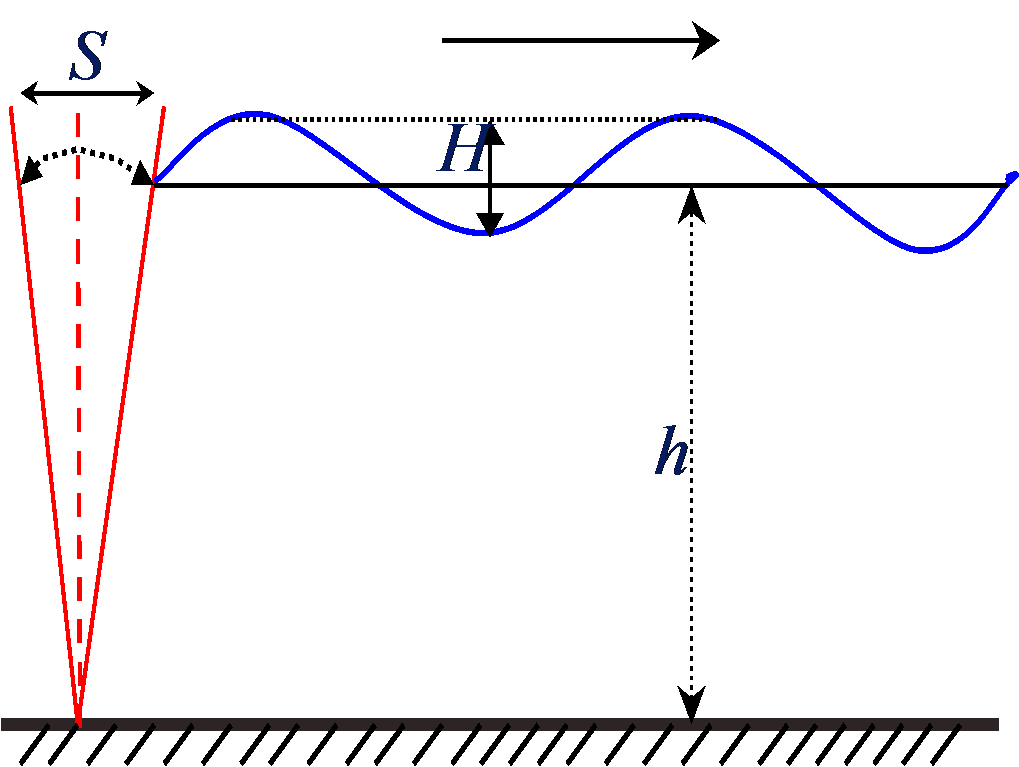}
\end{center}
\caption{A schematic diagram for a flap-type wavemaker.}             \label{flapscheme}
\end{figure}

For shallow-water waves, simple theory for the propagation of a wave profile produced by a wavemaker was first developed by Galvin in 1964. He reasoned that the displaced water by a wavemaker equals the volume of the top part from the propagated wave. Consider a flap-type wavemaker with a fixed bottom end and possesses a maximum stroke~$S$. The water depth in the towing tank is~$h$. The displaced water volume by the flap deviation is $\frac{1}{2}\,S\,h$. See Figure~\ref{flapscheme}. Meanwhile, the top part of water volume is $\int_{0}^{\frac{1}{2}\,L} \frac{1}{2}\,H \sin (k\,x) \, \mathrm{d}x$, where $k = 2\pi/L$ is the wavenumber. Hence,
\begin{align}
\int_{0}^{\frac{1}{2}\,L} \frac{1}{2} H \sin (k\,x) \, \mathrm{d}x
&= \int_{0}^{\frac{1}{2}\,L} \frac{1}{2} H \sin \left(\frac{2\, \pi}{L} x \right) \, \mathrm{d}x \\ 
&= \frac{H\,L}{2\,\pi } \\ 
&= \frac{H}{k}
\end{align}
By equating the two volumes, we have
\begin{equation*}
\frac{1}{2}\,S\,h = \frac{H}{k} = H\,\frac{L}{2\,\pi} = \frac{H}{2} \left(\frac{L}{2} \right) \, \frac{2}{\pi}
\end{equation*}
where the factor $2/\pi$, also known as the area factor, indicates the ratio between the area formed by the wave profile with the rectangle circumscribed it. From the relationship $\frac{1}{2}\,S\,h = \frac{H}{k}$, we can find out the ratio between the wave height~$H$ and the stroke~$S$, namely
\begin{equation*}
\frac{H}{S} = \frac{1}{2} \,k\,h.
\end{equation*}
This relationship is only valid for the shallow-water waves, i.e., for $ k\,h < \frac{1}{10}\pi$.

\section{Linear Wavemaker Theory}

According to the assumption that water is an incompressible fluid, then from the result of equation~\eqref{vpk} in Section~\ref{kontinuitas}, we have $\nabla \cdot \mathbf{u} = 0$. Additionally, another adopted assumption is the absence of water whirlpool, in other words, the water flow is irrotational. Based on this assumption, from the explanation in Section~\ref{bernoulli}, we have obtained $\mathbf{u} = \nabla\,\phi$. Combining these two equations, we obtain $\nabla \cdot \nabla \,\phi = \nabla^{2}\phi = 0$. This equation is known as Laplace's equation\footnote[1]{Laplace, Marquis Pierre Simon de (1749-1827). \ French mathematical-physicist who contributed to the study of celestial mechanics, particularly in explaining the orbit of planets Jupiter and Saturn. He also developed an idea in using potential and orthogonal functions, as well as introduced their integral transformations. He also played an important role in the development of probability theory.}, and it plays a role as a governing differential equation for the velocity potential $\phi$. In the two-dimensional Cartesian coordinates--the $x$- and $z$-axes in the horizontal and vertical directions, respectively--Laplace's equation can be written as follows:
\begin{equation}
\nabla^2\phi = \frac{\partial^2\phi}{\partial x^2} + \frac{\partial^2\phi}{\partial z^2} =0,
\end{equation}
and it is valid for $-h \leq z \leq \eta(x,t)$, and $0 \leq x \leq \infty$.

The boundary value problem in this context is the one for two-dimensional wave propagation in an ideal fluid. We will solve the problem using several boundary conditions that are suitable for the situation in the towing tank. We have at least three boundary conditions as auxiliary equations in solving the governing equation. What follows is the boundary conditions related to the physical condition of the towing tank.

\subsection{Lateral boundary condition at the wavemaker\\ (\emph{pseudo-boundary})}

Let the function describing a horizontal movement at the wavemaker surface be given as follows:
\begin{equation}
F(x,\,z,\,t) = x - \frac{1}{2} \, S(z) \, \sin (\omega \, t) = 0,
\end{equation}
then by applying a total differential to $F(x,\,z,\,t)$, we obtain
\begin{equation}
\frac{\mathrm{D}F}{\mathrm{D}t} = \frac{\partial F}{\partial t} + \frac{\partial F}{\partial x} \frac{\mathrm{d}x}{\mathrm{d}t} + \frac{\partial F}{\partial z} \frac{\mathrm{d}z}{\mathrm{d}t} 
\end{equation}
\begin{align}
0 &= -\frac{1}{2} \, \omega \, S(z) \, \cos (\omega \, t) + v - \frac{1}{2} \, w \, \frac{\mathrm{d}S(z)}{\mathrm{d}z} \, \sin (\omega \, t) \\
v - \frac{1}{2} \, w \, \frac{\mathrm{d}S(z)}{\mathrm{d}z} \, \sin (\omega \, t) &= \frac{1}{2} \, \omega \, S(z) \, \cos (\omega \, t) \qquad 
\textmd{on} \quad F(x,\,z,\,t) = 0   \label{phix}
\end{align}
where $v$ and $w$ are the velocity components in the $x$ and $z$ directions, respectively. (Remember that $\mathbf{u} = (v, w)$ is the two-dimensional velocity vector.)
Equation~\eqref{phix} can also be written as follows:
\begin{equation}
\phi_{x} = \frac{1}{2} \left[\phi_{z} \,S\,^{\prime}(z) \, \sin (\omega\, t) + \omega \, S(z) \, \cos \: (\omega\, t) \right].
\end{equation}

For the sufficiently small stroke movement $S(z)$ and stroke velocity, we could linearize equation~\eqref{phix} by neglecting the second term of the left-hand side. Similar to what is applied to a free surface, we can express conditions at the moving lateral boundary in terms evaluated at the mean position, i.e., at $x = 0$. We proceed it by expanding the condition as a truncated Taylor series, or more precisely, in a truncated Maclaurin series
\begin{align}
\displaystyle \left. \left(\phi_{x} - \frac{1}{2} \, \omega \, S(z) \, \cos (\omega\,t) \right) \right|_{x = \frac{1}{2} \, S(z) \, \sin (\omega\,t)} = \left.\left(\phi_{x} - \frac{1}{2} \, \omega \, S(z) \, \cos (\omega\, t) \right) \right|_{x=0} & \nonumber \\
+ \frac{1}{2} \, S(z) \, \sin (\omega\,t) \, \frac{\partial}{\partial x} \left. \left(\phi_{x} - \frac{1}{2} \, \omega \, S(z) \, \cos (\omega\, t) \right) \right|_{x=0} & + \dots.
\end{align}
It should be obvious that only the first term from the above expansion is linear in $\phi_{x}$ and $S(z)$, while other terms can be neglected since they are assumed to be tiny. Therefore, the final lateral boundary condition as a consequence of the linearization process is the following equation:
\begin{equation}
u(0,\,z,\,t) = \phi_{x} = \frac{1}{2} \, \omega \, S(z) \, \cos(\omega t).
\end{equation}

\subsection{Boundary condition at the bottom of the towing tank (kinematic condition)}

Since there is no water flowing through the bottom of the towing thank, then the fluid velocity in the vertical direction at the bottom is zero
\begin{equation}
\phi_{z} = \frac{\partial\phi}{\partial z} = 0, \qquad \textmd{at} \quad z = -h.
\end{equation}

\subsection{Boundary conditions at the water surface}

\begin{itemize}[leftmargin=1.2em]
\item \textbf{Free-surface kinematic boundary condition}
\end{itemize}

At every boundary, irrespective of whether it is free, such as at the water surface, or fixed, such as at the bottom of a pond, the fluid velocity must satisfy a number of physical requirements. Under the influence of a force, this boundary might experience a deformation. All requirements which act upon the water particle kinematics are known as kinematics boundary conditions. At each fluid surface or interface, there exists no flow passing through the interface, otherwise, the interface itself does not exist in the first place. This occurrence is obvious for the case of an impermeable fixed surface, such as a sheet pile seawall~\cite{drg}.

Let the water surface function be given as follows:
\begin{equation}
G(x,\,z,\,t) = z - \eta(x,\,t) = 0,
\end{equation}
where $\eta(x,\,t)$ denotes the wave elevation, also known as the water surface elevation, which indicates the distance of a point in a wave surface from the mean free surface.
Then, the total differential of the above function gives  
\begin{align}
\frac{\mathrm{D}G}{\mathrm{D}t} &= \frac{\partial G}{\partial t} + \frac{\partial G}{\partial x} \, \frac{\mathrm{d}x}{\mathrm{d}t} + \frac{\partial G}{\partial z} \, \frac{\mathrm{d}z}{\mathrm{d}t} \\
0 &= -\frac{\partial \eta}{\partial t} + \left(-\frac{\partial \eta}{\partial x} \right) \,v + w
\end{align}
or
\begin{equation}
\eta_{t} = w - v \,\eta_{x}.  \label{eta}
\end{equation}
Since $\mathbf{u} = (v,\,w)= (\phi_{x}, \phi_{z}) = \nabla \phi$, then equation~\eqref{eta} can be written as follows:
\begin{equation}
\eta_{t} = \phi_{z} - \phi_{x} \,\eta_{x}.
\end{equation}

Now define $\varphi(x,\,t)= \left. \phi(x,\,z,\,t)\right|_{z=\eta(x,\,t)} = \phi(x,\,\eta(x,\,t),\,t)$. By differentiating it with respect to $x$, we now have $\displaystyle \frac{\partial\varphi}{\partial x} = \frac{\partial\phi}{\partial x} + \frac{\partial\phi}{\partial\eta} \,\frac{\partial\eta}{\partial x}$. We write a notation $\frac{\partial \phi}{\partial \eta} = \gamma$. Hence, this equation becomes $\varphi_{x} = \phi_{x} + \gamma\,\eta_{x}$. By substituting $\phi_{\eta} = \frac{\partial \phi}{\partial \eta} = \gamma$ dan $\phi_{x} = \varphi_{x} - \gamma \,\eta_{x}$ to the equation above we obtain the following equation
\begin{align}
\eta_{t} &= \gamma - \left(\varphi_{x} - \gamma\, \eta_{x} \right) \, \eta_{x} = \gamma - \varphi_{x} \,\eta_{x} + \gamma \, \eta_{x}^2 \\
\eta_{t} &= \left(1 + \eta_{x}^2 \right) \, \gamma - \varphi_{x} \,\eta_{x}.
\end{align}
This is the free surface kinematic boundary condition, a condition that states the fact that there is no fluid particle passes through the free surface.

\begin{itemize}[leftmargin=1.2em]
\item \textbf{Free-surface dynamic boundary condition}
\end{itemize}

A ``free'' water surface, such as the interface between air and water, cannot sustain the pressure difference along the boundary and thus must respond to receive a uniform pressure. Dynamic boundary condition describes a pressure distribution that acts upon the boundaries such as free surface and interface.

From Bernoulli's equation for unsteady flow, we already have the equation $\displaystyle \frac{\partial \mathbf{u}}{\partial t} + \nabla \left(\frac{p}{\rho} + g\,z + \frac{1}{2} \left|\mathbf{u}\right|^2 \right) = 0$. By assuming that the free atmospheric pressure above the water layer bounded by only the free surface, this equation can be written as follows:
\begin{equation}
\frac{\partial {\phi}}{\partial t} + g\,\eta + \frac{1}{2} \left|\nabla \phi \right|^2 = 0, \qquad \textmd{at} \quad z = \eta(x,\,t).  \label{bernou}
\end{equation}

We first calculate $\frac{1}{2} \left|\nabla \phi \right|^2$ to find out its representation in the form of its partial derivatives. We also know that 
\begin{align}
\frac{1}{2} \left|\nabla \phi \right|^2 
&= \frac{1}{2} \, \left(\nabla \phi \cdot \nabla \phi \right) 
 = \frac{1}{2} \, \left[\,\left(\phi_{x}, \, \phi_{z} \right) \cdot \left(\phi_{x}, \, \phi_{z} \right) \,\right]  \nonumber \\
&= \frac{1}{2} \, \left(\phi_{x}^2 + \phi_{z}^2 \right) = \frac{1}{2} \,\left(\phi_{x}^2 + \gamma^2 \right).
\end{align}
By substituting $\phi_{x} = \varphi_{x} - \gamma \, \eta_{x}$ to this equation, we arrive at 
\begin{align}
\frac{1}{2} \left|\nabla \phi \right|^2 
&= \frac{1}{2} \left[\left(\varphi_{x} - \gamma \, \eta_{x} \right)^2 + \gamma^2 \right]  \nonumber \\
&= \frac{1}{2} \left(\varphi_{x}^2 - 2 \gamma \, \varphi_{x} \, \eta_{x} + \gamma^2 \, \eta_{x}^2 + \gamma^2 \right).
\end{align}
Thus, $\displaystyle \frac{1}{2} \left|\nabla \phi \right|^2 = \frac{1}{2} \, \varphi_{x}^2 - \gamma \, \varphi_{x} \, \eta_{x} + \frac{1}{2} \,\left(1 + \eta_{x}^2 \right) \, \gamma^2$.

From $\varphi(x,\,t) = \phi(x, \, \eta(x,\,t), \,t)$, a partial differentiation with respect to $t$ gives $\displaystyle \frac{\partial \varphi}{\partial t} =
\frac{\partial\phi}{\partial t} + \frac{\partial\phi}{\partial\eta} \frac{\partial\eta}{\partial t}$ \quad $\Longrightarrow$ \quad $\varphi_{t} = \phi_{t} +
\gamma \,\eta_{t}$ \quad $\Longrightarrow$ \quad $\phi_{t} = \varphi_{t} - \gamma\,\eta_{t}$.

Substituting to Bernoulli's equation~\eqref{bernou} yields
\begin{equation}
\varphi_{t} - \gamma\,\eta_{t} + \frac{1}{2}\,\varphi_{x}^2 - \gamma \,\varphi_{x} \, \eta_{x} + \frac{1}{2}\,\left(1 + \eta_{x}^2 \right) \,\gamma^2 + g\,\eta = 0.
\end{equation}
To obtain a simplified form, we substitute $\eta_{t} = \left(1 + \eta_{x}^2 \right) \,\gamma - \varphi_{x}\,\eta_{x}$ which is obtained from the kinematic boundary condition. Consequently,
\begin{align}
\varphi_{t} - \gamma\, \left[\left(1 + \eta_{x}^2 \right) \,\gamma - \varphi_{x}\,\eta_{x} \right] + \frac{1}{2}\,\varphi_{x}^2 - \gamma\,\varphi_{x}\,\eta_{x} + \frac{1}{2}\,\left(1 + \eta_{x}^2 \right) \,\gamma^2 + g\,\eta &= 0 \\
\varphi_{t} - \left(1 + \eta_{x}^2 \right) \,\gamma^2 + \varphi_{x}\,\eta_{x}\,\gamma + \frac{1}{2}\,\varphi_{x}^2 - \gamma\,\varphi_{x}\,\eta_{x} + \frac{1}{2} \,\left(1 + \eta_{x}^2 \right) \,\gamma^2 + g\,\eta &= 0 \\
\varphi_{t} - \frac{1}{2} \,\left(1 + \eta_{x}^2 \right) \,\gamma^2 + \frac{1}{2} \,\varphi_{x}^2 + g\,\eta &= 0.
\end{align}
Finally, we obtain $\displaystyle \varphi_{t} = \frac{1}{2} \,\left(1 + \eta_{x}^2 \right) \,\gamma^2 -\frac{1}{2} \,\varphi_{x}^2 - g\,\eta$. This equation is known as the free surface dynamic boundary condition.

For shallow-water wave, we are more interested in studying and using the linearized form of the free surface kinematic and dynamic boundary conditions since these simplify the problem when we solve the governing differential equations with the prescribed boundary conditions.

Consider again equation~\eqref{eta} as the free surface kinematic boundary condition
\begin{equation}
\frac{\partial \eta}{\partial t} =  w \, \bigg|_{z = \eta} - \frac{\partial \eta}{\partial x} \, (v) \bigg|_{z=\eta}
\end{equation}
or
\begin{equation}
\frac{\partial \eta}{\partial t} = \left. \left(\frac{\partial \phi}{\partial z} \right) \right|_{z = \eta} - \frac{\partial \eta}{\partial x} \, \left. \left(\frac{\partial \phi}{\partial x} \right) \right|_{z = \eta}.
\end{equation}
Neglecting the nonlinear terms, we obtain a simplified form of the free surface kinematic boundary condition, given as follows:
\begin{equation}
\left. \left(\frac{\partial \phi}{\partial z} \right) \right|_{z = 0} = \frac{\partial\eta}{\partial t}.
\end{equation}
Similarly, we also have a simplified form of the free surface dynamic boundary condition, given as follows:
\begin{equation}
\left. \left(\frac{\partial \phi}{\partial t} \right) \right|_{z = 0} = -g \,\eta. 			\label{permukaan}
\end{equation}

\begin{figure}[h]
\begin{center}
\includegraphics[scale = 3]{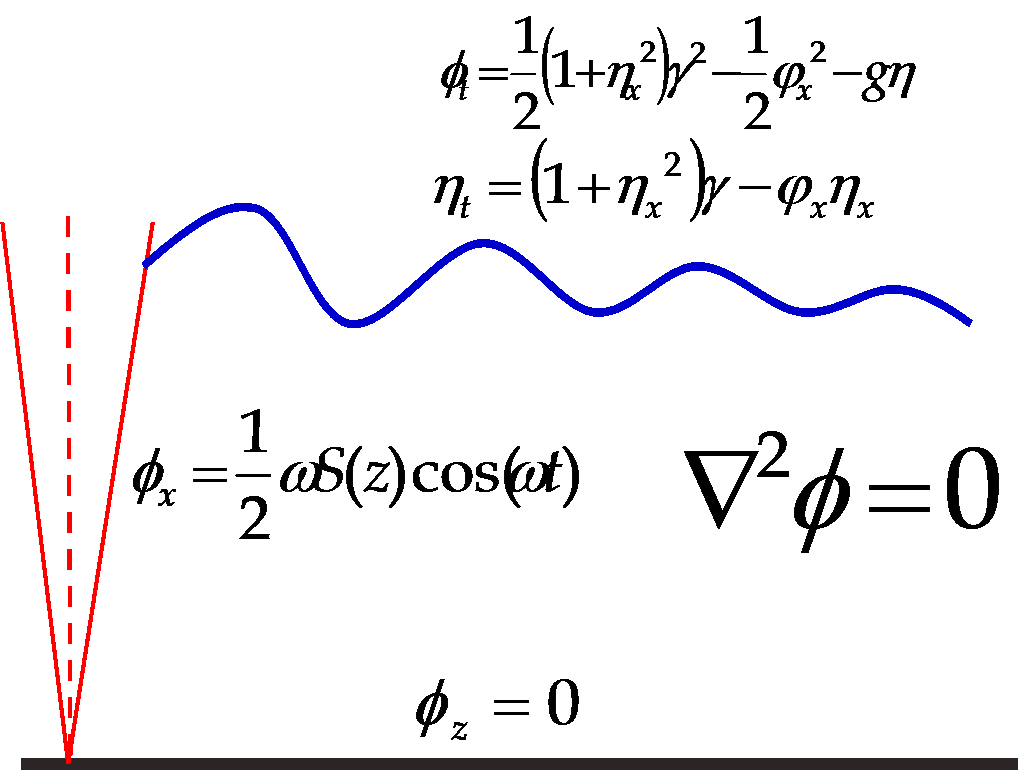}
\end{center}
\caption{A schematic diagram for a wavemaker with a governing equation and its boundary conditions.}                 \label{wavemakerscheme}
\end{figure}

A schematic diagram for a wavemaker with a governing equation and its boundary conditions is displayed in Figure~\ref{wavemakerscheme}.

\section{Solutions for the Governing Differential Equation}

The following is an Ansatz for the general solution of velocity potential $\phi$
\begin{align}
\phi(x,\,z,\,t) &= A_{p} \: \cosh \, k_{p}\,(h+z) \: \sin\,(k_{p}\,x - \omega\,t) + (A\,x + B) \nonumber \\
& \quad + C \: e^{-k_{s}\,x} \: \cos k_{s}\,(h + z) \: \cos \, (\omega\, t).   \label{ansatz}
\end{align}
For wavemaker problem, the coefficient $A$ above must equal to zero since there exists no possible uniform flow through the wavemaker and we can also set the coefficient $B$ to be zero as well without influencing the velocity field. The remainder terms must satisfy the two linearized free surface boundary conditions.
It would also be useful to combine these two boundary conditions to a simpler form.
\begin{figure}[h]
\begin{center}
\includegraphics[scale = 0.8]{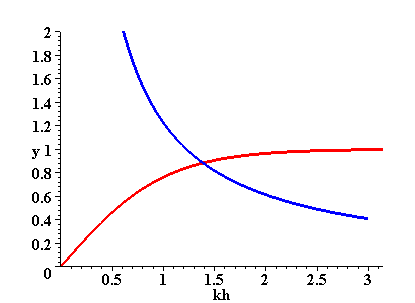}
\end{center}
\caption[A graph for the progressive wave dispersion relationship.]{A graph for the progressive wave dispersion relationship. This graph illustrates a single root $k_{p}\,h$. The horizontal axis label \textbf{kh} indicates $k_{p}\,h$.}
\label{gambar2}
\end{figure}

Deriving $\phi$ with respect to $t$, we obtain
\begin{equation}
\frac{\partial\phi}{\partial t} = -\, \omega \, A_{p} \: \cosh \, k_{p} \,(h + z) \: \cos \,(k_{p}\,x - \omega\, t) 
- \, \omega \, C \, e^{-k_{s}\,x} \: \cos k_{s}\,(h + z) \:\sin (\omega\, t).
\end{equation}
The second derivative of $\phi$ with respect to $t$ yields
\begin{align}
\frac{\partial^2 \phi}{\partial t^2} &= -\,\omega^2 \, A_{p} \: \cosh \,k_{p} \,(h + z) \: \sin \,(k_{p}\,x - \omega\, t)
- \, \omega^2 \, C \, e^{-k_{s}\,x} \: \cos k_{s} \, (h + z) \: \cos \, (\omega\, t) \nonumber \\
&= -\,\omega^2\,\phi.
\end{align}
From the free surface kinematic and dynamic boundary conditions, we have
\begin{equation}
\frac{\partial \eta}{\partial t} = -\,\frac{1}{g} \, \frac{\partial^2 \phi}{\partial t^2} = -\, \frac{\omega^2}{g}\,\phi = -\,\frac{\partial \phi}{\partial z}, \qquad \textmd{at} \quad z = 0,
\end{equation}
or
\begin{equation}
\frac{\partial \phi}{\partial z} - \frac{\omega^2}{g} \,\phi = 0, \qquad \textmd{at} \quad z = 0.
\end{equation}
By substituting the Ansatz~\eqref{ansatz} to this equation, we obtain
\begin{equation}
\omega^2 = g \, k_{p} \: \tanh \left(k_{p}\,h \right)	  \label{progresif}
\end{equation}
and
\begin{equation}
\omega^2 = -g \, k_{s} \: \tan \left(k_{s}\,h \right).  \label{berdiri}
\end{equation}
The former~\eqref{progresif} is a dispersion relationship for the progressive wave. By rewriting this equation as follows:
\begin{equation}
\frac{\omega^2\,h}{g\,k_{p}\,h} = \tanh \left(k_{p}\,h \right)
\end{equation}
and sketch each term with respect to $k_{p}\,h$ for a particular value of $\displaystyle \frac{\omega^2\,h}{g}$, then we are able to discover the solution for the dispersion relationship~\eqref{progresif} as depicted in Figure~\ref{gambar2}.

\begin{figure}[h]
\begin{center}
\includegraphics[scale = 0.8]{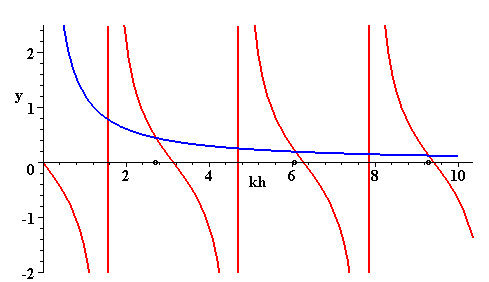}
\end{center}
\caption[A graph for the standing wave dispersion relationship.]{A graph for the standing wave dispersion relationship. This graph illustrates infinitely many compound roots $k_{s}(n)\,h$. The horizontal axis label \textbf{kh} means $k_{s}\,h$.} 				\label{gambar3}
\end{figure}

The latter~\eqref{berdiri}, which relates $k_{s}$ with the wavemaker frequency $\omega$, determines the wavenumber for the standing wave with an amplitude decaying exponentially as it travels far away from the wavemaker. By rewriting this equation as follows:
\begin{equation}
\frac{\omega^2 \, h}{g \, k_{s} \, h} = -\tan \left(k_{s}\,h\right)
\end{equation}
we can sketch its graph and observe that it possesses an infinite numbers of solution. The solutions of this problem can be observed from the graph for its dispersion relationship, as depicted in Figure~\ref{gambar3}.

Each solution is expressed in terms of $k_{s}(n)$, where $n \in \mathbb{N}$. The final form for the velocity potential is given as follows:
\begin{align}
\phi(x,\,z,\,t) &= A_{p} \: \cosh \, k_{p}\,(h + z) \: \sin \,(k_{p}\,x - \omega\,t) \nonumber \\
& \quad + \, \sum_{n = 1}^{\infty} C_{n} \: e^{-k_{s}(n) \, x} \: \cos \, k_{s}(n)\,(h + z) \: \cos (\omega \,t).  \label{lastpotensial}
\end{align}
The first term indicates the progressive wave produced by the wavemaker, while the series terms expresses decaying standing waves as they travel far away from the wavemaker.

To obtain a complete wave solution, we need to determine the coefficients $A_{p}$ and $C_{n}$. We can calculate these values using the wavemaker lateral boundary condition, namely
\begin{equation}
u(0,\,z,\,t) = \left. \left(\frac{\partial \phi}{\partial x} \right) \right|_{x = 0} = \frac{1}{2} \: \omega \, S(z) \: \cos (\omega\,t).
\end{equation}
By finding the first derivative of velocity potential $\phi$ with respect to $x$ and evaluating it at $x = 0$, we attain
\begin{align}
\frac{1}{2} \: \omega \,S(z) \: \cos (\omega\,t) &= A_{p} \, k_{p} \: \cosh \, k_{p}\,(h + z) \: \cos (\omega\,t) \nonumber \\
& \quad - \, \sum_{n = 1}^{\infty} C_{n} \, k_{s}(n) \: \cos \, k_{s}(n) \,(h + z) \: \cos (\omega \,t)
\end{align}
or
\begin{equation}
\frac{1}{2} \: \omega \, S(z) = A_{p} \, k_{p} \: \cosh \, k_{p}\,(h + z) - \sum_{n = 1}^{\infty} C_{n} \, k_{s}(n) \: \cos \, k_{s}(n) \,(h + z).  \label{fourier}
\end{equation}
We now possess a function in the $z$-variable which equals to a series of trigonometric functions on the right-hand side, a situation similar to a Fourier series.
We also have the fact that the set of functions $\left\{ \cosh \,k_{p}\,(h + z), \; \cos \left[ k_{s}(n)\,(h + z) \right] \right \}_{n = 1}^{\infty}$ forms a complete harmonic series orthogonal functions. Any arbitrary continuous function can be expanded in terms of the series.

To find the coefficients $A_p$, therefore, we multiple the above equation with $\cosh \,k_{p} \,(h + z)$ and integrate it with respect to the variable $z$ from $-h$ until $0$. We now acquire
\begin{align}
\int_{-h}^{0} \: \frac{1}{2} \: \omega \, S(z) & \cosh \, k_{p} \,(h + z) \, \mathrm{d}z 
= \int_{-h}^{0} \: A_{p} \, k_{p} \: \cosh \, k_{p}^2 \,(h + z) \, \mathrm{d}z \nonumber \\
&- \int_{-h}^{0} \; \sum_{n = 1}^{\infty} C_{n} \, k_{s}(n) \: \cos \left[k_{s}(n) \, (h + z) \right] \: \cosh \,k_{p}\,(h + z) \, \mathrm{d}z.
\end{align}
Applying the orthogonality property, the last term vanishes and consequently
\begin{equation}
A_{p} = \frac{\displaystyle \frac{1}{2} \: \omega \: \int_{-h}^{0} \: S(z) \: \cosh \, k_{p} \,(h + z) \, \mathrm{d}z}
{\displaystyle k_{p} \: \int_{-h}^{0} \: \cosh \,k_{p}^2 \, (h + z) \,dz}.
\end{equation}
For a flap-type wavemaker, the stroke function $S$ can be specifically expressed as follows:
\begin{equation}
S(z) = S \, \left(1 + \frac{z}{h} \right).
\end{equation}
Employing a simple calculus, we could express the coefficient $A_{p}$ explicitly without the integral sign, given as follows:
\begin{equation}
A_{p} = \frac{2 \, \omega \, S}{k_{p}^2\,h} \; \frac{k_{p} \, h \: \sinh \,k_{p} \,h - \cosh \,k_{p}\,h + 1}{\sinh \, 2 \, k_{p} \, h + 2 \,k_{p} \,h}.  \label{firstconst}
\end{equation}

Similarly, we can obtain the coefficient $C(n)$ by multiplying~\eqref{fourier} with $\cos \,[k_{s}(n) \, (h + z)]$ and integrate it with respect to the variable $z$ over the water depth from $z = -h$ until $z = 0$. We attain
\begin{equation}
C_{n} = \frac{\displaystyle -\frac{1}{2} \: \omega \: \int_{-h}^{0} \: S(z) \: \cos \, k_{s}(n)\,(h + z) \, \mathrm{d}z}
{\displaystyle k_{s}(n) \: \int_{-h}^{0} \: \cos^2 \, [k_{s}(n) \, (h + z)] \,\mathrm{d}z}
\end{equation}
or, by employing a little bit of integration process, we arrive at the following result:
\begin{equation}
C_{n} = \frac{-2 \, \omega \, S}{[k_{s}(n)]^2 \, h} \; \frac{[k_{s}(n) \, h] \: \sin \, [k_{s}(n) \ ,h] + \cos \,[k_{s}(n) \, h]}{\sin \, [2 \, k_{s}(n) \, h] + 2 \, k_{s}(n) \, h}. 			\label{nextconst}
\end{equation}

\begin{figure}[h]
\vspace*{-0.1cm}
\begin{center}
\includegraphics[scale = 0.8]{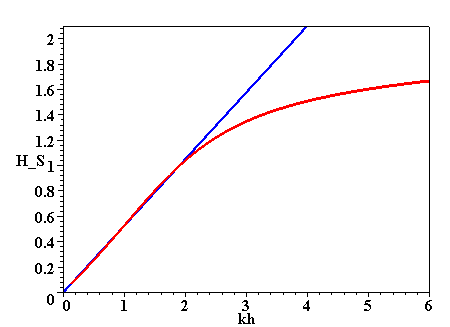}
\end{center}
\caption[A graph for the ratios between the wave height $H$ and the stroke $S$ with respect to the relative depth $k_{p}\,h$.]{A graph for the ratios between the wave height $H$ and the stroke $S$ with respect to the relative depth $k_{p}\,h$. The horizontal axis label \textbf{kh} refers to $k_{p}\,h$, while the vertical axis label \textbf{H\_S} refers to $H/S$.}			\label{gambar1}
\end{figure}

The wave height $H$ for the progressive wave can be determined by evaluating $\eta$ sufficiently far from the wavemaker
\begin{align}
\eta = -\frac{1}{g} \left. \left(\frac{\partial \phi}{\partial t} \right) \right|_{z = 0} 
&= \frac{A_{p}}{g} \, \omega \: \cosh \, k_{p}\,h \: cos \,(k_{p} \, x - \omega \, t)  \nonumber \\
&= \frac{H}{2} \: \cos \, (k_{p} \, x - \omega \, t) \qquad \text{for} \quad x \gg \, h.
\end{align}
By substituting the values $A_p$ obtained previously, we can acquire the ration between the wave height $H$ and the stroke $S$. It reads
\begin{equation}
\frac{H}{S} = 4 \: \left(\frac{\sinh \, k_{p} \, h}{k_{p} \, h} \right) \; \left(\frac{k_{p} \, h \: \sinh \, k_{p}\,h - \cosh \, k_{p} \, h + 1}{\sinh \, 2 \, k_{p} \, h + 2 \, k_{p} \, h} \right). 		 \label{ratio}
\end{equation}
The graph for the ratios between $H$ and $S$ for the (shallow-water wave) simplified theory and for a more complete linear wavemaker theory are depicted in Figure~\ref{gambar1}.

\chapter{Numerical Simulation}  \label{numeric}

In this chapter, we will discuss several results by selecting particular values from the wave-related quantities.
Using the computer software \emph{Maple V Release 5}, we could also observe an animation for the resulting monochromatic waves when we move the wavemaker with a particular stroke.

\section{Calculating the Progressive Wavenumber} 		\label{NR1}

Consider again equation~\eqref{progresif} which expresses the dispersion relationship for progressive waves
\begin{equation}
\omega^2 = g\,k_{p}\: \tanh\: k_{p}\,h.
\end{equation}
Since we are not able to calculate the exact value of the wavenumber $k_p$, we need to calculate it numerically. We use the following values: monochromatic frequency $\omega = 2$~rad/s, gravitational acceleration constant $g = 9.8$~m/s$^2$, and the water depth of the towing tank $h = 3$~m. Employing the Newton-Raphson iteration, we seek the value $k_p$ given an initial guess $p_0$ by finding the roots of $f(k_{p}) = 0$, where $f(k_{p}) = \omega^2 - g\, k_{p} \: \tanh \: k_{p} \, h$. Please see Appendix~B for more information on the algorithm for the Newton-Raphson iteration. By choosing error values of $\delta = 10^{-9}$, $\epsilon = 10^{-9}$, and {\ttfamily small} $= 10^{-9}$, as well as the maximum iteration $= 5000$, then using the initial guess $p_{0} = 0.5$, we obtain $k_{p} = 0,4624593666$. For $\omega = 1$, the obtained value for the progressive wavenumber is $k_{p} = 0,1943823443$. See again the graph in Figure~\ref{gambar2}.

\section{Calculating Standing Wave Wavenumbers} \label{NR2}

Consider again equation~\eqref{berdiri} which expresses the dispersion relationship for standing waves
\begin{equation}
\omega^2 = -g \, k_{s} \: \tan \: k_{s} \,h.
\end{equation}
Similar as previously, since we cannot calculate exact values of the wavenumber $k_s$, we calculate them numerically. We take identical values as previously: monochromatic frequency $\omega = 2$~rad/s, gravitational acceleration constant $g = 9.8$~m/s$^2$, and the water depth of the towing tank $h = 3$~m. Employing the Newton-Raphson iteration, we seek the values $k_s$ given an initial guess $p_0$ by finding the roots of $g(k_{s}) = 0$, where $g(k_{s}) = \omega^2 + g \, k_{s} \: \tan \: k_{s}\,h$. Please consult Appendix~B for more detailed information on the algorithm for the Newton-Raphson iteration. The selected error values are $\delta = 10^{-5}$, $\epsilon = 10^{-5}$, and {\ttfamily small} $= 10^{-5}$ with the maximum iteration of 10000. Since the graph in the dispersion relationship produces infinitely many intersections, i.e., the values of standing wave wavenumbers, then the values for an initial guess also vary according to the computational need, $p_{0} = 1, 2, 3, \cdots$. For a practical purpose, we only select several successive wavenumber values, given as follows:
\begin{align*}
k_{s}(1) &= 0.906121231; \\
k_{s}(2) &= 2.028197863;  \\
k_{s}(3) &= 3.097926261;  \\
k_{s}(4) &= 4.156159228;  \\
k_{s}(5) &= 5.209926525;  \\
k_{s}(6) &= 6.261487236;  \\
k_{s}(7) &= 7.311794624;  \\
k_{s}(8) &= 8.361321437;  \\
k_{s}(9) &= 9.410329031.
\end{align*}
See again the graph in Figure~\ref{gambar3}.

Similarly, using various values for the initial guess, we obtain the following wavenumber values corresponding to the standing waves for $\omega = 1$:
\begin{align*}
k_{s}(1) &= 1.013758189;	\\
k_{s}(2) &= 2.078040121;	\\
k_{s}(3) &= 3.130732073;	\\
k_{s}(4) &= 4.180655871.
\end{align*}

\section{Monochromatic Wave Profiles}

\subsection{Two Wave Profiles with Distinct Wave Height $H$}

Consider again the velocity potential equation~\eqref{lastpotensial} describing a water wave produced by the wavemaker
\begin{align}
\phi(x,\,z,\,t) &= A_{p} \: \cosh \, k_{p} \, (h + z) \: \sin \, (k_{p}\,x - \omega\,t) \nonumber \\
& \quad + \, \sum_{n=1}^{\infty} C_{n} \: e^{-k_{s}(n) \, x} \: \cos \, k_{s}(n) \, (h + z) \: \cos (\omega \,t).
\end{align}
Using equations~\eqref{firstconst} and~\eqref{ratio}, we can express the coefficient $A_p$ in terms of $\omega$, $H$, $k_{p}$, and $h$, i.e.,
\begin{equation}
A_{p} = \frac{\omega \, H}{2 \, k_{p} \: \sin \: k_{p}\,h}.
\end{equation}
By substituting the values $\omega = 2$, $k_{p} = 0,4624593666$, and $h = 3$, then
\begin{itemize}[leftmargin=1.2em]
\item for $H = 0.5 $, we have $A_{p} = 1.099621133$; and
\item for $H = 0.75$, we have $A_{p} = 1.649431700$.
\end{itemize}

To obtain the maximum stroke $S$, we use equation~\eqref{ratio}, i.e.,
\begin{equation}
S = \left(\frac{H \, k_{p} \, h}{4 \, \sinh \, k_{p} \, h} \right) \; 
\left(\frac{\sinh \, 2 \, k_{p} \, h + 2 \, k_{p} \, h}{k_{p} \,h \: \sinh \, k_{p} \, h - \cosh \, k_{p} \, h + 1} \right).
\end{equation}
Substituting the values we have chosen, it yields
\begin{itemize}[leftmargin=1.2em]
\item for $H = 0.5 $, we obtain $S = 0.2908632665$; and
\item for $H = 0,75$, we obtain $S = 0.4362948995$.
\end{itemize}

\begin{figure}[!htbp]
\begin{center}
\includegraphics[scale = 0.7]{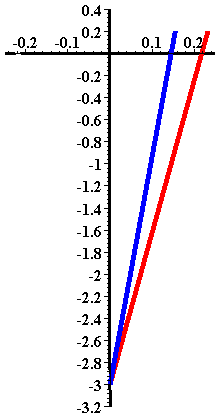}
\end{center}
\caption{Maximum strokes from the wavemaker flap which produce two wave-profiles with different wave height.}	\label{gambar4}
\end{figure}

The wavemaker movement profile when it reaches the maximum stroke can be viewed in Figure~\ref{gambar4}.~

To find the coefficients for the standing wave $C_n$, we use~\eqref{nextconst}, namely
\begin{equation}
C_{n} = \frac{-2 \, \omega \, S}{[k_{s}(n)]^2 \, h} \; \frac{[k_{s}(n) \, h] \: \sin \, [k_{s}(n) \, h] 
+ \cos \, [k_{s}(n) \, h]}{\sin \, [2 \, k_{s}(n) \,h ] + 2 \, k_{s}(n) \,h}.
\end{equation}
We have already selected four values of $k_{s}(n)$. For larger values of $n$, the effect of $C_n$ to the wave profile solution is relatively small so that it can be neglected.
\begin{itemize}[leftmargin=1.2em]
\item For $S = 0.2908632665$, the coefficient values are \ $C_{1} = 0.08013595964$, $C_{2} = 0.00976252019$, $C_{3} = 0.001714039296$, and $C_{4} = 0.001110132786$.
\item For $S = 0.4362948995$, the coefficient values are \ $C_{1} = 0.1202039395 $, $C_{2} = 0.01464378027$, $C_{3} = 0.002571058938$, and $C_{4} = 0.001665199177$.
\end{itemize}

\begin{figure}[h]
\begin{center}
\includegraphics[scale = 0.8]{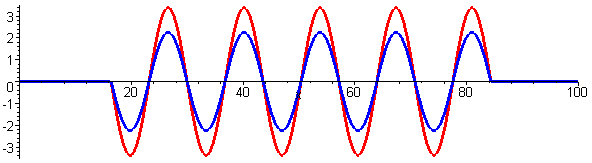}
\end{center}
\caption{Shapes of surface water wave elevation for two distinct wave heights.} 	\label{gambar5}
\end{figure}

Using equation~\eqref{permukaan}
\begin{equation}
\eta(x,\,t) = -\frac{1}{g} \: \left. \left(\frac{\partial \phi}{\partial t} \right) \right|_{z = 0}
\end{equation}
we are able to discover an equation for the shape of the formed water wave elevation. We can also display this surface elevation using computer software. The simulation result for the water surface with two distinct wave heights is displayed in Figure~\ref{gambar5}.

\subsection{Two Wave Profiles with Distinct Angular Frequency $\omega$}

In this subsection, we implement an identical technique as in the previous subsection. The water depth in the towing tank is also $h = 3$~m and the desired wave height profile is $H = 0.5$~m. The angular frequencies are $\omega = 2$ and $\omega = 1$. Employing the Newton-Raphson iteration toward both the progressive wave and the standing wave dispersion relationships, we could attain the values $k_{p}$ and $k_{s}(n)$ for these frequencies. We have obtained these results in Sections~\ref{NR1} and~\ref{NR2}. We tabulate them again in this section. Additionally, we also list the coefficients for the velocity potential related to the values of $\omega$, i.e., $A_{p}$ and $C_{n}$. 

\begin{figure}[h]
\begin{center}
\includegraphics[scale = 0.8]{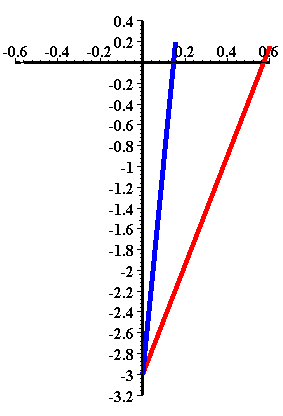}
\end{center}
\caption{The maximum strokes from the flap which produce two wave-profiles with distinct frequency.}   		\label{gambar6}
\end{figure}

\begin{figure}[h]
\begin{center}
\includegraphics[scale = 0.8]{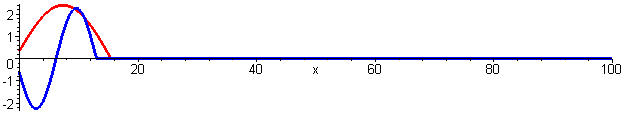}
\end{center}
\caption{Water wave surface graphs at $t = 3$.}   \label{gambar7}
\end{figure}

\begin{figure}[h]
\begin{center}
\includegraphics[scale = 0.8]{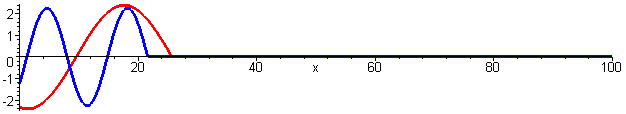}
\end{center}
\caption{Water wave surface graphs at $t = 5$.}   \label{gambar8}
\end{figure}

\begin{figure}[h]
\begin{center}
\includegraphics[scale = 0.8]{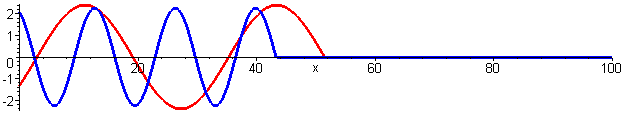}
\end{center}
\caption{Water wave surface graphs at $t = 10$.}  \label{gambar9}
\end{figure}

\begin{figure}[b!]
\begin{center}
\includegraphics[scale = 0.8]{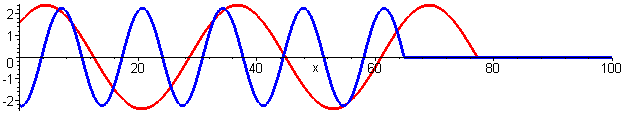}
\end{center}
\caption{Water wave surface graphs at $t = 15$.}  \label{gambar10}
\end{figure}

\begin{figure}[h]
\begin{center}
\includegraphics[scale = 0.8]{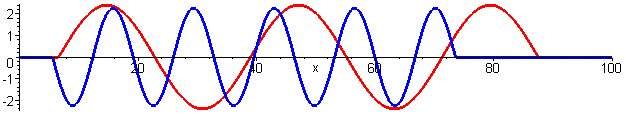}
\end{center}
\caption{Water wave surface graphs at $t = 17$.}   \label{gambar11}
\end{figure}

\begin{figure}[h]
\begin{center}
\includegraphics[scale = 0.8]{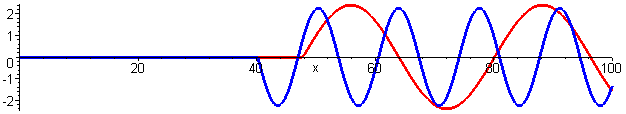}
\end{center}
\caption{Water wave surface graphs at $t = 25$.}   \label{gambar12}
\end{figure}

\begin{figure}[h]
\begin{center}
\includegraphics[scale = 0.8]{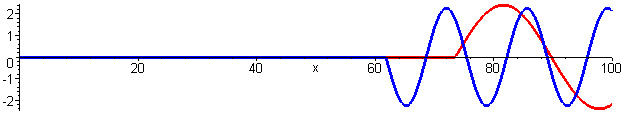}
\end{center}
\caption{Water wave surface graphs at $t = 30$.}   \label{gambar13}
\end{figure}

\begin{figure}[b!]
\begin{center}
\includegraphics[scale = 0.8]{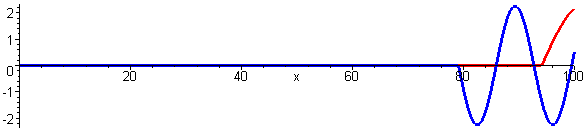}
\end{center}
\caption{Water wave surface graphs at $t = 34$.}   \label{gambar14}
\end{figure}

The following provides results obtained by the computer software \emph{Maple V Release 5}.
\begin{description}[leftmargin=1.3em]
\item[\textbf{a.}]  For $\omega = 2$.
\begin{itemize}[leftmargin=0.8em]
\item A wavenumber for the progressive wave $k_{p} = 0.4624593666$.

\item Wavenumbers for the standing wave:\\
\hspace*{1em} $k_{s}(1) = 0.9061212307$, \qquad \qquad $k_{s}(2) = 2.028197863$, \\
\hspace*{1em} $k_{s}(3) = 3.0979262610$, \qquad \qquad $k_{s}(4) = 4.156159228$.

\item The coefficient $A_{p} = 1.099621133$.

\item The maximum stroke $S = 0.2908632665$.

\item The coefficients $C_{n}$ are \\
\hspace*{1em} $C_{1} = 0.801359596400$, \qquad \qquad $C_{2} = 0.009762520190$, \\
\hspace*{1em} $C_{3} = 0.001714039296$, \qquad \qquad $C_{4} = 0.001110132786$.
\end{itemize}

\item[\textbf{b.}] For $\omega = 1$.
\begin{itemize}[leftmargin=0.8em]
\item A wavenumber for the progressive wave $k_{p} = 0.1943823443$.

\item Wavenumbers for the standing wave: \\
\hspace*{1em} $k_{s}(1) = 1.013758189$, \qquad \qquad $k_{s}(2) = 2.078040121$, \\
\hspace*{1em} $k_{s}(3) = 3.130732073$, \qquad \qquad $k_{s}(4) = 4.180655871$.

\item The coefficient $A_{p} = 2.335633604$.

\item The maximum stroke $S = 1.140576790$.

\item The coefficients $C_{n}$ are \\
\hspace*{1em} $C_{1} = 0.212585384200$, \qquad \qquad $C_{2} = 0.0043694275640$, \\
\hspace*{1em} $C_{3} = 0.007018430814$, \qquad \qquad $C_{4} = 0.0005323318426$.
\end{itemize}
\end{description}
The flap deviation profile for the maximum strokes is displayed in Figure~\ref{gambar6}.
The wave profiles evolution in time is presented in Figures~\ref{gambar7}--\ref{gambar14}.
\newpage
\vfill
\vfill
\vfill
\newpage
\vfill
\vfill
\vfill

\chapter{Conclusion and Suggestion}

\section{Conclusion}

The following presents conclusion for this final project report:
\begin{itemize}[leftmargin=1.5em]
\item[\textbf{a.}] We have adopted ideal fluid assumptions for wave generation modeling.
\item[\textbf{b.}] We have implemented linear wave theory for solving the governing differential equations obtained from part (a).
\item[\textbf{c.}] We were able to derive the relationship among the wavenumber, wave height, and wavemaker stroke from the theory developed in part (b).
\item[\textbf{d.}] For shallow-water waves, both the simplified and the linear wave theories produce an identical result, while for deeper-water waves, both theories provide different results.
\end{itemize}

\section{Suggestion}

The following presents suggestion for another final project or further research: 
\begin{itemize}[leftmargin=1.5em]
\item[\textbf{a.}] The wave theory can be extended to nonlinear theory and three-dimensional wavemaker theory.
\item[\textbf{b.}] The type of waves can be expanded to include random and irregular waves.
\item[\textbf{c.}] Adding a beach can be utilized to replace a semi-infinite domain.
\end{itemize}

\appendix

\chapter{Material Derivative}

The operator
\begin{equation}
\frac{\mathrm{D}}{\mathrm{D}t} = \frac{\partial}{\partial t} + \mathbf{u}\cdot \nabla  			\label{op}
\end{equation}
is known as the material derivative, total derivative, substantial derivative, or Lagrange derivative. In general $\displaystyle \frac{\mathrm{D}}{\mathrm{D}t}$ is a vector operator. As a consequence, $\displaystyle \frac{\mathrm{D}}{\mathrm{D}t}$ components acts on a vector will not be equal with the ones act on a scalar component of a vector, except in the Cartesian coordinates system. The material derivative can also be applied to a scalar quantity, such as temperature. Physically, it states the rate of change of that quantity with respect to time, where the observer moves along with the measured fluid at a particular location in space and instant time when that derivative is evaluated.

This operator is the total derivative with respect to time acting on a fluid element, which can be viewed as follows. Let $\xi$ be a spatial coordinate describing the fluid at an initial fixed instant $t = 0$, let also $\mathbf{x}$ be the spatial coordinate providing the location during time $t$ of a fluid element for which it is $\xi$ when $t = 0$. We have $\mathbf{x} = \mathbf{x}(\xi,\tau)$. The Eulerian derivative with respect to time is given by
\begin{equation}
\frac{\partial}{\partial t} = \left. \frac{\partial}{\partial t} \right|_{\text{fixed} \; \mathbf{x}},
\end{equation}
while the derivative which plays a role as Lagrangian derivative is given by
\begin{equation}
\frac{\mathrm{D}}{\mathrm{D}t} = \left. \frac{\partial}{\partial t}\right|_{\text{fixed} \; \xi}.
\end{equation}
Employing the Chain Rule, we obtain
\begin{equation*}
\frac{\mathrm{D}}{\mathrm{D}t} 
= \left. \frac{\partial}{\partial t} \right|_{\text{fixed} \; \xi} 
= \left. \frac{\partial}{\partial t} \right|_{\text{fixed} \; \mathbf{x}} + 
  \left. \frac{\partial x_i}{\partial t} \right|_{\text{fixed} \; \xi} \frac{\partial}{\partial t},
\end{equation*}
which results equation~\eqref{op} since $\displaystyle \left. \frac{\partial \mathbf{x}}{\partial t} \right|_{\xi} = \mathbf{u}$.

This operator is also used in the Navier-Stokes equation which states the total acceleration of a particle as follows:
\begin{equation}
\mathbf{a} = \frac{\mathrm{D}\mathbf{u}}{\mathrm{D}t} = 
\frac{\partial \mathbf{u}}{\partial t} + \left(u \frac{\partial \mathbf{u}}{\partial x} + v \frac{\partial \mathbf{u}}{\partial y} + 
w \frac{\partial \mathbf{u}}{\partial z} \right).  			\label{nv}
\end{equation}
The first term on the right-hand side of~\eqref{nv} indicates a local acceleration, which is zero for the steady-state flow. The other terms express convective acceleration, which show that the fluid flow has a different velocity at a different position. For more detailed explanation, please consult~\cite{fac,hwf}.

\chapter[Newton-Raphson Iteration]{An Algorithm for the Newton-Raphson Iteration}


\ovalbox{
\begin{minipage}{15.0cm}
\vspace{0.5cm}
\bfseries Searching the root of $f(x) = 0$ given an initial guess $p_0$ using the iteration\newline $\displaystyle p_{n} = p_{n - 1} - \frac{f(p_{n - 1})}{f'(p_{n - 1})}$ for $n = 1, 2, 3, \dots$. \vspace{0.5cm}
\end{minipage}
} 
\vspace{0.5cm} \\
\textbf{Algorithm:}\\
{\ttfamily $\delta:= 10^{-6}$, \qquad $\epsilon:= 10^{-6}$, \qquad small$:= 10^{-6}$ \\
\{ several error values, can be adjusted according to the need \} \\
maks$:= 99$ \qquad \qquad \{ the maximum number of iterations \} \\
kond$:=  0$ \qquad \qquad \{ the condition for a loop termination \} \\
INPUT $P_0$ \qquad \qquad \{ $P_0$ must be sufficiently close with the desired root \} \\
$Y_0:=F(P_0)$ \qquad \qquad \{ calculating a value of the function \} \\
\newline
DO FOR $N:= 1$ TO maks UNTIL kond $\neq$ 0 \\
\hspace*{.2in} $Df:= F^{\prime}(P_0)$ \qquad \qquad \{ calculating the derivative \} \\
\hspace*{.5in} IF $Df = 0$ THEN \qquad \qquad 	\\	\{ checking whether there exists any division by zero \} \\
\hspace*{1in} kond$:= 1$ \\
\hspace*{1in} $Dp:= 0$ \\
\hspace*{.8in} ELSE \\
\hspace*{1in} $Dp:= Y_0/Df$ \\
\hspace*{.5in} ENDIF \\
\hspace*{.2in} $P_1:= P_0 - Dp$ \qquad \qquad \{ new iteration \} \\
\hspace*{.2in} $Y_1:= F(P_1)$   \qquad \qquad \{ the value of a new function \} \\
\hspace*{.2in} RelErr$:= 2\ast|Dp|/(|P1| + \texttt{small})$ \qquad \qquad \{ relative error \} \\
\hspace*{.2in} IF RelErr $< \delta$ AND $|Y_1| < \epsilon$ THEN \\
\hspace*{.5in} IF kond $\neq$ 1 THEN kond$:= 2$ \qquad \qquad \{ checking convergence \}  \\
\hspace*{.2in} $P_0:= P_1$; \qquad $Y_0:= Y_1$  \qquad \qquad \{ replacing with new values \} \\
\newline
PRINT `The value of the $n$-th iteration is' $P_1$ \qquad \qquad \{ ouput \} \\
PRINT `The successive iteration differs by' $Dp$  \\
PRINT `The value of $f(x)$ is' $Y1$ \\
IF kond$= 0$ THEN \\
\hspace*{.2in} PRINT `The number of maximum iteration has been exceeded.' \\
IF kond$= 1$ THEN \\
\hspace*{.2in} PRINT `There exists a division by zero.' \\
IF kond$= 2$ THEN \\
\hspace*{.2in} PRINT `A root has been found with the desired error.'
}

\chapter[Hydrodynamics Laboratories]{Hydrodynamics Laboratories Around the World}

This appendix lists hydrodynamic laboratories in various parts of the world. Some of them are used commercially and for marine structure defense. For detailed information, please consult~\cite{csk} or the International Towing Tank Conference association. Otherwise indicated, the size of the tank refers to the length, width, and depth, respectively.
\vfill
\begin{itemize}[leftmargin=1.5em]
\item[\textbf{a.}] Institute of Marine Dynamics Towing Tank, St. John's, Newfoundland, Canada \\
\begin{tabular}{@{}ll}
Facility: & Deep-water tank \\
Tank size:& 200~m $\times$ 12~m $\times$ 7~m \\
Carrier speed:& 10 m/s \\
Wave type:& Regular and irregular; 1~m \\
Wavemaker:& Double flap-type \\
Beach:& Surging wave bottom.
\end{tabular}
\vfill

\item[\textbf{b.}] Offshore Model Basin, Escondido, California, United States (US) \\
\begin{tabular}{@{}ll}
Tank size:& 90~m $\times$ 14.6~m $\times$ 4.6~m \\
Deeper part:& Circular hole with 9~m depth \\
Carrier speed:& 6 m/s \\
Wave type:& Regular and irregular; 0.74~m \\
Wavemaker:& Single-flap board \\
Beach:& Metal shaved.
\end{tabular}
\vfill

\item[\textbf{c.}] Offshore Technology Research Center, Texas A\&M, College Station, Texas, US \\
\begin{tabular}{@{}ll}
Tanks size:& 45.7~m $\times$ 30.5~m $\times$ 5.8~m \\
Deeper part:& 16.7~m hole with adjustable floor \\
Wavemaker:& Flap-type with hydraulic hinge control \\
Maximum wave height:& 80~cm \\
Period range:& 0.5--4.0~s \\
Beach:& Metal panel
\end{tabular}

\item[\textbf{d.}] David Taylor Research Center, Bethesda, Maryland, US \\
\begin{tabular}{@{}ll}
Facility:& Maneuvering and Seakeeping Facilities (MASK) \\
Tank size:& 79.3~m $\times$ 73.2~m $\times$ 6.1~m \\
Wavemaker:& Total of 21 pneumatic-type \\
Wave:& Multidirectional, regular and irregular; maximum height 0.6~m;\\
     & and wavelength 0.9--12.2~m \\
Beach:& Wave absorber \\
Carrier speed:& 7.7 m/s.
\end{tabular}

\begin{tabular}{@{}ll}
Facility:& Deep-Water Basin \\
Tank size:& 846~m $\times$ 15.5~m $\times$ 6.7~m \\
Wave:& Maximum height 0.6~m and wavelength 1.5--12.2~m \\
Carrier speed:& 10.2 m/s. 
\end{tabular}

\begin{tabular}{@{}ll}
Facility:& High-Speed Basin \\
Tank size:& 79.3~m $\times$ 73.2~m $\times$ 6.1~m \\
Wave:& Maximum height 0.6~m and wavelength 0.9--12.2~m \\
Carrier speed:& 35.8--51.2 m/s.
\end{tabular}

\item[\textbf{e.}] Maritime Research Institute (MARIN), the Netherlands \\
\begin{tabular}{@{}ll}
Facility:& Seakeeping Basin \\ 
Tank size:& 100~m $\times$ 24.5~m $\times$ 2.5~m \\
Deeper part:& Hole with 6~m depth \\
Wave:& Regular and irregular; maximum height 0.3~m; and period range 0.7--3.0~s \\
Carrier speed:& 4.5 m/s.
\end{tabular}

\begin{tabular}{@{}ll}
Facility:& Wave and Current Basins \\
Tank size:& 60~m $\times$ 40~m $\times$ 1.2~m \\
Deeper part:& Hole with 3~m depth \\
Wave:& Regular and irregular \\
Carrier speed:& 3 m/s \\
Speed range:& 0.1--0.6 m/s.
\end{tabular}

\begin{tabular}{@{}ll}  
Facility:& Deep-Water Towing Tank \\
Tank size:& 252~m $\times$ 10.5~m $\times$ 5.5~m \\
Carrier speed:& 9 m/s.
\end{tabular}

\begin{tabular}{@{}ll}
Facility:& High-Speed Towing Tank \\
Tank size:& 220~m $\times$ 4~m $\times$ 4~m \\
Wavemaker:& Hydraulic flap-type \\
Wave:& Regular and irregular; maximum height 0.4~m; and period range 0.3--5~s \\
Carrier:& Motor and jet controls \\
Carrier speed:& 15~m/s and 30~m/s \\
Beach:& Circular arc lattices.
\end{tabular}

\item[\textbf{f.}] Danish Maritime Institute, Lyngby, Denmark \\
\begin{tabular}{@{}ll}
Tank size:& 240~m $\times$ 12~m $\times$ 5.5~m \\
Wavemaker:& Hydraulic double-flap type controlled numerically \\
Wave:& Regular and irregular; maximum wave 0.4~m; and period range 0.5--7~s \\
Carrier speed:& 0--11 m/s (accurate $\pm$ 2\%).
\end{tabular}

\item[\textbf{g.}] Danish Hydraulic Institute, Horsholm, Denmark \\
\begin{tabular}{@{}ll}
Tank size:& 30~m $\times$ 20~m $\times$ 3~m \\
Deeper part:& 12~m in the middle \\
Wavemaker:& 60 Hydraulic flaps controlled by a mini-computer on one side \\
Wave:& Maximum height $\cong$ 0.6~m and period range $\cong$ 0.5--4~s.
\end{tabular}

\item[\textbf{h.}] Norwegian Hydrodynamic Laboratory (MARINTEK), Trondheim, Norway \\
\begin{tabular}{@{}ll}
Facility:& The Ocean Basin \\
Tank size:& 80~m $\times$ 50~m $\times$ 10~m \\
Wavemaker:& Hinged double-flap type, 144 self-control; Hydraulic controlled hinge-type \\
Wave:& Regular and irregular; maximum height 0.9~m \\
Wave speed:& Maximum speed 0.2 m/s.
\end{tabular}
\end{itemize}


\end{document}


\newpage
\renewcommand{\thepage}{\roman{page}}
\setcounter{page}{0}
\begin{center}
\vspace*{1cm}
\LARGE 
\textbf{TEORI PEMBANGKIT GELOMBANG DUA-DIMENSI TIPE FLAP}\\
\vspace{.5 in} 
\large 
\textbf{Tugas Akhir}\\ 
\textbf{Diajukan sebagai syarat untuk memenuhi Sidang Sarjana Jurusan Matematika}\\
\vspace{4cm} 
\large 
\textbf{oleh :}\\ 
\Large 
\textbf{Natanael Karjanto}\\ 
\textbf{10197045}\\ 
\vspace{4cm}
\begin{figure}[h]
\begin{center}
\includegraphics[scale = 0.8]{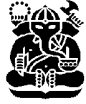}
\end{center}
\end{figure}
\normalsize 
\textbf{JURUSAN MATEMATIKA}\\ 
\textbf{FAKULTAS MATEMATIKA DAN ILMU PENGETAHUAN ALAM} \\ 
\textbf{INSTITUT TEKNOLOGI BANDUNG}\\ 
\textbf{2001}
\end{center}

\newpage
\setcounter{page}{0}
\begin{center}
\vspace*{1cm}
\LARGE \textbf{TEORI PEMBANGKIT GELOMBANG DUA-DIMENSI TIPE FLAP}\\
\vspace{0.5 in} \large 
\textbf{Lembar Pengesahan}\\ \vspace{1 in}
\Large 
\textbf{Telah diperiksa dan disetujui oleh Pembimbing :}\\
\vspace{1.25 in} 
\textbf{\underline{Dr. Andonowati}}\\
\textbf{NIP. 131803263}\\ \vspace{1.5 in} \Large 
\textbf{Penilai \ (Penguji) :}\\
\Large \vspace{1.25 in}
\textbf{\underline{Prof. Dr. M. Ansjar}} \hfill \textbf{\underline{Dr. Wono Setya Budhi}}\\ 
\hspace*{0.5cm} \textbf{NIP. 130143972} \hfill \textbf{NIP. 131284801} \hspace*{1cm}
\end{center}

\newpage
\begin{verse}
\setcounter{page}{0}
\vspace*{2cm}
\begin{center}
\emph{\normalsize Untuk yang terkasih papa, mama, dan adik perempuanku.}
\end{center}
\end{verse}

\def\abstractname{\LARGE \textbf{Abstrak}}
\begin{abstract}
\setcounter{page}{5}

{\normalsize Pemodelan secara matematis pembangkitan gelombang arah
tunggal disampaikan dalam laporan tugas akhir ini. Pemodelannya
mencakup persamaan Laplace dalam kolam air dengan setengah batas
tak hingga, syarat batas dinamika dan kinematika, syarat batas
lateral di pembangkit gelombang, dan dinding tetap di dasar kolam.
Modelnya diterapkan pada \textit{wavemaker} tipe \textit{flap}
yang sering digunakan pada suatu \textit{towing tank} dalam
laboratorium hidrodinamika. Untuk menyederhanakan permasalahan,
beberapa asumsi diterapkan, yakni bahwa air adalah fluida yang
bersifat ideal. Teori pembangkit gelombang linear digunakan dan
tipe gelombang monokromatik (frekuensi tunggal) diamati. Kaitan
antara bilangan gelombang, ketinggian gelombang, dan simpangan
\textit{wavemaker} juga akan diturunkan. \par} \vspace*{1cm}

\noindent
{\normalsize \textbf{Kata-kata kunci:} fluida ideal, pembangkit
gelombang (\textit{wavemaker}), persamaan Laplace, syarat batas
kinematika permukaan bebas, syarat batas dinamika permukaan bebas,
teori pembangkit gelombang linear, gelombang monokromatik,
bilangan gelombang, ketinggian gelombang, dan simpangan pembangkit
gelombang. \addcontentsline{toc}{chapter}{Abstrak} \par}
\end{abstract}

\begin{center}
\huge {\textbf{Kata Pengantar}}
\end{center}
\setcounter{page}{6}

\normalsize {\Huge \textbf{T}}erima kasih kepada pembaca yang
dengan senang hati meluangkan waktu untuk membuka tugas akhir ini.%
\addcontentsline{toc}{chapter}{Kata Pengantar} Namun sebelumnya,
izinkanlah penulis untuk menyampaikan puji dan syukur ke hadirat
Tuhan, Pencipta alam semesta beserta segala isinya. Atas
berkat dan rahmat-Nyalah penulis dapat menyelesaikan tugas akhir
sekaligus mengakhiri jenjang pendidikan pada tahap sarjana.
Penulis juga tidak lupa mengucapkan terima kasih yang
sebesar-besarnya karena telah mendukung penulis untuk
menyelesaikan studi di institusi tercinta ini, terutama kepada :

\begin{enumerate}
\item[\textbf{1.}]  \textbf{Dr. Andonowati}, yang telah bersedia dan dengan
sabar menjadi pembimbing tugas akhir penulis. \textit{Merci, Mom
!}

\item[\textbf{2.}]  \textbf{Prof. Dr. M. Ansjar} dan \textbf{Dr. Wono Setya
Budhi}, yang telah bersedia menguji dan menilai penulis pada
seminar tugas akhir tanggal 15 Desember 2000 lalu. Terima kasih
juga kepada Pak Ansjar yang telah mempercayakan penulis sebagai
asisten grader Metode Matematika tahun 2000. Teristimewa untuk Pak
Wono yang telah dengan sabar mengajari Maple dan \LaTeX \ pada
penulis sehingga dapat merampungkan tugas akhir ini.

\item[\textbf{3.}]  \textbf{Dr. Nana Nawawi Gaos}, yang telah menjadi dosen
wali akademik penulis selama menempuh pendidikan di Tahap
Persiapan Bersama 1997/1998 dan Semester Pendek 1998.

\item[\textbf{4.}]  \textbf{Dr. Ahmad Muchlis}, yang telah menjadi dosen
wali akademik penulis selama menempuh pendidikan di tahap Sarjana
Muda dan tahap Sarjana. (\textit{Thanks for motivating me to
finish my study in \textsf{three and half a year}}.)

\item[\textbf{5.}]  \textbf{Warsoma Djohan M.Si.}, yang telah mempercayai
penulis menjadi asisten grader Kalkulus I TPB-04 tahun 1999 dan
asisten tutorial Kalkulus I TPB-07 tahun 2000 lalu. Juga tidak lupa untuk Bu \textbf{%
Jalina Widjaja} yang telah mempercayai penulis menjadi asisten
Matematika Rekayasa, Matriks \& Ruang Vektor, serta asisten
temporal Kalkulus I, dan Mba \textbf{Nuning Nuraini} yang juga
tidak ragu mengajak penulis menjadi asisten tutorial Kalkulus II.
Tak terlewatkan juga Pak \textbf{Koko Martono} yang telah
memberikan mandat asisten grader Kalkulus Peubah Banyak dan Fungsi
Kompleks serta pelatihan dasar menulis artikel ilmiah. Demikian
juga dengan staf pengajar lainnya di jurusan ini yang telah
memberikan kontribusi kematangan berpikir pada penulis.

\item[\textbf{6.}]  Kedua orang tua penulis, papa dan mama, yang telah
memberikan dukungan materi dan doa restu sehingga penulis menjadi
seorang sarjana. Juga adikku tercinta yang telah memberikan
dorongan dan semangat untuk rajin belajar. \setcounter{page}{7}

\item[\textbf{7.}]  \textbf{Hadi Susanto}, sebagai rekan kuliah
penulis, sering melewatkan saat-saat bersama baik dalam suka
maupun duka, teristimewa sebelum keberangkatannya ke Belanda.
\newline (\textsl{Ik wil U bedanken omdat U mij heel goed heeft
gemotiveend om extra hard te studeren. \textbf{Bedankt, Hadi !}})

\item[\textbf{8.}] \textbf{Maykel, Luis, Dina, Ety, Sica, Anna, Sondang,
Rilyovira}, dan teman-teman angkatan 97 yang lainnya yang tidak
dapat penulis sebutkan satu per satu. \textit{Thanks for our
friendship.} Juga pada teman-teman angkatan 95, 96, 98, dan 99
yang telah banyak membantu penulis dalam menempuh jenjang
pendidikan di ITB ini.

\item[\textbf{9.}] Pak \textbf{Toto Nusantara (MA-S3)}, Mas
\textbf{Lylye Sulaeman (P4M)}, dan \textbf{Surya (MA96)} atas
bantuannya dalam \TeX, \LaTeX, dan \textit{Scientific Work Place}
sehingga tugas akhir ini dapat selesai diketik.

\item[\textbf{10.}] \textbf{Wili (MA97), Henry (FI97),
Albert(TF97), Wila (TI97), Aan (TK97), Faiq (TG97), Krshna (EL97),
Fitra (TA98), Dindin (TL98), Hidayat (FA98), Dwi Susanti (FA98),
Mia (KI99), Erika (FA99), Dwi Hesti (TG2000)} dan teman-teman
serta junior lainnya yang satu almamater dengan penulis (SMUN 4
Bandung), yang telah memotivasi dan sangat mendukung penulis untuk
menyelesaikan studi tidak lebih dari 8 semester. \textsl{I am
grateful to you all !}
\end{enumerate}

\textsl{\textbf{``Experientia est optima rerum magistra."}} Itulah
pepatah dalam bahasa Latin yang kurang lebih berarti, "Pengalaman
adalah guru yang terbaik." Seperti peristiwa-peristiwa lainnya
dalam kehidupan, menyelesaikan studi di tahap sarjana dan
sekaligus juga merampungkan tugas akhir ini adalah suatu
pengalaman tersendiri yang unik dan menarik bagi penulis, menjadi
'guru' dalam suka dan duka. Ada banyak hal yang penulis dapatkan
setelah mengerjakan tugas akhir ini, pengetahuan akademis pada
khususnya dan kedewasaan serta kematangan berpikir pada umumnya.

Dengan penuh kerendahan pikiran, penulis menyadari bahwa secara
individu, penulis hanyalah satu dari banyak orang yang datang dan
pergi, yang berupaya menorehkan prestasi dalam jenjang pendidikan
di Jurusan Matematika ITB. Walaupun tidak dapat memberikan yang
terbaik untuk kemajuan Jurusan ini, penulis telah mengerahkan
sekuat tenaga untuk merampungkan tugas akhir ini.
Penulis juga berharap bahwa karya yang 'kecil' ini dapat menjadi \textbf{%
\textit{\mbox{\boldmath $\ll$}cr\mbox{\boldmath $\grave{e}$}me de
la cr\mbox{\boldmath $\grave{e}$}me\mbox{\boldmath$\gg$}}} dari
semua karya yang telah penulis buat, meskipun ada masih banyak
kekurangan di berbagai segi.

Oleh karena itu, penulis mohon maaf yang sebesar-besarnya kepada
para pembaca dan pengguna tugas akhir ini. Akhir kata, semoga
'karya' ini dapat bermanfaat bagi para pembaca pada khususnya dan
untuk kemajuan ilmu Matematika pada umumnya.

\vspace{0.4 in} \hspace*{2 in} Bandung, Medio Januari 2001 \vspace{0.4 in}%
\newline
\hspace*{3 in}\textbf{$\mathcal{N}$}atanael
\textbf{$\mathcal{K}$}aryanto

\newpage
\pagestyle{headings} \setcounter{page}{9}
\addcontentsline{toc}{chapter}{Daftar Isi} \tableofcontents

\newpage
\addcontentsline{toc}{chapter}{Daftar Gambar} \listoffigures

\chapter{Pendahuluan}

\pagenumbering{arabic}

\section{Latar Belakang dan Rumusan Masalah}

\subsection{Latar Belakang}

Seiring dengan peningkatan taraf kemajuan suatu bangsa, semakin meningkat
jugalah kebutuhan akan penelitian di bidang sains dan teknologi. Di negeri
kita, yang lebih dari dua per tiga wilayahnya adalah lautan, memiliki sumber
daya alam yang berkualitas. Sejak zaman dahulu, negeri kita dikenal sebagai
negara bahari, atau negeri kelautan. Seraya ilmu pengetahuan dan teknologi
berkembang dengan pesat, penelitian di bidang ini juga perlu mendapatkan
perhatian yang khusus.

Di bidang teknik kelautan (\textit{ocean engineering}), adanya kebutuhan
untuk membuat model matematis mengenai hal-hal yang terjadi di lepas pantai
mendorong banyak penelitian dan riset di bidang ini. Lebih jauh lagi, ini
juga mendorong kerja sama antara berbagai disiplin ilmu yang
mengkolaborasikan sains dan teknologi. Matematika sebagai \textit{``\,The
Queen and Servant of Science\,"} memiliki keampuhan dalam memodelkan banyak
hal yang terjadi di alam. Dalam tugas akhir ini, akan dibahas mengenai
masalah teoretis pembangkitan gelombang dalam sebuah \textit{towing tank}.%
\footnote[1]{\textit{Towing tank} adalah suatu fasilitas dalam laboratorium
hidrodinamika berbentuk kolam berisi air dengan dilengkapi pembangkit
gelombang di satu sisi dan penyerap gelombang di sisi lain. Beberapa
contohnya dapat dilihat di Apendiks C.}

Pemodelan pembangkit gelombang dalam laboratorium hidrodinamika ini sangat
berguna untuk pengujian kapal yang akan berlayar di laut bebas. Sebagai
contoh, kapal yang akan berlayar di Laut Jawa haruslah diuji dengan
gelombang laut yang terjadi di Laut Jawa juga. Dengan demikian, diperoleh
gambaran besarnya kekuatan gelombang sehingga mendorong untuk membuat kapal
yang cukup kuat untuk berlayar di sana.

\subsection{Rumusan Masalah}

Berdasarkan latar belakang di atas, rumusan masalah yang penulis
ajukan adalah bagaimana membuat model matematis dari gelombang
yang terjadi di laut yang jauh dari pantai dengan menggunakan
teori pembangkitan gelombang yang dapat dilakukan di dalam
laboratorium hidrodinamika. Kita ingin mengetahui besarnya
simpangan pembangkit gelombang yang harus digerakkan apabila
diinginkan profil gelombang dengan ketinggian tertentu. Selain
itu, kita juga ingin mengetahui profil gelombang yang terjadi
apabila pembangkit gelombang tadi digerakkan dengan frekuensi
tertentu. Hal ini dapat dijadikan sebagai sebuah model matematika
mengenai gelombang air yang terjadi pada \textit{towing tank}.

\section{Ruang Lingkup Kajian}

Berdasarkan rumusan masalah di atas, ruang lingkup kajian yang
akan penulis bahas adalah membuat model matematika sederhana
mengenai pembangkit gelombang dua dimensi tipe \textit{flap} pada
sebuah \textit{towing tank}. Dalam memodelkan masalah ini, penulis
menggunakan beberapa asumsi yang relevan sehingga akan
mempermudah penyelesaian masalah. Agar lebih spesifik, gelombang
yang dihasilkan diasumsikan gelombang monokromatik (mempunyai
frekuensi tunggal). Setelah itu, dengan bantuan perangkat lunak
komputer, akan dibuat juga simulasi sederhana yang menggambarkan
evolusi gelombang yang terjadi setelah selang waktu tertentu.

\section{Tujuan Penulisan}

Tujuan subyektif penulisan laporan tugas akhir ini adalah sebagai
syarat untuk memenuhi persyaratan Sidang Sarjana Jurusan
Matematika, Fakultas Matematika dan Ilmu Pengetahuan Alam, ITB
yang diselenggarakan pada tanggal 24 Januari 2001.\newline

Tujuan obyektif penulisan laporan tugas akhir ini adalah untuk memperdalam
bidang ilmu yang sedang penulis tekuni, yakni Matematika, terutama yang
berkaitan dengan penerapan dalam masalah fisis. Selain itu, dengan
memodelkan permasalahan yang datang dari luar disiplin ilmu matematika,
wawasan penulis mengenai keterkaitan ilmu pengetahuan dan teknologi serta
berbagai disiplin ilmu akan semakin luas.\newline

\section{Anggapan Dasar}

Untuk membuat suatu model dari teori pembangkit gelombang, bergantung pada
beberapa faktor berikut :

\begin{itemize}
\item  Sifat fluida yang diasumsikan yang akan membawa pada perumusan
matematis.

\item  Adanya suatu persamaan diferensial pengatur (\textit{governing
differential equation}) dan beberapa syarat batas.

\item  Asumsi berkenaan tipe dan sifat gelombang yang dihasilkan.
\end{itemize}

\section{Hipotesis}

Jika kita mengasumsikan bahwa air adalah salah satu jenis fluida yang
termasuk fluida ideal, kita bisa memperoleh suatu persamaan diferensial
parsial yang memiliki beberapa syarat batas sebagai persamaan diferensial
pengatur (\textit{governing differential equation}), dan asumsi lainnya
bahwa gelombang yang dihasilkan mempunyai tipe monokromatik dan bersifat
periodik, maka kita dapat membuat sebuah teori pembangkit gelombang linear
dua dimensi. \newline

\section{Metodologi Pengerjaan}

\subsection{Metode}

Metode yang penulis gunakan adalah metode analisis teoretis, yaitu
menganalisis secara teori beberapa model yang berkaitan dengan pembangkit
gelombang, termasuk menurunkan beberapa persamaan yang berkaitan dengan
teori ini. Di samping hal ini, penulis juga menggunakan metode analisis
komputasi, yaitu menganalisis dan memecahkan beberapa permasalahan dengan
bantuan komputer.

\subsection{Teknik Pengumpulan Data}

Teknik pengumpulan data yang penulis lakukan adalah studi
kepustakaan dan bimbingan secara spesifik dengan dosen pembimbing
tugas akhir. Studi kepustakaan mencakup mempelajari secara
mandiri beberapa bahan rujukan yang berkaitan dengan Mekanika
Fluida dan Pemodelan Matematika, khususnya yang berkaitan dengan
teori pembangkit gelombang. Beberapa bahan rujukan tersebut dapat
dilihat dalam Daftar Pustaka. Bimbingan tugas akhir mencakup
penugasan bahan yang harus dipelajari, mendiskusikannya apabila
ada kesulitan, dan mencoba melakukan \textit{problem-solving}
secara bersama-sama. Selain itu, untuk memperlengkapi pengalaman,
dilakukan juga presentasi secara berkala atas kemajuan tugas
akhir yang telah dibuat.

\section{Sistematika Pembahasan}

Di dalam bab berikutnya, akan dibahas secara singkat mengenai sifat-sifat
aliran fluida ideal. Bab ini menjelaskan sifat fisis fluida, penurunan
persamaan kontinuitas, persamaan Euler, dan persamaan Bernoulli dengan
menggunakan hukum kekekalan massa dan momentum, serta Teorema Transport.
Beberapa asumsi dan persamaan yang dibahas dalam bab ini akan digunakan
dalam bab berikutnya.

Di dalam bab yang ketiga, sebagai bagian utama tugas akhir ini, akan diulas
inti permasalahan yang telah dikemukakan pada rumusan masalah dan ruang
lingkup kajian pada bab ini. Selain teori pembangkit gelombang yang
disederhanakan, dibahas juga teori pembangkit gelombang linear yang lebih
lengkap dengan menyelesaikan persamaan Laplace sebagai persamaan pengatur (%
\textit{governing equation}) dengan syarat batas lateral di pembangkit
gelombang, syarat batas di dasar \textit{towing tank}, dan syarat batas di
permukaan air. Syarat batas di permukaan ini mencakup syarat batas
kinematika dan dinamika.

Apabila dalam bab-bab sebelumnya didominasi oleh bagian teoretis dan
penurunan rumus, maka di dalam bab yang keempat akan dijelaskan mengenai
bagian simulasi dengan bantuan perangkat lunak komputer, yakni \textbf{%
\textit{Maple V Release 5}}. Bab ini didahului dengan perhitungan bilangan
gelombang progresif dan bilangan gelombang berjalan. Setelah itu, akan
ditampilkan beberapa gambar yang berkaitan dengan gerakan pembangkit
gelombang dan gelombang monokromatik yang dihasilkan.

Di dalam bab yang terakhir, akan diberikan suatu kesimpulan terhadap tugas
akhir yang telah penulis kerjakan. Di samping itu, ditulis juga saran-saran
praktis bagi teman-teman yang lain ataupun siapapun juga yang berminat untuk
melakukan penelitian di bidang ini, mengerjakan, dan memperdalam kembali
bahan tugas akhir ini.

\chapter{Aliran Fluida Ideal}

\section{Sifat Fisis Fluida}

Fluida merupakan objek yang menarik untuk diamati karena kemampuannya untuk
bergerak. Fluida adalah suatu zat yang tidak mengalami hambatan untuk
mengalami perubahan bentuk (\textit{deformation}) dan akan terus berubah
bentuk ketika diberikan tekanan atau gaya. Fluida tidak memiliki bentuk yang
tetap dan selalu mengikuti bentuk ruang yang ditempatinya. Ketika mengalami
pengaruh gaya, fluida akan mengalami perubahan bentuk secara terus-menerus,
yang disebut mengalir (\textit{flow}).

Fluida dapat dikelompokkan menjadi cairan dan gas. Cairan memiliki sifat tak
termampatkan secara relatif (\textit{relatively incompressible}) dan
mempunyai permukaan bebas (\textit{free surface}). Gas memiliki sifat dapat
dimampatkan (\textit{readily compressible}) dan tidak mempunyai permukaan
bebas. Di dalam bab ini, fluida yang akan dibicarakan adalah cairan,
khususnya air. Sifat air yang akan dibahas adalah yang termasuk dalam
\textit{fluida ideal}, yakni fluida yang tak termampatkan (\textit{%
incompressible}) dan yang pengaruh kekentalannya diabaikan atau
fluida tak kental (\textit{inviscid}). Pemilihan asumsi fluida
ideal ini akan digunakan sebagai dasar pada masalah aliran yang
akan dibahas belakangan.\cite{rs,wlh}

\section{Hukum Kekekalan Massa dan Momentum}

Hukum-hukum kekekalan dalam Fisika yang berlaku pada sistem partikel dapat
diterapkan pada fluida, karena fluida adalah kumpulan dari partikel. Dengan
dasar inilah kita memusatkan perhatian pada sekumpulan partikel-partikel
fluida atau suatu \textit{volume material} fluida sehingga kita selalu
menguji kelompok partikel yang sama.

Untuk memudahkan penulisan simbol, kita nyatakan koordinat Kartesius dengan
indeks (1,\,2,\,3) dengan kesepakatan bahwa $x=x_{1},\: y=x_{2},$ dan $%
z=x_{3}.$ Demikian pula untuk komponen-komponen kecepatan, yakni $u=u_{1},\:
v=u_{2},$ dan $w=u_{3}.$

\subsection{Hukum Kekekalan Massa}

Hukum kekekalan massa menyatakan bahwa, "\textit{Massa total dari suatu
sistem yang dibangun oleh sekumpulan partikel adalah selalu tetap, tidak
berkurang ataupun bertambah}."

Dengan kata lain, tidak ada massa suatu fluida yang dapat diciptakan ataupun
dimusnahkan; perubahan massa dalam suatu daerah adalah semata-mata sebagai
akibat dari aliran massa yang melalui suatu batas. Berdasarkan pembatasan
ini, kita definisikan volume fluida adalah $V(t)$, yang dibatasi oleh
permukaan $S$. Jika fluida mempunyai kerapatan $\rho$, maka massa total
fluida dalam volume tersebut diberikan oleh integral $\displaystyle{%
\int\int\int} \rho \;dV $. Dalam permasalahan kita, hukum kekekalan massa
menyatakan bahwa massa total fluida yang mempunyai kerapatan $\rho$ dan
volume $V$ adalah konstan. Dengan kata lain, hukum kekekalan massa
memberikan syarat bahwa integral tersebut adalah konstan, yakni
\begin{equation}
\mathop{\int\int\int}_{V} \rho \; dV = \textmd{konstan},
\end{equation}
atau
\begin{equation}
\frac{d}{dt}\Bigg(\mathop {\int\int\int}_{V} \rho \; dV \Bigg)= 0.
\label{ms}
\end{equation}

\subsection{Hukum Kekekalan Momentum}

Hukum kekekalan momentum menyatakan bahwa, "\textit{Momentum total dari
suatu sistem yang dibangun oleh sekumpulan partikel yang saling berinteraksi
adalah tetap, asalkan tidak ada gaya luar yang bekerja pada sistem tersebut}."

Dengan cara yang serupa, kerapatan momentum partikel fluida sama dengan
vektor $\rho\,\mathbf{u}$, dengan komponen $\rho\,u_{i}$. Dalam permasalahan
kita, hukum kekekalan momentum memberikan syarat bahwa jumlah semua gaya
yang bekerja pada volume fluida sama dengan perubahan momentum fluida
tersebut. Berdasarkan kerangka acuan Newton, hukum kekekalan momentum dapat
dinyatakan sebagai :
\begin{equation}
\frac{d}{dt}\; \Bigg(\mathop{\int\int\int}_{V} \rho\, u_{i} \; dV\Bigg) = %
\mathop{\int\int}_{S} \tau_{ij} \; n_{j} \; dS + \mathop{\int\int\int}_{V}
F_{i} \; dV,  \label{mt}
\end{equation}
dengan\newline
$\rho\,u_{i} \equiv$ komponen kerapatan momentum partikel fluida; \newline
$\tau_{ij} \equiv$ tensor tekanan yang bekerja pada volume fluida; \newline
$n_{j} \equiv$ komponen vektor normal satuan pada permukaan fluida; dan
\newline
$F_{i} \equiv$ gaya luar yang bekerja pada partikel fluida. \newline
Integral permukaan di atas adalah komponen ke-i dari gaya-gaya permukaan
yang bekerja pada permukaan $S$, dan integral volume yang terakhir adalah
penjumlahan dari gaya luar (\textit{body force}), seperti yang berkaitan
dengan gaya gravitasi.

Dengan menggunakan \textbf{Teorema Divergensi}\footnote[1]{%
Di negara-negara Barat, teorema ini dikenal dengan istilah \textit{Teorema
Gauss} atau \textit{Teorema Green}, sedangkan di wilayah Blok Timur, teorema
ini dikenal dengan \textit{Teorema Ostrogradsky}, diambil berdasarkan nama
seorang matematikawan Rusia. \cite{tgb}}, suku pertama ruas kanan persamaan (%
\ref{mt}) dapat dinyatakan dengan integral lipat tiga divergensi suatu medan
vektor atas daerah yang melingkupi permukaan tersebut
\begin{equation}
\mathop{\int\int}_{S} \mathbf{Q}\cdot\mathbf{n} \; dS = \mathop{\int\int\int}%
_{V} \nabla\cdot\mathbf{Q} \; dV.  \label{tdg}
\end{equation}
Apabila dituliskan dalam komponen, maka persamaan di atas menjadi
\begin{equation}
\mathop{\int\int}_{S} Q_{i}\; n_{i} \; dS = \mathop{\int\int\int}_{V} \frac{%
\partial Q_{i}}{\partial x{i}}\; dV.  \label{td}
\end{equation}
Di sini, \textbf{Q} adalah sembarang medan vektor yang kontinu dan
terdiferensialkan dalam volume $V$, dan vektor normal satuan $\mathbf{n}$
adalah vektor normal arah luar yang keluar dari $V$ pada permukaan $S$.

Dengan menggunakan (\ref{td}) untuk mentransformasikan permukaan integral
pada (\ref{mt}), kita peroleh
\begin{equation}
\frac{d}{dt}\Bigg( \mathop{\int\int\int}_{V} \rho\; u_{i} \;dV \Bigg)= %
\mathop{\int\int\int}_{V} \Bigg(\frac{\partial \tau_{ij}}{\partial x_{j}}+
F_{i}\Bigg)\;dV  \label{tr}
\end{equation}

Persamaan (\ref{ms}) dan (\ref{tr}) masing-masing menyatakan hukum kekekalan
massa dan hukum kekekalan momentum untuk fluida, yang diintegralkan atas
sembarang volume material $V(t)$.

\section{Teorema Transport (\textit{Transport Theorem})}

Misalkan bentuk umum integral volume adalah
\begin{equation}
I(t)=\mathop{\int\int\int}_{V(t)} f(\mathbf{x},t)\; dV.
\end{equation}
Di sini $f$ adalah sembarang fungsi skalar terdiferensialkan yang bergantung
pada posisi $\mathbf{x}$ dan waktu $t$ yang diintegrasikan atas volume $V(t)$%
, yang juga dapat berubah terhadap waktu. Dengan demikian, permukaan batas $%
S $ volume ini akan berubah terhadap waktu, dan kecepatan normalnya
dinyatakan dengan $U_{n}.$

Dengan cara yang lazim digunakan dalam Kalkulus dasar kita perhatikan
perbedaan
\begin{equation}
\Delta I = I(t+\Delta t)-I(t) = \mathop{\int\int\int}_{V(t+\Delta t)} f(%
\mathbf{x},t+\Delta t)\;dV - \mathop{\int\int\int}_{V(t)}f(\mathbf{x},t)\;dV.
\label{di}
\end{equation}
Dari deret Taylor fungsi $f(\mathbf{x},t+ \Delta t)$, kita punya
\begin{equation}
f(\mathbf{x},t+\Delta t)=f(\mathbf{x},t)+\Delta t \;\frac{\partial f(\mathbf{%
x},t)}{\partial t}+\frac{1}{2!}\; (\Delta t)^2\; \frac{\partial ^2 f(\mathbf{%
x},t)}{\partial t^2}+ \cdots .
\end{equation}
Dengan mengabaikan orde suku kedua yang sebanding dengan $(\Delta t)^2$,
kita peroleh
\begin{equation}
f(\mathbf{x},t+\Delta t)=f(\mathbf{x},t)+\Delta t \;\frac{\partial f(\mathbf{%
x},t)}{\partial t}.
\end{equation}
Analisis yang serupa dapat kita terapkan pada volume material V(t)
\begin{equation}
V(t+\Delta t)=V(t)+ \Delta t \;\frac{dV}{dt}+\frac{1}{2!}\;(\Delta t)^2\;
\frac{d^2V}{dt^2} + \cdots .
\end{equation}
Dengan mengabaikan suku kedua dan suku-suku berikutnya, kita peroleh
\begin{equation}
\Delta V = V(t+\Delta t)-V(t).
\end{equation}
Jadi, $V(t+\Delta t)$ berbeda dari $V(t)$ dengan suatu volume tipis $\Delta
V $ yang termuat di dalam permukaan-permukaan $S(t+\Delta t)$ dan $S(t)$
yang berbatasan dan sebanding dengan $\Delta t.$ Dari persamaan (\ref{di})
kita memperoleh bahwa
\begin{eqnarray}
\Delta I &=& \mathop{\int\int\int}_{V+\Delta V}\Bigg(f+\Delta t\; \frac{%
\partial f}{\partial t}\Bigg)\;dV - \mathop{\int\int\int}_{V}f\; dV \\
&=&\Delta t\mathop{\int\int\int}_{V} \frac{\partial f}{\partial t}\;dV + %
\mathop{\int\int\int}_{\Delta V}f \;dV + O \bigg[(\Delta t)^2\bigg],
\end{eqnarray}
dengan suku terakhir menyatakan galat orde dua yang sebanding dengan $%
(\Delta t)^2$.

Untuk mengevaluasi integral atas volume kecil $\Delta V$, kita perhatikan
bahwa daerah yang tipis ini mempunyai ketebalan yang sama dengan jarak
antara $S(t)$ dan $S(t+\Delta t)$. Ketebalan ini sama dengan komponen normal
dari jarak yang dijalani oleh $S(t)$ dalam waktu $\Delta t$ yang sama dengan
hasil kali $U_{n}\;\Delta t$. Jadi, integral pada suku kedua di atas hanya
memberikan kontribusi orde pertama, yang sebanding dengan $\Delta t$.
Derajat keakuratan integran $f$ dapat diasumsikan konstan sepanjang daerah
tipis dalam arah normal permukaan $S$. Dengan mengintegrasikan hanya arah
ini saja, kita mempunyai
\begin{equation}
\Delta I= \triangle t \mathop{\int\int\int}_{V}\frac{\partial f}{\partial t}%
\;dV + \mathop{\int\int}_{S}(U_{n}\;\triangle t)\;f\;dS + O\bigg[(\triangle
t)^2\bigg].
\end{equation}
Akhirnya, kita mendapatkan hasil yang diinginkan dengan membagi kedua ruas
pada persamaan di atas dan mengambil limit selang waktunya menuju nol
\begin{equation}
\frac{dI}{dt}= \displaystyle\lim_{\triangle t \rightarrow 0}\; \frac{%
\triangle I }{\triangle t}= \mathop{\int\int\int}_{V}\frac{\partial f}{%
\partial t}\;dV + \mathop{\int\int}_{S}f\;U_{n}\;dS  \label{tt}
\end{equation}

Persamaan (\ref{tt}) dikenal sebagai \textbf{Teorema Transport} atau \textbf{%
Persamaan Transport} (\textit{Transport Theorem, Transport Equation}).
Integral permukaan pada persamaan ini menyatakan transpor besaran $f$ yang
keluar dari volume $V$, sebagai hasil dari gerakan batasnya. Dalam kasus
khusus ketika $S$ tetap dan $U_{n}=0$, persamaan (\ref{tt}) tereduksi ke
dalam bentuk yang lebih sederhana ketika diferensial yang keluar dari dalam
tanda integral dibenarkan. Untuk penjelasan lebih lengkap, bisa dilihat di
\cite{njn}.

Dalam kasus lain yang khususnya menarik adalah sebagai berikut. Karena
volume material $V$ selalu dibentuk oleh partikel-partikel fluida yang sama,
maka permukaan $S$ bergerak dengan kecepatan normal yang sama dengan
kecepatan fluida sendiri dan $U_{n}=\mathbf{{u}\cdot{n}=u_{i}\;n_{i}}$.
Dalam kasus ini, dengan menggunakan Teorema Divergensi pada persamaan (\ref
{tdg}), maka persamaan (\ref{tt}) dapat dituliskan menjadi
\begin{eqnarray}
\frac{dI}{dt}&=&\frac{d}{dt}\Bigg(\mathop{\int\int\int}_{V(t)}f\;dV\Bigg)
\nonumber \\
&=&\mathop{\int\int\int}_{V(t)}\frac{\partial f}{\partial t}\;dV + %
\mathop{\int\int}_{S}fu_{i}n_{i}\;dS \\
&=& \mathop{\int\int\int}_{V(t)}\frac{\partial f}{\partial t}\;dV + %
\mathop{\int\int\int}_{V(t)}\frac{\partial}{\partial x_{i}}\;\bigg(f u_{i}%
\bigg)\;dV \\
&=& \mathop{\int\int\int}_{V(t)}\Bigg[\frac{\partial f}{\partial t }+\frac{%
\partial}{\partial x_{i}}\;\bigg(fu_{i}\bigg)\Bigg]\;dV  \label{fdv}
\end{eqnarray}

\section{Persamaan Kontinuitas}

\label{kontinuitas}

Persamaan ini berkaitan erat dengan hukum kekekalan massa dan Teorema
Transport karena ia diturunkan dari kedua persamaan tersebut. Perhatikan
kembali persamaan (\ref{ms}) yang menyatakan hukum kekekalan massa. Dengan
memanfaatkan hasil Teorema Transport pada persamaan (\ref{fdv}) kita
memperoleh
\begin{eqnarray}
\frac{d}{dt}\Bigg(\mathop {\int\int\int}_{V} \rho \; dV\Bigg) = %
\mathop{\int\int\int}_{V}\Bigg[\frac{\partial \rho}{\partial t }+\frac{%
\partial}{\partial x_{i}}\;\bigg(\rho\, u_{i}\bigg)\Bigg]\;dV = 0.
\label{ce}
\end{eqnarray}
Karena integral terakhir dievaluasi pada waktu sesaat yang tetap, perbedaan
bahwa $V$ adalah volume material tidaklah perlu pada taraf/keadaan ini.
Lebih jauh, volume tersebut dapat disusun oleh sembarang kelompok
partikel-partikel fluida. Dengan demikian, integran di atas sama dengan nol
pada seluruh fluida. Jadi, integral volume pada persamaan (\ref{ce}) dapat
diganti dengan suatu persamaan diferensial parsial yang menyatakan hukum
kekekalan massa dalam bentuk sebagai berikut :
\begin{eqnarray}
\frac{\partial \rho}{\partial t}+\frac{\partial}{\partial x_{i}}%
\;(\rho\,u_{i})=0
\end{eqnarray}
Apabila dinyatakan dengan komponen kecepatan tiga dimensi, persamaan
tersebut mempunyai bentuk :
\begin{eqnarray}
\frac{\partial \rho}{\partial t}+\frac{\partial \rho u}{\partial x}+\frac{%
\partial \rho v}{\partial y}+\frac{\partial \rho w }{\partial z}=0,
\label{ko}
\end{eqnarray}
atau bila ditulis dalam bentuk diferensial :
\begin{eqnarray}
\frac{\partial\rho}{\partial t}+\nabla\cdot\ (\rho\:\mathbf{u})=0
\end{eqnarray}
Persamaan penting inilah yang dikenal sebagai \textit{kondisi untuk hukum
kekekalan massa} atau \textbf{persamaan kontinuitas} \textit{untuk suatu
aliran fluida termampatkan}. Persamaan (\ref{ko}) menggambarkan rata-rata
perubahan rapat massa pada suatu titik tetap sebagai hasil dari perubahan
pada vektor kecepatan massa $\rho\:\mathbf{u}$. Dengan mengekspansi
suku-suku yang memuat hasil kali kerapatan dan komponen kecepatan, kita bisa
menurunkan bentuk lain dari persamaan kontinuitas.
\begin{equation}
\frac{\partial \rho}{\partial t}+u\;\frac{\partial \rho}{\partial x}+\rho\;%
\frac{\partial u}{\partial x}+v\;\frac{\partial \rho}{\partial y}+\rho\;%
\frac{\partial v}{\partial y}+w\;\frac{\partial \rho}{\partial z}+\rho\;%
\frac{\partial w}{\partial z}=0 \\
\end{equation}
\begin{equation}
\Bigg(\frac{\partial \rho}{\partial t}+u\;\frac{\partial \rho}{\partial x}%
+v\;\frac{\partial \rho}{\partial y}+w\;\frac{\partial \rho}{\partial z}%
\Bigg)+\rho\;\Bigg(\frac{\partial u}{\partial x}+\frac{\partial v}{\partial y%
}+\frac{\partial w}{\partial z}\Bigg)=0.  \label{eks}
\end{equation}
Dengan menggunakan operator turunan material atau turunan total (lihat
Apendiks A), persamaan (\ref{eks}) dapat dituliskan menjadi
\begin{equation}
\frac{D\rho}{Dt}+\rho\;(\nabla \cdot\mathbf{u}) = 0.  \label{mat}
\end{equation}
Jika alirannya tunak (\textit{steady}), yakni tidak bergantung waktu, maka $%
\displaystyle \frac{\partial \rho}{\partial t}=0$ sehingga persamaan
kontinuitas menjadi $\nabla \cdot(\rho\;\mathbf{u}) = 0.$ Dengan menggunakan
asumsi bahwa air adalah fluida yang tak termampatkan {\itshape%
(incompressible)}, yakni mempunyai rapat massa $\rho$ yang konstan, maka
persamaan kontinuitasnya (\ref{eks}) tereduksi menjadi
\begin{equation}
\frac{\partial u}{\partial x}+\frac{\partial v}{\partial y}+\frac{\partial w%
}{\partial z}=0.
\end{equation}
Atau dalam bentuk vektor, yang bisa juga diturunkan dari persamaan (\ref{mat}%
), didapat
\begin{equation}
\nabla\cdot\ \mathbf{u}=0.  \label{vpk}
\end{equation}
Hubungan inilah yang dikenal sebagai \textbf{kondisi tak termampatkan} (%
\textit{condition of incompressibility}). Kondisi ini menyatakan fakta bahwa
keseimbangan antara aliran yang keluar (\textit{outflow}) dan aliran yang
masuk (\textit{inflow}) untuk suatu elemen volume atau volume material
adalah nol pada tiap saat.

\section{Persamaan Euler}

Berbeda dengan persamaan kontinuitas yang berkaitan dengan hukum kekekalan
massa, persamaan ini berkaitan erat dengan hukum kekekalan momentum dan
Teorema Transport karena ia diturunkan dari kedua persamaan tersebut.
Perhatikan kembali persamaan (\ref{tr}) yang menyatakan hukum kekekalan
momentum. Dengan memanfaatkan hasil Teorema Transport pada persamaan (\ref
{fdv}) kita memperoleh
\begin{eqnarray}
\mathop{\int\int\int}_{V}\Bigg[\frac{\partial \rho u_{i}}{\partial t }+\frac{%
\partial}{\partial x_{j}}\;\bigg(\rho u_{i}\,u_{j}\bigg)\Bigg]\;dV = %
\mathop{\int\int\int}_{V} \Bigg(\frac{\partial \tau_{ij}}{\partial x_{j}}+
F_{i}\Bigg)\;dV.  \label{eu1}
\end{eqnarray}
Karena integral di atas sama-sama dilakukan atas volume yang sama, maka
persamaan (\ref{eu1}) haruslah dipenuhi untuk integrannya saja, yakni
\begin{eqnarray}
\frac{\partial}{\partial t}(\rho\,u_{i})+\frac{\partial}{\partial x_{j}}\;%
\bigg(\rho\,u_{i}\:u_{j}\bigg)=\frac{\partial \tau_{ij}}{\partial x_{j}}+
F_{i}.  \label{eu2}
\end{eqnarray}
Selanjutnya, jika turunan hasil kali di ruas kiri dari persamaan ini
diekspansi dengan aturan rantai, maka kita peroleh
\begin{eqnarray}
u_{i}\frac{\partial \rho}{\partial t}+\rho\frac{\partial u_{i} }{\partial t}%
+u_{i}u_{j}\frac{\partial \rho}{\partial x_{j}}+\rho u_{j}\frac{\partial
u_{i}}{\partial x_{j}}+\rho u_{i}\frac{\partial u_{j}}{\partial x_{j}}=\frac{%
\partial \tau_{ij}}{\partial x_{j}}+F_{i}.
\end{eqnarray}
Dengan asumsi bahwa fluida yang diamati adalah tak termampatkan (\textit{%
incompressible}) dan mempunyai rapat massa $\rho$ yang konstan serta dari
hasil persamaan kontinuitas pada (\ref{vpk}), kita sampai pada \textbf{%
persamaan Euler}\footnote[1]{\textbf{Euler, Leonhard
(1707-83).}$\;$ Matematikawan Swiss dengan pengertian tenang yang
memberikan sumbangan dasar yang sangat berarti bagi banyak cabang
matematika dan penerapannya: persamaan diferensial, deret tak
hingga, analisis kompleks, mekanika dan hidrodinamika, serta
kalkulus variasi; dia juga sangat berpengaruh dalam mempromosikan
penggunaan dan pengertian analisis.\cite{jrs}}, yang juga dikenal
sebagai \textbf{persamaan momentum}:
\begin{eqnarray}
\frac{\partial u_{i} }{\partial t}+u_{j}\;\frac{\partial u_{i}}{\partial
x_{j}}=\frac{1}{\rho}\;\Bigg(\frac{\partial \tau_{ij}}{\partial x_{j}}+F_{i}%
\Bigg).  \label{tum}
\end{eqnarray}
Dengan menggunakan turunan material (lihat kembali Apendiks A), persamaan
Euler dapat dinyatakan menjadi
\begin{eqnarray}
\frac{D u_{i}}{Dt}=\frac{1}{\rho}\;\Bigg(\nabla \cdot \tau_{i}+F_{i}\Bigg).
\end{eqnarray}

\section{Persamaan Bernoulli}

\label{bernoulli}

Kita tinjau kembali persamaan (\ref{tum}). Dalam aliran yang tidak ada
gesekan (\textit{frictionless flow}), tidak terdapat tegangan geser dan
tegangan normal hanyalah tekanan yang \textit{isotropik}. Kita sepakati
tegangan normal (\textit{normal stress}) $\tau_{11}, \tau_{22},$ dan $%
\tau_{33}$ berarah positif jika mereka adalah \textit{tension}. Karena kita
telah mengasumsikan bahwa air adalah fluida yang tak kental (\textit{%
inviscid fluid}), maka tensor tegangan \mbox{\boldmath
$\tau$} hanya mempunyai komponen-komponen normal dari tekanan. Untuk
penjelasan lebih lanjut, silakan lihat di \cite{hwf,kjp,njn}. Lalu, kita
bentuk menjadi
\begin{equation}
\tau_{11} = \tau_{22} = \tau_{33} = -\,p,
\end{equation}
sehingga persamaan momentum menjadi
\begin{equation}
\frac{\partial u_{i} }{\partial t}+u_{j}\;\frac{\partial u_{i}}{\partial
x_{j}}=\frac{1}{\rho}\;\Bigg(\frac{\partial p}{\partial x_{j}}+F_{i}\Bigg),
\end{equation}
atau dalam notasi vektor
\begin{equation}
\frac{D \mathbf{u}}{Dt}=\frac{1}{\rho}\;\Bigg(-\nabla p + \mathbf{F}\Bigg),
\end{equation}
dengan $\mathbf{u}=(u,\,v,\,w), \quad \mathbf{F}=(0,\,0,\,-\rho\,g)$, dan $p$
adalah tekanan. Jadi,
\begin{equation}
\frac{\partial \mathbf{u}}{\partial t}+(\mathbf{u}\cdot\nabla)\;\mathbf{u}%
=-\nabla\;\Bigg(\frac{p}{\rho}+ g\,z\Bigg).
\end{equation}
Dari identitas vektor, kita punya
\begin{equation}
\mathbf{u}\times(\nabla\times\mathbf{u})= (\mathbf{u}\cdot\nabla)\;\mathbf{u}%
-\nabla\;\Bigg( \frac{1}{2}\mid\mathbf{u}\mid^2\Bigg).
\end{equation}
Apabila diasumsikan bahwa air adalah fluida yang tak berputar (\textit{%
irrotational})\footnote[1]{%
Fluida \textit{irrotational} di sini memaksudkan tidak adanya pusaran arus (%
\textit{eddy}) atau pusaran air (\textit{whirlpool}).}, yakni $\nabla \times
\mathbf{u} = \mathbf{0} \label{curl}$, maka $\displaystyle(\mathbf{u}%
\cdot\nabla)\;\mathbf{u}=\nabla\;\Bigg(
\frac{1}{2}\mid\mathbf{u}\mid^2\Bigg).$ Selain itu, karena asumsi ini, maka
vektor kecepatan \textbf{u} dapat dinyatakan sebagai gradien dari skalar
potensial kecepatan, yakni $\mathbf{u}=\nabla\,\phi$. (Untuk keterangan yang
terperinci, silakan lihat di \cite{drg} dan \cite{lld}). Ini dilakukan
dengan tujuan untuk mempermudah kerja kita, karena kita berurusan dengan
besaran skalar yang lebih mudah dianalisis daripada besaran vektor.

Dengan demikian,
\begin{equation}
\frac{\partial \,\nabla \phi }{\partial t}+\nabla \;\Bigg(\frac{1}{2}\mid
\mathbf{u}\mid ^{2}\Bigg)=-\nabla \;\Bigg(\frac{p}{\rho }+g\,z\Bigg),
\end{equation}
atau
\begin{equation}
\nabla \;\Bigg(\frac{\partial \phi }{\partial t}+\frac{1}{2}\mid \mathbf{u}%
\mid ^{2}+\frac{p}{\rho }+g\,z\Bigg)=0.  \label{ber1}
\end{equation}
Apabila persamaan (\ref{ber1}) diintegrasikan terhadap seluruh
ruang kita mendapatkan \textbf{persamaan
Bernoulli}\footnote[7]{\textbf{Bernoulli, Daniel (1700-82).}$\;$
Ilmuwan kelahiran Belanda ini adalah anggota keluarga Swiss
terkenal yang terdiri atas sekitar 10 matematikawan (ayah, putra,
paman, sepupu), dikenal baik karena karyanya mengenai aliran
fluida dan teori kinetik gas; persamaannya untuk aliran fluida
pertama kali diterbitkan pada 1783; dia juga menekuni bidang
astronomi dan kemagnetan, serta ilmuwan pertama yang memecahkan
persamaan Ricatti.\cite{jrs}}
\begin{equation}
\frac{\partial \phi }{\partial t}+\frac{1}{2}\mid \mathbf{u}\mid ^{2}+\frac{p%
}{\rho }+g\,z=\textmd{konstan}\equiv f(t).  \label{ber2} \\
\end{equation}
\vspace*{1cm}
\begin{figure}[h]
\begin{center}
{Bagan keterkaitan fluida ideal} \vspace*{1cm} \\
\includegraphics[scale = 4]{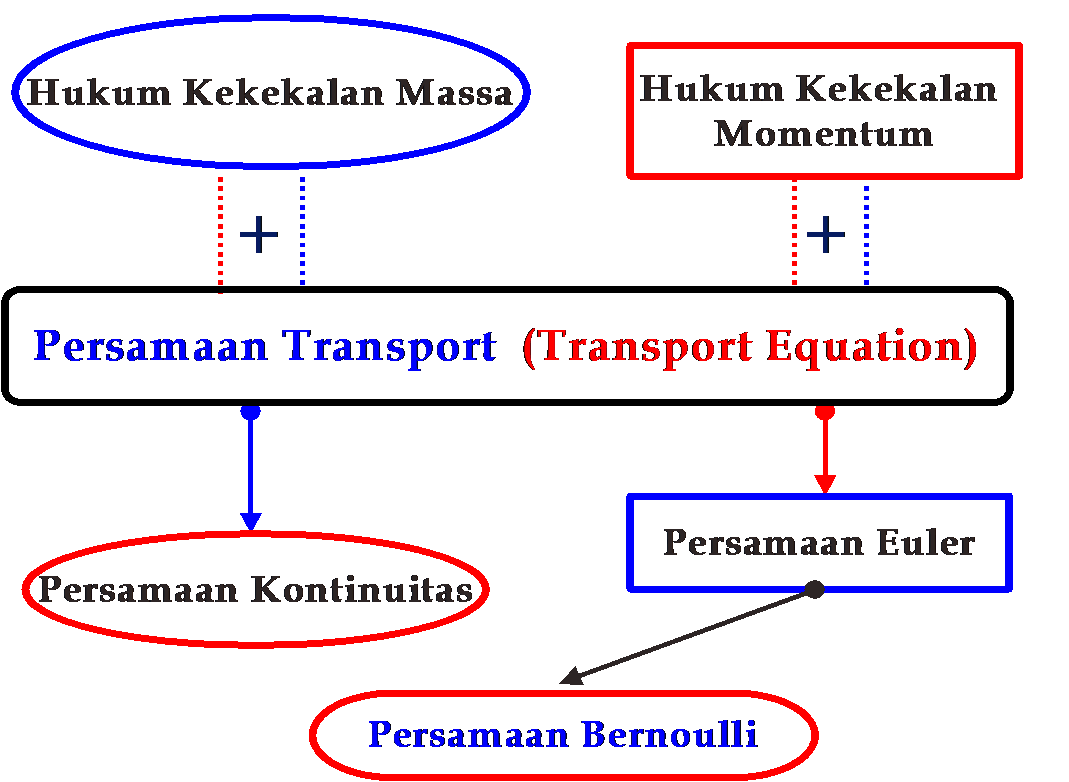}
\end{center}
\end{figure}

\chapter{Teori Pembangkit Gelombang Dua Dimensi}

Kita akan membandingkan dua teori pembangkit gelombang yang telah diselidiki
dan dibuat model matematisnya. Bagian yang pertama adalah teori yang
disederhanakan yang berangkat dari ide dasar berkenaan volume air yang
dipindahkan oleh pembangkit gelombang tipe \textit{'flap'} pada sebuah
tangki atau kolam air yang cukup panjang (\textit{towing tank}). Kemudian,
bagian yang kedua, akan diulas mengenai teori yang lebih terperinci yang
dihasilkan oleh pembangkit gelombang tipe {\itshape'flap'} di salah satu
ujung \textit{towing tank} tadi. Dari hasil kedua teori ini, kita akan
mengamati perbandingan antara amplitudo gelombang yang dihasilkan $H$ dengan
besarnya simpangan maksimum pembangkit gelombang $S$ (\textit{stroke}).

Di dalam pembahasan tugas akhir ini, teori yang lebih terperinci tersebut
didasarkan atas {\itshape'teori gelombang linear'\ } yang dibahas dan
dikembangkan oleh Airy\footnote[1]{\textbf{Airy, Sir George Biddel (1801-92).%
}$\;$ Matematikawan dan fisikawan Inggris, yang menjadi astronom
kerajaan (Astronomer Royal) selama 46 tahun; dia memberikan
sumbangan pada teori-teori cahaya dan, tentulah, pada astronomi,
tetapi juga pada gravitasi, kemagnetan dan suara, sebagaimana
halnya pada perambatan gelombang secara umum dan pada teori
pasang-surut air pada khususnya. \cite{jrs}} sekitar 160 tahun
yang lalu, dalam menganalisis perilaku gelombang air laut.
\cite{aph}

\section{Teori Pembangkit Gelombang yang\\ Disederhanakan}

Di dalam air dangkal, teori sederhana untuk profil perambatan gelombang yang
dihasilkan oleh suatu pembangkit gelombang pertama kali dikembangkan oleh
Galvin pada tahun 1964, yang bernalar bahwa air yang dipindahkan oleh suatu
pembangkit gelombang seharusnya sama dengan volume puncak dari bentuk
gelombang yang merambat. Pertimbangkan sebuah pembangkit gelombang tipe flap
yang ujung bawahnya terikat dan mempunyai simpangan maksimum \textit{stroke}
$S$. Kedalaman air dalam \textit{towing tank} adalah $h$. Volume air yang
dipindahkan oleh simpangan flap adalah $\displaystyle\frac{1}{2}\,S\,h$.
Lihat Gambar 3.1. Volume air dalam suatu puncak gelombang adalah $%
\displaystyle\int_{0}^{\frac{1}{2}\,L}\frac{1}{2}\,H\sin \:k\,x\:dx$ dengan $k=%
\displaystyle\frac{2\,\pi }{L}$ adalah bilangan gelombang. Dengan demikian,
\begin{eqnarray}
\displaystyle\int\limits_{0}^{\frac{1}{2}\,L}\frac{1}{2}\:H\:\sin
\:(k\,x)\:dx
&=&\displaystyle\int_{0}^{\frac{1}{2}\,L}\frac{1}{2}\:H\:\sin \Bigg(\frac{%
2\,\pi }{L}x\Bigg)\:dx \\ \vspace{0.5cm}
&=&\displaystyle\frac{H\,L}{2\,\pi } \\ \vspace{0.5cm}
&=&\displaystyle\frac{H}{k}
\end{eqnarray}
Dengan menyamakan kedua volume,
\[
\frac{1}{2}\,S\,h=\frac{H}{k}=H\,\frac{L}{2\,\pi }=\frac{H}{2}\Bigg(\frac{L}{%
2}\Bigg)\:\frac{2}{\pi }
\]
dengan $\displaystyle\frac{2}{\pi }$ adalah faktor yang menyatakan
perbandingan antara luas daerah yang dibentuk oleh profil gelombang dengan
persegipanjang yang melingkupinya (yang disebut juga dengan istilah {\textbf{%
faktor luas.}}) Dari persamaan $\displaystyle\frac{1}{2}\,S\,h=\frac{H}{k}$,
kita bisa mengetahui besarnya perbandingan ketinggian gelombang $H$ dengan
simpangan $S$, yaitu
\[
\frac{H}{S}=\frac{1}{2}\,k\,h.
\]
Hubungan ini berlaku hanya untuk daerah air dangkal, yakni $\displaystyle k\,h<\frac{1}{10}\pi$.
\begin{figure}[h]
\begin{center}
\includegraphics[scale = 3]{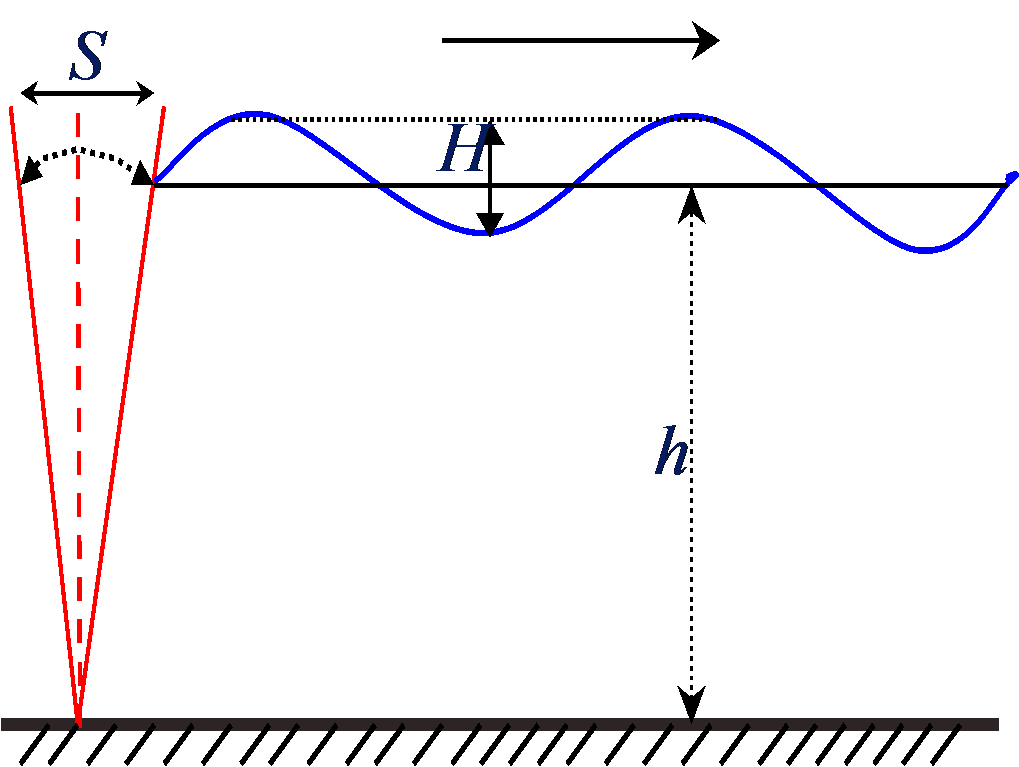}
\end{center}
\caption{Skema pembangkit gelombang tipe flap.}
\end{figure}

\section{Teori Pembangkit Gelombang Linear}

Berdasarkan asumsi bahwa air adalah fluida yang tak termampatkan (\textit{%
incompressible}), maka dari hasil di sub-bab \ref{kontinuitas}, persamaan (%
\ref{vpk}), kita telah mempunyai $\nabla\cdot\mathbf{u}=0$. Selain itu,
asumsi lain yang digunakan adalah tidak adanya pusaran air (\textit{whirpool}%
), dengan kata lain, aliran air tak berputar
(\textit{irrotational}). Berdasarkan asumsi ini, dari penjelasan
di sub-bab \ref{bernoulli}, kita telah mendapatkan
$\mathbf{u}=\nabla\,\phi$. Apabila kedua persamaan ini
digabungkan, maka kita peroleh
$\nabla\cdot\nabla\,\phi=\nabla^{2}\phi=0$. Persamaan ini dikenal
dengan persamaan Laplace,\footnote[1]{\textbf{Laplace, Marquis
Pierre Simon de (1749-1827).}$\;$ Matematika-fisikawan Perancis
yang memberikan sumbangan pada studi mengenai celestial mekanika
dan, pada khususnya, menjelaskan orbit planet Yupiter dan
Saturnus; dia mengembangkan ide dalam menggunakan fungsi potensial
dan fungsi ortogonal, serta memperkenalkan transformasi
integralnya; dia juga memainkan peranan penting dalam perkembangan
teori peluang.} yang merupakan persamaan diferensial pengatur
(\textit{governing differential equation}) untuk potensial
kecepatan $\phi$. Dalam koordinat kartesius untuk dua dimensi
\textbf{---} arah horizontal (sumbu $x$) dan arah vertikal (sumbu
$z$) \textbf{---} persamaan Laplace dapat dituliskan sebagai
\begin{eqnarray}
\nabla^2\phi = \frac{\partial^2\phi}{\partial x^2}+\frac{\partial^2\phi}{%
\partial z^2}=0,
\end{eqnarray}
dan persamaan ini berlaku pada daerah $-h \,\leq\, z \, \leq\, \eta(x,\,t)$,
\quad $0 \,\leq\, x\, \leq \,\infty$.

Masalah nilai batas yang terkait di sini adalah persoalan nilai batas untuk
perambatan gelombang dua dimensi dalam suatu fluida ideal. Kita akan
memecahkan persamaan tersebut dengan beberapa syarat batas yang sesuai
dengan kondisi \textit{towing tank}. Setidaknya, kita memiliki tiga buah
kondisi syarat batas yang akan membantu kita untuk memecahkan persamaan
penuntun di atas. Berikut ini adalah syarat batas yang berkaitan dengan
kondisi \textit{towing tank}.

\subsection{Syarat batas lateral di pembangkit gelombang\\ (\textit{pseudo-boundary})}

Misal fungsi yang menggambarkan pergeseran horizontal pada permukaan
pembangkit gelombang adalah
\begin{eqnarray}
F(x,\,z,\,t) = x - \frac{1}{2}\,S(z)\:\sin \:\omega\, t = 0,
\end{eqnarray}
maka dengan melakukan diferensial total pada $F(x,\,z,\,t)$, kita peroleh
\begin{equation}
\frac{DF}{Dt} = \frac{\partial F}{\partial t} + \frac{\partial F}{\partial x}
\frac{dx}{dt} + \frac{\partial F}{\partial z} \frac{dz}{dt}
\end{equation}
\begin{equation}
0 = -\frac{1}{2}\:\omega\: S(z)\: \cos\: \omega\, t + v - \frac{1}{2}\,w\:
\frac{dS(z)}{dz}\: \sin\: \omega\, t
\end{equation}
\begin{equation}
v - \frac{1}{2}\:w\: \frac{dS(z)}{dz}\: \sin\: \omega\, t = \frac{1}{2}%
\:\omega\: S(z)\: \cos\: \omega\, t \quad \textmd{pada}
\;F(x,\,z,\,t)=0
\end{equation}
dengan $v$ dan $w$ masing-masing adalah komponen kecepatan pada arah $x$ dan
$z$. (Ingat bahwa \textbf{u} = (v,w) adalah vektor kecepatan dua dimensi.)
Atau, persamaan ini juga dapat dituliskan sebagai
\begin{equation}
\phi_{x}=\frac{1}{2}\:\Bigg(\phi_{z}\:S\,^{\prime}(z)\:\sin\: \omega\,
t+\omega\: S(z)\:\cos\: \omega\, t\Bigg).
\end{equation}

Untuk pergeseran $S(z)$ yang cukup kecil dan kecepatan yang juga
kecil, kita dapat melinearkan persamaan tersebut dengan
mengabaikan suku kedua di ruas kiri. Seperti yang akan dilakukan
pada permukaan bebas, akan lebih tepat apabila kita mengungkapkan
syarat pada batas lateral yang bergerak dalam suku-suku yang
dievaluasi di posisi rata-ratanya, yakni pada $x = 0.$ Untuk
melakukan hal ini, kita ekspansikan syarat tersebut dalam deret
Taylor terpotong (lebih spesifik, deret Maclaurin terpotong).
\[
\displaystyle\left.\Bigg(\phi_{x}-\frac{1}{2}\,\omega\,S(z)\,\cos \omega\,t%
\Bigg)\right|_{x=\frac{1}{2}\,S(z)\,\sin\,\omega\,t} =\left.\Bigg(\phi_{x}-%
\frac{1}{2}\,\omega\,S(z)\,\cos\,\omega\, t\Bigg)\right|_{x=0}
\]
\begin{equation}
\displaystyle +\frac{1}{2}\,S(z)\,\sin\,\omega\,t\: \frac{\partial}{\partial
x}\left.\Bigg(\phi_{x}-\frac{1}{2}\,\omega\,S(z)\,\cos\,\omega\, t\Bigg)%
\right|_{x=0} + \dots.
\end{equation}
Jelaslah, hanya suku pertama dari ekspansi di atas yang linear terhadap $%
\phi_{x}$ dan S(z), sedangkan suku-suku yang lain dapat dibuang karena
diasumsikan sangat kecil. Dengan demikian, syarat batas lateral akhir
sebagai akibat dari proses linearisasi adalah persamaan
\begin{equation}
u(0,\,z,\,t)=\phi_{x}=\frac{1}{2}\;\omega\; S(z)\;\cos(\omega t).
\end{equation}

\subsection{Syarat batas di dasar \textit{towing tank} (persyaratan
kinematika)}

Karena tidak ada aliran air yang menembus ke dasar \textit{towing tank},
maka kecepatan fluida pada arah vertikal adalah sama dengan nol, yakni
\begin{equation}
\phi_{z} = \frac{\partial\phi}{\partial z} = 0, \quad
\textmd{pada} \quad z = -h.
\end{equation}

\subsection{Syarat batas di permukaan air}

\begin{itemize}
\item  \textbf{Syarat batas kinematika permukaan bebas}
\end{itemize}

Pada setiap batas, apakah ia bebas, seperti pada permukaan air,
ataupun tidak bebas (\textit{fixed}), seperti pada dasar kolam,
beberapa persyaratan fisis haruslah dipenuhi oleh kecepatan
fluida. Di bawah pengaruh gaya, batas ini bisa saja mengalami
perubahan bentuk (deformasi). Semua persyaratan ini yang bekerja
pada kinematika partikel air disebut dengan \textit{syarat batas
kinematika} (\textit{kinematics boundary conditions}). Pada setiap
permukaan fluida atau perbatasan fluida (\textit{interface}),
jelaslah bahwa tidak akan terdapat aliran yang melewati
\textit{interface}, jika tidak demikian, maka tak terdapat
\textit{interface}. Hal seperti ini sangatlah jelas dalam kasus
permukaan tetap kedap air (\textit{impermeable fixed surface})
seperti suatu lembar timbunan dinding laut (\textit{sheet pile
seawall}).\cite{drg}

Misalkan fungsi yang menggambarkan permukaan air adalah
\begin{equation}
\displaystyle G(x,\,z,\,t) = z \,-\, \eta\,(x,\,t) = 0,
\end{equation}
dengan $\eta(x,\,t)$ adalah ketinggian gelombang, yakni jarak suatu titik
pada permukaan gelombang dari permukaan bebas rata-rata atau sering dikenal
dengan istilah \textit{elevasi permukaan air.} Maka, diferensial total
persamaan di atas akan memberikan
\begin{equation}
\frac{DG}{Dt} = \frac{\partial G}{\partial t} + \frac{\partial G}{\partial x}%
\; \frac{dx}{dt} + \frac{\partial G}{\partial z}\; \frac{dz}{dt}
\end{equation}
\begin{equation}
0 = -\frac{\partial \eta}{\partial t} + \Bigg(-\frac{\partial \eta}{\partial
x}\Bigg)\:v + w
\end{equation}
atau
\begin{equation}
\eta_{t} = w - v\,\eta_{x}  \label{eta}
\end{equation}
Karena $\mathbf{u}=(v,\,w)= (\phi_{x},\phi_{z})=\nabla\phi$, maka persamaan (%
\ref{eta}) dapat dituliskan menjadi
\begin{equation}
\eta_{t} = \phi_{z} - \phi_{x}\,\eta_{x}.
\end{equation}

Definisikan $\varphi(x,\,t)= \left.\phi(x,\,z,\,t)\right|_{z=\eta(x,\,t)} =
\phi(x,\,\eta(x,\,t),\,t)$, dengan melakukan diferensiasi terhadap $x$ kita
sekarang mempunyai $\displaystyle\frac{\partial\varphi}{\partial x} = \frac{%
\partial\phi}{\partial x} + \frac{\partial\phi}{\partial\eta}\:\frac{%
\partial\eta}{\partial x}$. Kita notasikan $\displaystyle \frac{\partial \phi%
}{\partial \eta} = \gamma$. Dengan demikian, persamaan tersebut menjadi $%
\varphi_{x} = \phi_{x} + \gamma\,\eta_{x}.$ Dengan mensubstitusi $%
\displaystyle\phi_{\eta}=\frac{\partial \phi}{\partial \eta}=\gamma$ dan $%
\phi_{x} = \varphi_{x} - \gamma\,\eta_{x}$ pada persamaan di atas, kita
dapatkan persamaan
\begin{equation}
\eta_{t} = \gamma - (\varphi_{x} - \gamma\, \eta_{x})\,\eta_{x} = \gamma -
\varphi_{x}\,\eta_{x} + \gamma\, \eta_{x}^2
\end{equation}
\begin{equation}
\eta_{t} = (1 + \eta_{x}^2)\,\gamma - \varphi_{x}\,\eta_{x}.
\end{equation}
Ini adalah syarat batas kinematika permukaan bebas, yaitu kondisi yang
menyatakan fakta bahwa tidak ada partikel fluida yang dapat melewati
permukaan bebas.

\begin{itemize}
\item  \textbf{Syarat batas dinamika permukaan bebas}
\end{itemize}

Permukaan air yang "bebas", seperti pada perbatasan
(\textit{interface}) udara dan air, tidak dapat menunjang
perbedaan dalam tekanan sepanjang perbatasan dan dengan demikian
harus merespon untuk memperoleh tekanan yang seragam. Syarat
batas dinamika menggambarkan distribusi tekanan yang bekerja pada
batas yang berupa permukaan bebas dan interface.

Dari persamaan Bernoulli untuk aliran tak tunak (\textit{unsteady}), kita
telah mempunyai persamaan $\displaystyle\frac{\partial \mathbf{u}}{\partial t%
}+\nabla\Bigg(\frac{p}{\rho}+g\,z+\frac{1}{2}\mid\mathbf{u}\mid^2\Bigg)=0.$
Dengan mengasumsikan bahwa tekanan atmosfer bebas di atas lapisan air
dibatasi hanya pada permukaan bebas, persamaan di atas dapat dituliskan
menjadi
\begin{equation}
\frac{\partial {\phi}}{\partial t}+ g\,\eta+\frac{1}{2}\mid\nabla\phi%
\mid^2=0, \quad \textmd{di} \quad z =\eta(x,\,t).  \label{bernou}
\end{equation}

Sebelumnya, kita hitung dahulu $\displaystyle \frac{1}{2}\mid
\nabla\phi\mid^2$ agar diketahui representasi dalam bentuk turunan
parsialnya. Kita tahu bahwa
\begin{eqnarray}
\frac{1}{2}\mid\nabla\phi\mid^2 &=& \frac{1}{2}\,\bigg(\nabla\phi\cdot\nabla%
\phi\bigg) =\frac{1}{2}\,\bigg[\,\bigg(\phi_{x},\,\phi_{z}\bigg)\cdot \bigg(%
\phi_{x},\,\phi_{z}\bigg)\,\bigg]  \nonumber \\
&=&\frac{1}{2}\,\bigg(\phi_{x}^2+\phi_{z}^2\bigg) =\frac{1}{2}\,\bigg(%
\phi_{x}^2+\gamma^2\bigg).
\end{eqnarray}
Dengan melakukan substitusi $\phi_{x}=\varphi_{x}-\gamma\,\eta_{x}$ pada
persamaan tersebut, kita sampai pada
\begin{eqnarray}
\frac{1}{2}\mid\nabla\phi\mid^2 &=& \frac{1}{2}\,\bigg[\,\bigg(\varphi_{x}-
\gamma\,\eta_{x}\bigg)^2+\gamma^2\bigg]  \nonumber \\
&=&\frac{1}{2}\,\bigg(\varphi_{x}^2-2\gamma\,\varphi_{x}\,\eta_{x}+
\gamma^2\,\eta_{x}^2+\gamma^2\bigg)
\end{eqnarray}
Jadi, $\displaystyle \frac{1}{2}\mid\nabla\phi\mid^2 = \frac{1}{2}%
\,\varphi_{x}^2 - \gamma\,\varphi_{x}\,\eta_{x} + \frac{1}{2}\,\bigg(%
1+\eta_{x}^2\bigg)\,\gamma^2$.\newline

Dari $\varphi(x,\,t)=\phi(x,\,\eta(x,\,t),\,t)$, diferensiasi parsial
terhadap t memberikan $\displaystyle\frac{\partial\varphi}{\partial t} =
\frac{\partial\phi}{\partial t} + \frac{\partial\phi}{\partial\eta}\frac{%
\partial\eta}{\partial t}$ \quad $\Longrightarrow \varphi_{t} = \phi_{t} +
\gamma\,\eta_{t}$ $\Longrightarrow \phi_{t} = \varphi_{t} - \gamma\,\eta_{t}$%
.

Dengan demikian, substitusi pada persamaan Bernoulli (\ref{bernou}) akan
menghasilkan
\begin{equation}
\varphi_{t} - \gamma\,\eta_{t} + \frac{1}{2}\,\varphi_{x}^2 -
\gamma\,\varphi_{x}\,\eta_{x} + \frac{1}{2}\,\bigg(1+\eta_{x}^2\bigg)%
\,\gamma^2 + g\,\eta=0.
\end{equation}
Untuk memperoleh bentuk yang lebih sederhana, kita substitusi $\eta_{t} = %
\bigg(1 + \eta_{x}^2\bigg)\,\gamma - \varphi_{x}\,\eta_{x}$ yang telah
diperoleh dari syarat batas kinematika. Akibatnya,
\begin{equation}
\varphi_{t} - \gamma\,\bigg[\,\bigg(1 + \eta_{x}^2\bigg)\,\gamma -
\varphi_{x}\,\eta_{x}\bigg] + \frac{1}{2}\,\varphi_{x}^2 -
\gamma\,\varphi_{x}\,\eta_{x} + \frac{1}{2}\,\bigg(1+\eta_{x}^2\bigg)%
\,\gamma^2 + g\,\eta=0,
\end{equation}
\begin{equation}
\varphi_{t} - \bigg(1 + \eta_{x}^2\bigg)\,\gamma^2 +
\varphi_{x}\,\eta_{x}\,\gamma + \frac{1}{2}\,\varphi_{x}^2 -
\gamma\,\varphi_{x}\,\eta_{x} + \frac{1}{2}\,\bigg(1+\eta_{x}^2\bigg)%
\,\gamma^2 + g\,\eta=0,
\end{equation}
\begin{equation}
\varphi_{t} - \frac{1}{2}\,\bigg(1+\eta_{x}^2\bigg)\,\gamma^2 + \frac{1}{2}%
\,\varphi_{x}^2 + g\,\eta=0.
\end{equation}
Akhirnya diperoleh, $\displaystyle\varphi_{t}=\frac{1}{2}\,\bigg(1+\eta_{x}^2%
\bigg)\,\gamma^2 -\frac{1}{2}\,\varphi_{x}^2 - g\,\eta$. Persamaan inilah
yang dikenal dengan syarat batas dinamika permukaan bebas.
\begin{figure}[h]
\begin{center}
\includegraphics[scale = 3]{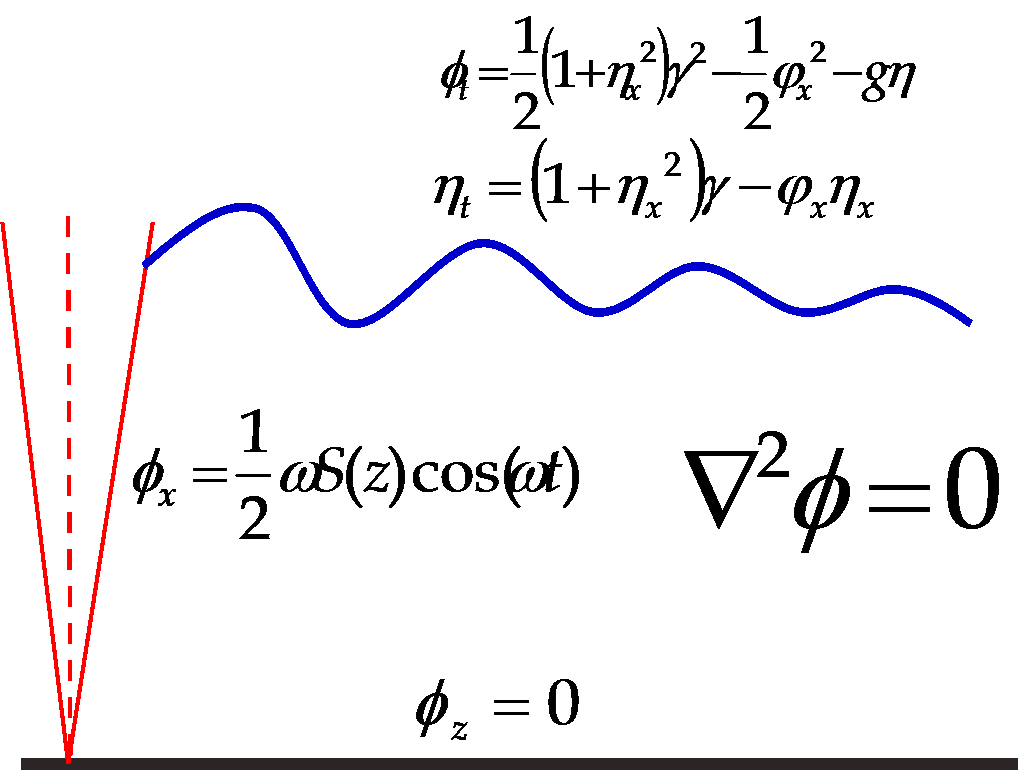}
\end{center}
\caption{Skema pembangkit gelombang dengan persamaan pengatur dan syarat-syarat batasnya.}
\end{figure}

Untuk air dangkal, kita lebih tertarik untuk mempelajari dan menggunakan
bentuk terlinearisasi dari syarat batas kinematika dan syarat batas dinamika
permukaan bebas agar lebih memudahkan dalam menyelesaikan persamaan
diferensial pengatur dengan syarat-syarat batas yang telah ditentukan.

Tinjau kembali persamaan (\ref{eta}) sebagai syarat batas kinematika
permukaan bebas,
\begin{equation}
\frac{\partial \eta}{\partial t}= \left.(w)\right|_{z=\eta}- \frac{\partial
\eta}{\partial x}\,\left.(v)\right|_{z=\eta}
\end{equation}
atau
\begin{equation}
\frac{\partial \eta}{\partial t}= \left.\Bigg(\frac{\partial \phi}{\partial z%
}\Bigg)\right|_{z=\eta} - \frac{\partial \eta}{\partial x}\, \left.\Bigg(%
\frac{\partial \phi}{\partial x}\Bigg)\right|_{z=\eta}
\end{equation}
Dengan mengabaikan suku-suku non-linear-nya, kita miliki bentuk yang lebih
sederhana dari syarat batas kinematika, yakni
\begin{equation}
\left.\Bigg(\frac{\partial \phi}{\partial z} \Bigg)\right|_{z=0}=\frac{%
\partial\eta}{\partial t}.
\end{equation}

Dengan cara serupa, kita juga mempunyai bentuk yang lebih sederhana dari
syarat batas dinamika permukaan bebas, yakni
\begin{equation}
\left. \Bigg(\frac{\partial \phi }{\partial t}\Bigg)\right| _{z=0}=-g\,\eta .
\label{permukaan}
\end{equation}

Skema pembangkit gelombang dengan \textit{governing equation} dan
beberapa syarat batas dapat dilihat pada Gambar 3.2.

\section{Solusi \textit{Governing Differential Equation}}

\begin{figure}[h]
\begin{center}
\includegraphics[scale = 0.8]{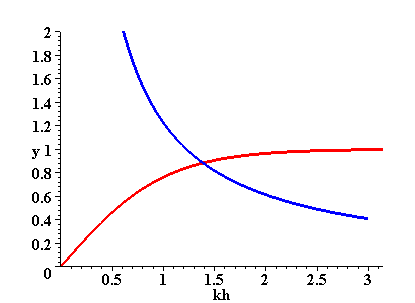}
\end{center}
\par
{Grafik ini mengilustrasikan akar tunggal $k_{p}\,h$. Label sumbu horizontal
\textbf{kh} memaksudkan $k_{p}\,h$.}
\caption{Grafik relasi dispersi gelombang progresif}
\label{gambar2}
\end{figure}

Berikut ini adalah solusi umum potensial kecepatan
\[
\phi(x,\,z,\,t)=A_{p}\:\cosh \,k_{p}\,(h+z)\:\sin\,(k_{p}\,x-\omega\,
t)+(A\,x+B)
\]
\begin{equation}
+ C\:e^{-k_{s}\,x}\:\cos k_{s}\,(h+z)\:\cos \,\omega\, t.
\end{equation}
Untuk permasalahan pembangkit gelombang, koefisien A di atas haruslah sama
dengan nol karena tidak ada aliran seragam yang mungkin melalui pembangkit
gelombang dan koefisien B di atas juga dapat diset menjadi nol juga tanpa
mempengaruhi medan kecepatan. Suku-suku sisanya haruslah memenuhi dua syarat
batas permukaan bebas yang terlinearisasi. Biasanya akan lebih bermanfaat
apabila menggabungkan kedua syarat batas tersebut menjadi bentuk yang lebih
sederhana.\newline
Dengan menurunkan $\phi$ terhadap t, kita peroleh
\[
\frac{\partial\phi}{\partial t}=-\,\omega\, A_{p}\:\cosh\,
k_{p}\,(h+z)\:\cos\,(k_{p}\,x-\omega\, t)
\]
\begin{equation}
-\,\omega\, C\,e^{-k_{s}\,x}\:\cos k_{s}\,(h+z)\:\sin \omega\, t.
\end{equation}
Turunan kedua $\phi$ terhadap t menghasilkan
\begin{eqnarray}
\frac{\partial^2\phi}{\partial t^2} &=& -\,\omega^2\, A_{p}\:\cosh\,
k_{p}\,(h+z)\:\sin\,(k_{p}\,x-\omega\, t)  \nonumber \\
& & -\, \omega^2\, C\,e^{-k_{s}\,x}\:\cos k_{s}\,(h+z)\:\cos\, \omega\, t
\nonumber \\
&=& -\,\omega^2\,\phi.
\end{eqnarray}
Tetapi dari syarat batas dinamika dan kinematika permukaan bebas kita punya
\begin{equation}
\frac{\partial \eta}{\partial t}=-\,\frac{1}{g}\,\frac{\partial^2\phi}{%
\partial t^2}= -\,\frac{\omega^2}{g}\,\phi = -\,\frac{\partial \phi}{%
\partial z}, \quad \textmd{pada} \quad z = 0,
\end{equation}
atau
\begin{equation}
\frac{\partial \phi}{\partial z}-\frac{\omega^2}{g}\,\phi=0, \quad \textmd{pada} \quad z = 0.
\end{equation}
Dengan mensubstitusi solusi yang telah kita asumsikan pada persamaaan ini,
maka dihasilkan
\begin{equation}
\omega^2 = g\,k_{p}\: \tanh\: k_{p}\,h  \label{progresif}
\end{equation}
dan
\begin{equation}
\omega^2 = -g\,k_{s} \:\tan\: k_{s}\,h.  \label{berdiri}
\end{equation}
Persamaan yang pertama adalah relasi dispersi untuk gelombang progresif.
Dengan menuliskan kembali persamaan tersebut menjadi
\begin{equation}
\frac{\omega^2\,h}{g\,k_{p}\,h}= \tanh\:k_{p}\,h
\end{equation}
dan mengeplot tiap suku terhadap $k_{p}\,h$ untuk nilai $\displaystyle \frac{%
\omega^2\,h}{g}$ tertentu, maka kita bisa melihat solusi untuk relasi
dispersi tersebut seperti diperlihatkan pada Gambar \ref{gambar2}.

\begin{figure}[h]
\caption{Grafik relasi dispersi gelombang berdiri} 				\label{gambar3}
\begin{center}
\includegraphics[scale = 0.8]{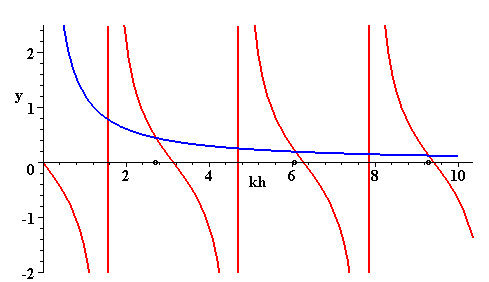}
\end{center}
\par
{Grafik ini mengilustrasikan terdapat tak hingga banyaknya akar majemuk $k_{s}(n)\,h$. Label sumbu horizontal \textbf{kh} memaksudkan $k_{s}\,h$.}
\end{figure}

Persamaan (\ref{berdiri}), yang mengaitkan $k_{s}$ dengan frekuensi
pembangkit gelombang, menentukan bilangan gelombang untuk gelombang berdiri
dengan amplitudo yang berkurang secara eksponensial seraya menjauh dari
pembangkit gelombang. Dengan menuliskan kembali persamaan tersebut sebagai
\begin{equation}
\frac{\omega^2\,h}{g\,k_{s}\,h}=-\tan\:k_{s}\,h,
\end{equation}
maka dengan mengeplot grafiknya kita bisa melihat bahwa ia mempunyai tak
hingga banyaknya penyelesaian. Penyelesaian persamaan ini dapat dilihat
dalam bentuk grafik relasi dispersi, seperti tampak pada Gambar~\ref{gambar3}.

Setiap penyelesaian akan dinyatakan dengan $k_{s}(n)$, dengan $n$ bilangan
asli. Bentuk akhir dari potensial kecepatan adalah
\[
\phi(x,\,z,\,t)=A_{p}\:\cosh\,k_{p}\,(h+z)\:\sin\,(k_{p}\,x-\omega\,t)
\]
\begin{equation}
+ \mathop {\sum}_{n=1}^{\infty} C_{n}\:e^{-k_{s}(n)\,x}\:\cos\,
k_{s}(n)\,(h+z)\:\cos\, \omega \,t.  \label{lastpotensial}
\end{equation}
Suku pertama menyatakan gelombang progresif yang dihasilkan oleh pembangkit
gelombang, sedangkan suku deret menyatakan gelombang berdiri yang semakin
meluruh seraya menjauh dari pembangkit gelombang.

Untuk memperoleh solusi gelombang yang lengkap, kita harus menentukan $A_{p}$
dan $C_{n}$. Nilai-nilai ini diperoleh dengan bantuan syarat batas lateral
di pembangkit gelombang, yakni
\begin{equation}
u(0,\,z,\,t)=\left.\Bigg(\frac{\partial \phi}{\partial x}\Bigg)\right|_{x=0}
= \frac{1}{2}\:\omega\,S(z)\:\cos\,\omega\,t
\end{equation}
Dengan menurunkan potensial kecepatan terhadap $x$ dan mengevaluasinya di $%
x=0$, kita memperoleh
\[
\frac{1}{2}\:\omega\,S(z)\:\cos\,\omega\,t =
A_{p}\,k_{p}\:\cosh\,k_{p}\,(h+z)\:\cos\,\omega\,t
\]
\begin{equation}
-\mathop {\sum}_{n=1}^{\infty} C_{n}\,k_{s}(n)\:\cos\,
k_{s}(n)\,(h+z)\:\cos\, \omega \,t,
\end{equation}
atau
\begin{equation}
\frac{1}{2}\:\omega\,S(z) = A_{p}\,k_{p}\:\cosh\,k_{p}\,(h+z)-\mathop {\sum}%
_{n=1}^{\infty} C_{n}\,k_{s}(n)\:\cos\, k_{s}(n)\,(h+z).  \label{fourier}
\end{equation}
Sekarang kita mempunyai sebuah fungsi $z$ yang sama dengan suatu deret
fungsi-fungsi trigonometri dari $z$ di ruas kanan, situasi yang mirip dengan
deret Fourier. Kita punya fakta bahwa himpunan fungsi-fungsi $\displaystyle%
\bigg\{\cosh\,k_{p}\,(h+z),\; \cos\, [k_{s}(n)\,(h+z)]\bigg\}_{n=1}^{\infty}$
membentuk suatu deret harmonik lengkap fungsi-fungsi ortogonal, dan
sembarang fungsi kontinu dapat diekspansi dalam suku-suku deret tersebut.

Dengan demikian, untuk mencari $A_{p},$ kalikan persamaan di atas dengan $%
\cosh\,k_{p}\,(h+z)$ dan diintegrasikan terhadap $z$ dari $-h$ sampai $0$.
Kita sekarang mempunyai
\[
\mathop {\int}_{-h}^{0}\: \frac{1}{2}\:\omega\,S(z)\:\cosh\,k_{p}\,(h+z)%
\,dz= \mathop {\int}_{-h}^{0}\:A_{p}\,k_{p}\:\cosh\,k_{p}^2\,(h+z)\,dz
\]
\begin{equation}
-\mathop {\int}_{-h}^{0}\; \mathop {\sum}_{n=1}^{\infty}
C_{n}\,k_{s}(n)\:\cos\,k_{s}(n)\,(h+z)\:\cosh\,k_{p}\,(h+z)\,dz.
\end{equation}
Dengan menggunakan sifat keortogonalan, suku terakhir akan sama dengan nol,
dan akibatnya
\begin{equation}
A_{p}=\frac{\frac{1}{2}\:\omega\:\int_{-h}^{0}\:S(z)\:\cosh\,k_{p}\,(h+z)\,dz%
} {k_{p}\:\mathop {\int}_{-h}^{0}\:\cosh\,k_{p}^2\,(h+z)\,dz}
\end{equation}
Untuk pembangkit gelombang tipe flap, fungsi simpangan $S$ dapat
dinyatakan secara spesifik sebagai
\begin{equation}
S(z)=S\,\Bigg(1+\frac{z}{h}\Bigg).
\end{equation}
Dengan menggunakan kalkulus sederhana kita bisa menyatakan koefisien $A_{p}$
tanpa menggunakan integral, yaitu
\begin{equation}
A_{p}=\frac{2\,\omega\,S}{k_{p}^2\,h}\;\frac{k_{p}\,h\:\sinh\,k_{p}\,h
-\cosh\,k_{p}\,h+1}{\sinh\,2\,k_{p}\,h+2\,k_{p}\,h}  \label{firstconst}
\end{equation}

\begin{figure}[h]
\caption{Grafik perbandingan ketinggian gelombang $H$ dengan simpangan $S$ terhadap kedalaman relatif $k_{p}\,h$.}
\label{gambar1}
\begin{center}
\includegraphics[scale = 0.8]{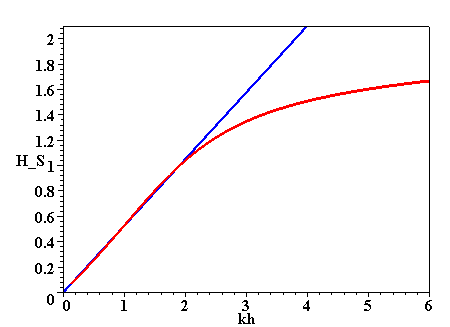}
\end{center}
\par
{Label sumbu horizontal \textbf{kh} memaksudkan $k_{p}\,h$, sedangkan label sumbu vertikal \textbf{H\_S} memaksudkan $\displaystyle\frac{H}{S}$.}
\end{figure}

Dengan cara yang serupa, kita bisa mendapatkan koefisien $C(n)$ dengan
mengalikan persamaan (\ref{fourier}) dengan $\cos\,[k_{s}(n)\,(h+z)]$ dan
mengintegrasikannya terhadap $z$ dari $-h$ sampai dengan 0 (sepanjang
kedalaman). Kita dapatkan
\begin{equation}
C_{n}=\frac{-\frac{1}{2}\:\omega\:\int_{-h}^{0}\:S(z)\:\cos\,k_{s}(n)\,(h+z)%
\,dz} {k_{s}(n)\:\int_{-h}^{0}\:\cos^2\,[k_{s}(n)\,(h+z)]\,dz}
\end{equation}
atau, dengan melakukan sedikit proses integrasi kita sampai pada hasil
berikut :
\begin{equation}
C_{n}=\frac{-2\,\omega\,S}{[k_{s}(n)]^2\,h}\;\frac{[k_{s}(n)\,h]\:\sin%
\,[k_{s}(n)\,h]+ \cos\,[k_{s}(n)\,h]}{\sin\,[2\,k_{s}(n)\,h]+2\,k_{s}(n)\,h}.
\label{nextconst}
\end{equation}
Tinggi gelombang untuk gelombang progresif ditentukan dengan mengevaluasi $%
\eta$ cukup jauh dari pembangkit gelombang.
\begin{eqnarray}
\eta=-\frac{1}{g}\left.\Bigg(\frac{\partial \phi}{\partial t}\Bigg)%
\right|_{z=0} &=& \frac{A_{p}}{g}\,\omega\: \cosh\,k_{p}\,h
\:cos\,(k_{p}\,x-\omega\,t)  \nonumber \\
&=&\frac{H}{2}\:\cos\,(k_{p}\,x-\omega\,t) \quad x\gg\, h
\end{eqnarray}
Dengan mensubstitusi nilai $A_{p}$ yang telah diperoleh, kita bisa
mendapatkan perbandingan tinggi gelombang $H$ dengan besarnya stroke $S$,
yaitu
\begin{equation}
\frac{H}{S}= 4\: \Bigg(\frac{\sinh\,k_{p}\,h}{k_{p}\,h}\Bigg)\; \Bigg(\frac{%
k_{p}\,h\:\sinh\,k_{p}\,h-\cosh\,k_{p}\,h+1}{\sinh\,2\,k_{p}\,h+ 2\,k_{p}\,h}%
\Bigg).  \label{ratio}
\end{equation}
Plot grafik perbandingan $H$ dan $S$, baik untuk teori yang disederhanakan
(air dangkal) maupun yang lebih lengkap (teori pembangkit gelombang linear)
dapat dilihat pada Gambar~\ref{gambar1}.

\chapter{Simulasi Numerik}

Pada bab ini, akan dibahas beberapa hasil yang diperoleh dengan memilih
nilai-nilai besaran tertentu. Dengan bantuan perangkat lunak \textit{Maple V
Release 5}, kita bisa melihat animasi gelombang monokromatik yang dihasilkan
apabila pembangkit gelombang digerakkan dengan simpangan tertentu.

\section{Menghitung Bilangan Gelombang Progresif} 		\label{NR1}

Kita perhatikan kembali persamaan (\ref{progresif}) yang menyatakan relasi
dispersi gelombang progresif :
\begin{equation}
\omega^2 = g\,k_{p}\: \tanh\: k_{p}\,h.
\end{equation}
Untuk memperoleh nilai bilangan gelombang $k_{p}$, kita tidak dapat
menghitungnya secara eksak, sehingga digunakan metode numerik untuk
menyelesaikannya. Beberapa nilai yang diberikan adalah frekuensi sudut
monokromatik $\omega = 2$ rad/detik, tetapan percepatan gravitasi $g = 9,8 $
m/detik$^2$, dan kedalaman \textit{towing tank} $h = 3$ m. Dengan
menggunakan iterasi \textit{Newton-Raphson}, kita akan mencari nilai $k_{p}$
apabila diberikan tebakan awal $p_{0}$ dengan mencari akar persamaan $%
f(k_{p})=0$, dengan $f(k_{p}) = \omega^2 - g\,k_{p}\: \tanh\: k_{p}\,h$.
Untuk mengetahui lebih lengkap mengenai algoritma iterasi \textit{%
Newton-Raphson}, dapat dilihat di Apendiks B. Dengan memilih nilai-nilai
galat $\delta = 10^{-9}, \quad \epsilon = 10^{-9},$ dan $kecil = 10^{-9}$
juga, serta iterasi maksimum = 5000, maka dengan menggunakan tebakan awal $%
p_{0} = 0,5$ kita memperoleh hasil $k_{p} = 0,4624593666$. Lihat grafiknya
pada Gambar \ref{gambar2}.

Untuk $\omega = 1$, nilai bilangan gelombang progresif yang diperoleh adalah
$k_{p} = 0,1943823443$.

\section{Menghitung Bilangan Gelombang Berdiri} \label{NR2}

Kita perhatikan kembali persamaan (\ref{berdiri}) yang menyatakan relasi
dispersi gelombang berdiri :
\begin{equation}
\omega^2 = -g\,k_{s}\: \tan\: k_{s}\,h.
\end{equation}
Untuk memperoleh nilai bilangan gelombang $k_{s}$, kita tidak dapat
menghitungnya secara eksak, sehingga digunakan metode numerik untuk
menyelesaikannya. Beberapa nilai yang diberikan adalah frekuensi sudut
monokromatik $\omega = 2$ rad/detik, tetapan percepatan gravitasi $g = 9,8 $
m/detik$^2$, dan kedalaman \textit{towing tank} $h = 3$ m. Dengan
menggunakan iterasi \textit{Newton-Raphson}, kita akan mencari nilai $k_{s}$
apabila diberikan tebakan awal $p_{0}$ dengan mencari akar persamaan $%
g(k_{s})=0$, dengan $g(k_{s}) = \omega^2 + g\,k_{s}\: \tan\: k_{s}\,h$.
Untuk mengetahui lebih lengkap mengenai algoritma iterasi \textit{%
Newton-Raphson}, dapat dilihat kembali di Apendiks B. Nilai-nilai galat yang
dipilih adalah $\delta = 10^{-5}, \quad \epsilon = 10^{-5},$ dan $kecil =
10^{-5}$ dengan menggunakan iterasi maksimum = 10000. Karena perpotongan
grafik pada relasi dispersi menghasilkan tak hingga banyaknya titik potong,
yakni nilai-nilai bilangan gelombang berdiri, maka digunakan tebakan awal
yang bervariasi sesuai dengan kebutuhan, $p_{0} = 1, 2, 3, \cdots.$ Untuk
keperluan, kita pilih beberapa bilangan gelombang tertentu saja yang
berurutan. Akhirnya, kita memperoleh hasil\newline
$k_{s}(1)=0,9061212307;\newline
\; k_{s}(2)=2,028197863;\newline
$ $\quad k_{s}(3)=3,097926261;\newline
\quad k_{s}(4)=4,156159228;\newline
$ $\quad k_{s}(5)=5,209926525;\newline
\quad k_{s}(6)=6,261487236;\newline
$ $\quad k_{s}(7)=7,311794624;\newline
\quad k_{s}(8)=8,361321437;\newline
$ $\quad k_{s}(9)=9,410329031$.\newline
Lihat kembali grafiknya pada Gambar \ref{gambar3}.

Dengan cara yang serupa seperti di atas, yakni menggunakan tebakan awal yang
bervariasi, maka untuk $\omega = 1$ kita memperoleh beberapa nilai bilangan
gelombang berdiri seperti berikut:\newline
$k_{s}(1)=1,013758189;\newline
\; k_{s}(2)=2,078040121;\newline
\; k_{s}(3)=3,130732073;\newline
\; k_{s}(4)=4,180655871$.\newline

\section{Profil Gelombang Monokromatik}

\subsection{Dua Profil Gelombang dengan Ketinggian \textit{\textbf{H}} Berbeda}

Perhatikan kembali persamaan (\ref{lastpotensial}), persamaan potensial
kecepatan yang menggambarkan gelombang air yang dihasilkan oleh pembangkit
gelombang
\[
\phi(x,\,z,\,t)=A_{p}\:\cosh\,k_{p}\,(h+z)\:\sin\,(k_{p}\,x-\omega\,t)
\]
\begin{equation}
+ \mathop {\sum}_{n=1}^{\infty} C_{n}\:e^{-k_{s}(n)\,x}\:\cos\,
k_{s}(n)\,(h+z)\:\cos\, \omega \,t.
\end{equation}
Dengan menggunakan persamaan (\ref{firstconst}) dan persamaan (\ref{ratio}),
kita bisa menyatakan konstanta $A_{p}$ dalam $\omega$, $H$, $k_{p}$, dan $h$, yakni :
\begin{equation}
A_{p}=\frac{\omega\,H}{2\,k_{p}\:\sin\:k_{p}\,h}
\end{equation}
Dengan mensubstitusi nilai $\omega = 2$, $k_{p} = 0,4624593666$, dan $h = 3$%
, maka :

\begin{itemize}
\item  untuk $H = 0,5$, kita punya $A_{p} = 1,099621133$; dan

\item  untuk $H = 0,75$, kita punya $A_{p} = 1,649431700$.
\end{itemize}

Untuk mendapatkan simpangan maksimum $S$, kita menggunakan persamaan (\ref
{ratio}), yakni
\begin{equation}
S = \Bigg(\frac{H\,k_{p}\,h}{4\,\sinh\,k_{p}\,h}\Bigg)\; \Bigg(\frac{%
\sinh\,2\,k_{p}\,h+2\,k_{p}\,h} {k_{p}\,h\:\sinh\,k_{p}\,h-\cosh\,k_{p}\,h+1}%
\Bigg).
\end{equation}
Dengan mensubstitusi kembali nilai-nilai yang telah kita punya, akibatnya :

\begin{itemize}
\item  untuk $H = 0,5$, kita peroleh $S = 0,2908632665$; dan

\item  untuk $H = 0,75$, kita peroleh $S = 0,4362948995$.
\end{itemize}

\begin{figure}[h]
\begin{center}
\includegraphics[scale = 0.7]{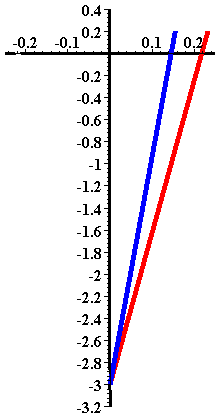}
\end{center}
\caption{Simpangan terjauh flap yang menghasilkan dua profil gelombang dengan ketinggian berbeda.}	\label{gambar4}
\end{figure}

Profil gerakan pembangkit gelombang ketika ia mencapai simpangan terjauh dapat dilihat pada Gambar~\ref{gambar4}.

Untuk mencari besarnya konstanta $C_{n}$ sebagai koefisien gelombang
berdiri, kita menggunakan persamaan (\ref{nextconst}), yaitu
\begin{equation}
C_{n}=\frac{-2\,\omega\,S}{[k_{s}(n)]^2\,h}\;\frac{[k_{s}(n)\,h]\:\sin%
\,[k_{s}(n)\,h]+ \cos\,[k_{s}(n)\,h]}{\sin\,[2\,k_{s}(n)\,h]+2\,k_{s}(n)\,h}.
\end{equation}
Nilai-nilai $k_{s}(n)$ yang telah kita punyai adalah sebanyak 4 buah, karena
untuk $n$ yang lebih besar pengaruh konstanta $C_{n}$ terhadap solusi
gelombang relatif kecil sehingga dapat diabaikan.

\begin{itemize}
\item  Untuk $S = 0,2908632665$, nilai\newline
$C_{1} = 0,08013595964$;\newline
$C_{2} = 0,00976252019$;\newline
$C_{3} = 0,001714039296$; \quad dan \newline
$C_{4} = 0,001110132786$.

\item  Untuk $S = 0,4362948995$, nilai\newline
$C_{1} = 0,1202039395$;\newline
$C_{2} = 0,01464378027$;\newline
$C_{3} = 0,002571058938$; \quad dan \newline
$C_{4} = 0,001665199177$.
\end{itemize}

\begin{figure}[h]
\begin{center}
\includegraphics[scale = 0.8]{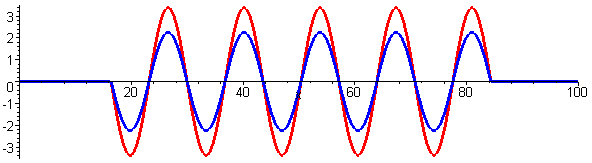}
\end{center}
\caption{Bentuk permukaan gelombang air (elevasi) yang dibentuk oleh dua gelombang dengan ketinggian yang berbeda.} 	\label{gambar5}
\end{figure}

Dengan menggunakan persamaan (\ref{permukaan}), yakni
\begin{equation}
\eta(x,\,t) =-\frac{1}{g}\:\left.\Bigg(\frac{\partial \phi}{\partial t}\Bigg)%
\right|_{z=0}
\end{equation}
kita dapat mencari persamaan elevasi permukaan gelombang air yang
terbentuk. Dengan bantuan perangkat lunak komputer kita juga bisa
menampilkan hasil plot permukaan. Hasil plot permukaan air dengan
dua ketinggian yang berbeda diperlihatkan pada Gambar~\ref{gambar5}.

\subsection{Dua Profil Gelombang dengan Frekuensi Sudut \textbf{$\protect \omega$} Berbeda}

Cara yang serupa seperti pada sub-bab sebelumnya diterapkan juga pada
sub-bab ini. Kedalaman \textit{towing tank} masih sama yaitu $h = 3$ m dan
profil ketinggian gelombang yang diinginkan adalah $H = 0,5$ m. Namun,
frekuensi sudut $\omega$ yang diberikan adalah masing-masing 2 dan 1. Dengan
menggunakan iterasi Newton-Raphson terhadap relasi dispersi gelombang
progresif maupun gelombang berdiri, kita bisa mendapatkan nilai $k_{p}$ dan $%
k_{s}(n)$ untuk $\omega = 1.$ Hasil-hasilnya telah diperoleh pada sub-bab
\ref{NR1} dan \ref{NR2}. Untuk lebih jelas, hasilnya dicantumkan kembali
pada sub-bab ini. Sebagai tambahan, dicantumkan juga nilai-nilai konstanta
potensial kecepatan yang berkaitan dengan nilai $\omega$, yakni $A_{p}$ dan $%
C_{n}$.

Berikut ini adalah hasil yang telah diperoleh dengan menggunakan bantuan
perangkat lunak Maple V Release 5.

\begin{description}
\item[\textbf{a.}]  Untuk $\omega = 2$

\begin{itemize}
\item[$\bullet$]  Nilai bilangan gelombang progresif $k_{p} = 0,4624593666$.

\item[$\bullet$]  Nilai-nilai bilangan gelombang berdiri \newline
$k_{s}(1)=0,9061212307;$ \quad $k_{s}(2)=2,028197863;$\newline
\quad $k_{s}(3)=3,0979262610;$ \quad $k_{s}(4)=4,156159228.$

\item[$\bullet$]  Nilai koefisien $A_{p} = 1,099621133$

\item[$\bullet$]  Besar simpangan maksimum $S = 0,2908632665$

\item[$\bullet$]  Nilai-nilai koefisien $C_{n}$ adalah \newline
$C_{1}=0,8013595964;$ \quad \quad $C_{2}=0,009762520190;$\newline
\quad $C_{3}=0,001714039296;$ \quad $C_{4}=0,001110132786.$
\end{itemize}

\item[\textbf{b.}]  Untuk $\omega = 1$

\begin{itemize}
\item[$\bullet$]  Nilai bilangan gelombang progresif $k_{p} = 0,1943823443$.

\item[$\bullet$]  Nilai-nilai bilangan gelombang berdiri \newline
$k_{s}(1)=1,013758189; \quad k_{s}(2)=2,078040121;\newline
\; k_{s}(3)=3,130732073; \quad k_{s}(4)=4,180655871$.\newline

\item[$\bullet$]  Nilai koefisien $A_{p} = 2,335633604$

\item[$\bullet$]  Besar simpangan maksimum $S = 1,140576790$

\item[$\bullet$]  Nilai-nilai koefisien $C_{n}$ adalah \newline
$\quad C_{1}=0,2125853842;$ \quad \quad $C_{2}=0,004369427564;$\newline
\quad $C_{3}=0,007018430814;$ \quad $C_{4}=0,0005323318426.$
\end{itemize}
\end{description}

\begin{figure}[h]
\begin{center}
\includegraphics[scale = 0.7]{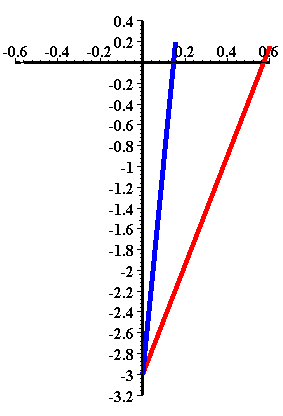}
\end{center}
\caption{Simpangan terjauh flap yang menghasilkan dua profil gelombang dengan frekuensi sudut berbeda.}   \label{gambar6}
\end{figure}

Profil gerakan \textit{flap} pada saat simpangan terjauh ditunjukkan pada Gambar~\ref{gambar6}.

\begin{figure}[h]
\begin{center}
\includegraphics[scale = 0.78]{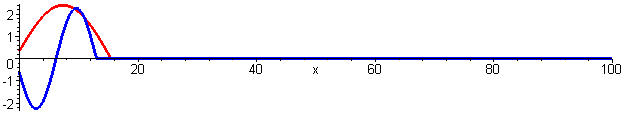}
\end{center}
\caption{Grafik permukaan gelombang air pada t = 3.}   \label{gambar7}
\end{figure}

\begin{figure}[h]
\begin{center}
\includegraphics[scale = 0.78]{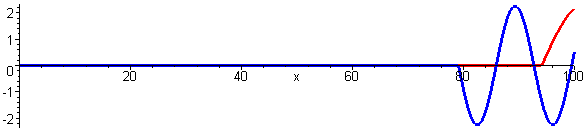}
\end{center}
\caption{Grafik permukaan gelombang air pada t = 34.}   \label{gambar14}
\end{figure}

Profil gelombang yang dihasilkan evolusinya terhadap waktu dapat dilihat pada Gambar~\ref{gambar7} sampai dengan Gambar~\ref{gambar14}.

\begin{figure}[h]
\begin{center}
\includegraphics[scale = 0.78]{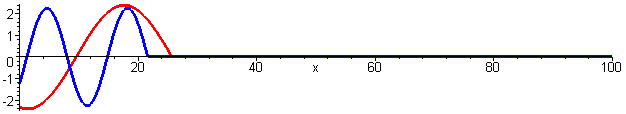}
\end{center}
\caption{Grafik permukaan gelombang air pada t = 5.}   \label{gambar8}
\end{figure}

\begin{figure}[h]
\begin{center}
\includegraphics[scale = 0.78]{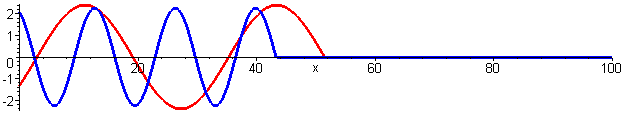}
\end{center}
\caption{Grafik permukaan gelombang air pada t = 10.}  \label{gambar9}
\end{figure}

\begin{figure}[h]
\begin{center}
\includegraphics[scale = 0.78]{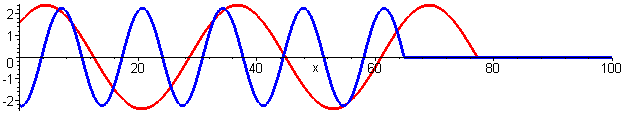}
\end{center}
\caption{Grafik permukaan gelombang air pada t = 15.}  \label{gambar10}
\end{figure}

\begin{figure}[h]
\begin{center}
\includegraphics[scale = 0.78]{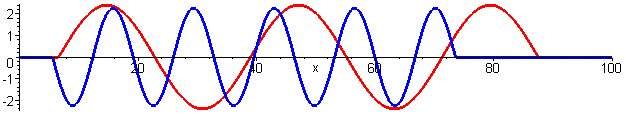}
\end{center}
\caption{Grafik permukaan gelombang air pada t = 17.}   \label{gambar11}
\end{figure}

\begin{figure}[h]
\begin{center}
\includegraphics[scale = 0.78]{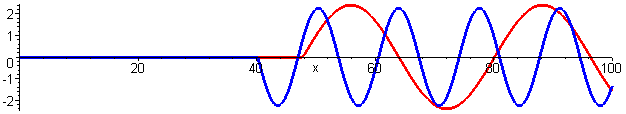}
\end{center}
\caption{Grafik permukaan gelombang air pada t = 25.}   \label{gambar12}
\end{figure}

\begin{figure}[h]
\begin{center}
\includegraphics[scale = 0.78]{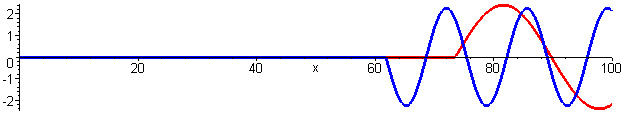}
\end{center}
\caption{Grafik permukaan gelombang air pada t = 30.}   \label{gambar13}
\end{figure}

\chapter{Kesimpulan dan Saran}

\section{Kesimpulan}

Berikut ini adalah beberapa kesimpulan yang diperoleh penulis setelah
mengerjakan tugas akhir mengenai pembangkit gelombang ini:

\begin{itemize}
\item[\textbf{a.}]  Asumsi fluida ideal diperlukan untuk memodelkan
permasalahan pembangkitan gelombang.

\item[\textbf{b.}]  Teori gelombang linear digunakan untuk
memperoleh pernyelesaian dari \textit{governing differential
equation} yang dihasilkan pada pemodelan pada bagian (a).

\item[\textbf{c.}]  Kaitan antara bilangan gelombang, ketinggian
gelombang, dan simpangan pembangkit gelombang diperoleh melalui
teori gelombang pada bagian (b).

\item[\textbf{d.}]  Untuk air dangkal, teori yang disederhanakan
dan teori gelombang linear memberikan hasil yang sama, sedangkan
untuk air yang lebih dalam, kedua teori tersebut memberikan hasil
yang berbeda.
\end{itemize}

\section{Saran}

Berikut ini adalah beberapa saran yang dapat dikembangkan untuk bahan tugas
akhir ataupun penelitian berikutnya :

\begin{itemize}
\item[\textbf{a.}]  Teori gelombang yang digunakan bisa menggunakan teori
gelombang non-linear dan bisa diperluas menjadi teori pembangkit gelombang
tiga dimensi.

\item[\textbf{b.}]  Tipe gelombang yang dihasilkan bisa dikembangkan menjadi
sembarang gelombang tak regular.

\item[\textbf{c.}]  Penambahan pantai (\textit{beach}) dapat
digunakan untuk menggantikan domain yang \textit{semi-infinite}.
\end{itemize}

\appendix

\chapter{Turunan Material}

Operator
\begin{equation}
\frac{D}{Dt}=\frac{\partial}{\partial t} + \mathbf{u}\cdot \nabla  \label{op}
\end{equation}
dikenal sebagai turunan \textit{material}, turunan \textit{total},
turunan \textit{substansial}, atau turunan \textit{Lagrange}.
Secara umum, $\displaystyle \frac{D}{Dt}$ adalah suatu operator
vektor. Akibatnya, komponen-komponen $\displaystyle \frac{D}{Dt}$
yang bekerja pada vektor tidaklah sama dengan yang bekerja pada
komponen skalar suatu vektor, kecuali dalam koordinat Kartesius.
Operator turunan material dapat juga diterapkan pada besaran
skalar, seperti temperatur. Secara fisis, ini menyatakan perubahan
besaran tersebut terhadap waktu, dengan pengamat bergerak bersama
fluida yang diukur pada lokasi tertentu di ruang dan waktu
\textit{instant} ketika turunan tersebut dievaluasi.

Operator ini adalah turunan total terhadap waktu yang bekerja pada elemen
fluida, yang dapat dilihat sebagai berikut. Misalkan $\xi $ adalah koordinat
ruang (\textit{spatial}) yang menggambarkan fluida pada keadaan (\textit{%
instant}) awal yang tetap, $t$ = 0, dan misalkan pula \textbf{x} adalah
koordinat ruang yang memberikan lokasi pada saat $t$ dari elemen fluida
bahwa pada saat $\xi $ ketika $t$ = 0. Tetapi \textbf{x} = \textbf{x($\xi ,t$%
)}. Turunan terhadap waktu (Eulerian) adalah
\begin{equation}
\frac{\partial}{\partial t} = \left. \frac{\partial}{\partial t} \right|_{%
\mathbf{x} \hskip 0.1cm {\textmd{{\footnotesize tetap}}}},
\end{equation}
di mana yang berperan sebagai turunan Lagrangian adalah
\begin{equation}
\frac{D}{Dt} = \left.\frac{\partial}{\partial t}\right|_{\xi \hskip 0.1cm {%
\textmd{{\footnotesize tetap}}}}.
\end{equation}
Dengan menggunakan aturan rantai,
\[
\frac{D}{Dt} = \left.\frac{\partial}{\partial t}\right|_{\xi \hskip 0.1cm {%
\textmd{{\footnotesize tetap}}}} = \left.\frac{\partial}{\partial t}\right|_{x %
\hskip 0.1cm {\textmd{{\footnotesize tetap}}}} + \left.\frac{\partial x_i}{%
\partial t}\right|_{\xi \hskip 0.1cm {\textmd{{\footnotesize tetap}}}} \frac{%
\partial}{\partial t},
\]
yang menghasilkan persamaan (\ref{op}), karena $\displaystyle\left. \frac{%
\partial\mathbf{x}}{\partial t}\right|_{\xi} = \mathbf{u }$.

Operator ini digunakan juga dalam persamaan \textit{Navier-Stokes}
yang menyatakan percepatan total suatu partikel sebagai
\begin{equation}
\mathbf{a}=\frac{D\mathbf{u}}{Dt}=\frac{\partial \mathbf{u}}{\partial t}+%
\Bigg(u\frac{\partial \mathbf{u}}{\partial x}+v\frac{\partial \mathbf{u}}{%
\partial y}+w\frac{\partial \mathbf{u}}{\partial z}\Bigg).  \label{nv}
\end{equation}

Di ruas kanan dari (\ref{nv}), suku pertama menyatakan percepatan lokal (%
\textit{local acceleration}) yang akan sama dengan 0 untuk aliran tunak (%
\textit{steady-state flow}) dan sisanya menyatakan percepatan konvektif (%
\textit{convective acceleration}). Suku percepatan konvektif menunjukkan
bahwa aliran fluida mempunyai kecepatan yang berbeda pada posisi yang
berbeda pula. Untuk penjelasan yang lebih lengkap, silakan lihat di \cite
{fac,hwf}.

\chapter{Algoritma Iterasi Newton-Raphson}

\fbox{\begin{minipage}{16cm} \vspace*{0.2cm} 
\textbf{Mencari akar \mbox{\boldmath $f(x)=0$} apabila diberikan
suatu tebakan awal \mbox{\boldmath $p_{0}$} dan
menggunakan iterasi \mbox{\boldmath $\displaystyle
p_{n}=p_{n-1}-\frac{f(p_{n-1})}{f'(p_{n-1})}$} \quad \hspace*{.1
in} \textmd{\textbf{untuk}} \ n = 1,2,...} \vspace*{0.2cm}
\end{minipage}} \vspace*{0.2cm} 
\newline
\textbf{Algoritma :}\newline
$\delta := 10^{-6}, \quad \epsilon := 10^{-6}, \quad kecil := 10^{-6}$ \quad \newline \{beberapa nilai galat, dapat diubah sesuai dengan kebutuhan\}\newline
maks := 99 \quad \quad \{jumlah iterasi maksimum\}\newline
kond := 0 \quad \quad \{kondisi untuk terminasi loop\}\newline
INPUT $P0$ \quad \quad \{$P0$ haruslah dekat dengan akar\} \newline
$Y0:=F(P0)$ \quad \quad \{menghitung nilai fungsi\}\newline
\newline
DO FOR $N$ := 1 TO maks UNTIL kond $\neq$ 0\newline
\hspace*{.2in} $Df := F^{\prime}(P0)$ \quad \quad \{menghitung turunan\}%
\newline
\hspace*{.5in} IF $Df = 0$ THEN \quad \quad \{mengecek pembagian oleh nol\}%
\newline
\hspace*{1in} kond := 1\newline
\hspace*{1in} $Dp := 0$\newline
\hspace*{.8in}ELSE \newline
\hspace*{1in} $Dp := Y0/Df$\newline
\hspace*{.5in}ENDIF\newline
\hspace*{.2in} $P1 := P0-Dp$ \quad \quad\{iterasi baru\}\newline
\hspace*{.2in} $Y1 := F(P1)$ \quad \quad \{nilai fungsi baru\}\newline
\hspace*{.2in} RelErr := 2*$|Dp|/(|P1|$+kecil) \quad \quad \{galat relatif\}%
\newline
\hspace*{.2in} IF RelErr $< \delta$ AND $|Y1|<\epsilon$ THEN \newline
\hspace*{.5in} IF kond $\neq$ 1 THEN kond := 2 \quad \quad \{mengecek kekonvergenan\} \newline 
\hspace*{.2in} $P0 := P1; \quad Y0 := Y1$ \quad \quad
\{mengganti dengan nilai baru\}\newline
\newline
PRINT 'Nilai iterasi ke-n adalah' $P1$ \quad \quad \{ouput\}\newline
PRINT 'Iterasi yang berurutan dibedakan sebesar' $Dp$ \newline
PRINT 'Nilai $f(x)$ adalah' $Y1$ \newline
IF kond = 0 THEN \newline
\hspace*{.2in} PRINT 'Jumlah iterasi maksimum telah terlewati.'\newline
IF kond = 1 THEN \newline
\hspace*{.2in} PRINT 'Pembagian dengan nol telah dilampaui.'\newline
IF kond = 2 THEN \newline
\hspace*{.2in} PRINT 'Akar telah ditemukan dengan galat yang diinginkan.'

\chapter{Laboratorium Hidrodinamika di Dunia}

Berikut ini adalah beberapa fasilitas laboratorium hidrodinamika
yang ada di berbagai negara di penjuru dunia. Banyak dari antara
mereka yang digunakan untuk ujicoba komersial dan pertahanan yang
berkaitan dengan struktur kelautan. Untuk keterangan terperinci,
bisa dilihat di \cite{csk} atau merujuk ke \textit{International
Towing Tank Conference}.

\begin{itemize}
\item[\textbf{a.}]  \textbf{\textit{Institute of Marine Dynamics Towing
Tank, St. John's, Newfoundland, Kanada}} \newline
\textbf{Kolam Air Dalam}\newline
Ukuran Kolam (Panjang, Lebar, dan Kedalaman): 200 m $\times$ 12 m $\times$ 7
m\newline
Kecepatan Pembawa : 10 m/detik\newline
Gelombang : Regular dan Tak regular; 1 m\newline
Pembangkit gelombang : Tipe flap-ganda\newline
Pantai : Memuat dasar berombak.

\item[\textbf{b.}]  \textbf{\textit{Offshore Model Basin, Escondido,
California, Amerika Serikat}}\newline
Ukuran Kolam : 90 m $\times$ 14,6 m $\times$ 4,6 m\newline
Bagian yang dalam : Lubang melingkar 9 m dalamnya\newline
Kecepatan Pembawa : 6 m/detik \newline
Gelombang : Regular dan Tak regular; 0,74 m \newline
Pembangkit gelombang : Papan flap-tunggal\newline
Pantai : Serutan logam

\item[\textbf{c.}]  \textbf{\textit{Offshore Technology Research Center,
Texas A \& M, College Station, Texas, Amerika Serikat}}\newline
Ukuran Kolam : 45,7 m $\times$ 30,5 m $\times$ 5,8 m\newline
Bagian yang dalam : 16,7 m lubang dengan lantai yang dapat disesuaikan%
\newline
Pembangkit gelombang : Tipe flap dengan kendali engsel hidrolik\newline
Ketinggian gelombang maksimum : 80 cm\newline
Kisaran frekuensi : 0,5 \textbf{--} 4,0 detik\newline
Pantai : Panel logam

\item[\textbf{d.}]  \textbf{\textit{David Taylor Research Center, Bethesda,
Maryland, Amerika Serikat}}

\begin{description}
\item  {$\bullet$} \textbf{Maneuvering and Seakeeping Facilities (MASK)} %
\vskip 0.2cm Ukuran Kolam : 79,3 m $\times$ 73,2 m $\times$ 6,1 m\newline
Pembangkit gelombang : Total sebanyak 21 pembangkit gelombang tipe pneumatic
\newline
Gelombang : Berarah banyak, regular dan tak regular; ketinggian maksimum 0,6
m; dan panjang gelombang 0,9 \textbf{--} 12,2 m \newline
Pantai : Penyerap gelombang\newline
Kecepatan pembawa : 7,7 m/detik\newline

\item  {$\bullet$} \textbf{Deep Water Basin} \vskip 0.2cm Ukuran Kolam : 846
m $\times$ 15,5 m $\times$ 6,7 m\newline
Gelombang : Ketinggian maksimum 0,6 m dan panjang gelombang 1,5 \textbf{--}
12,2 m \newline
Kecepatan pembawa : 10,2 m/detik\newline

\item  {$\bullet$} \textbf{High Speed Basin} \vskip 0.2cm Ukuran Kolam :
79,3 m $\times$ 73,2 m $\times$ 6,1 m\newline
Gelombang : Ketinggian maksimum 0,6 m; dan panjang gelombang 0,9 \textbf{--}
12,2 m \newline
Kecepatan pembawa : 35,8 \textbf{--} 51,2 m/detik
\end{description}

\item[\textbf{e.}]  \textbf{\textit{Maritime Research Institute, Belanda
(MARIN)}}

\begin{description}
\item  {$\bullet$} \textbf{Seakeeping Basin} \vskip 0.2cm Ukuran Kolam : 100
m $\times$ 24,5 m $\times$ 2,5 m\newline
Bagian yang dalam : Lubang sedalam 6 m\newline
Gelombang : Regular dan tak regular; ketinggian maksimum 0,3 m; dan kisaran
frekuensi 0,7 \textbf{--} 3,0 detik \newline
Kecepatan pembawa : 4,5 m/detik\newline

\item  {$\bullet$} \textbf{Wave and Current Basin} \vskip 0.2cm Ukuran Kolam
: 60 m $\times$ 40 m $\times$ 1,2 m\newline
Bagian yang dalam : Lubang sedalam 3 m\newline
Gelombang : Regular dan tak regular\newline
Kecepatan pembawa : 3 m/detik\newline
Kisaran kecepatan : 0,1 \textbf{--} 0,6 m/detik \newline

\item  {$\bullet$} \textbf{Deep Water Towing Tank} \vskip 0.2cm Ukuran Kolam
: 252 m $\times$ 10,5 m $\times$ 5,5 m\newline
Kecepatan pembawa : 9 m/detik\newline

\item  {$\bullet$} \textbf{High Speed Towing Tank} \vskip 0.2cm Ukuran Kolam
: 220 m $\times$ 4 m $\times$ 4 m\newline
Pembangkit gelombang : Tipe flap hidrolik\newline
Gelombang : Regular dan tak regular; ketinggian maksimum 0,4 m; dan kisaran
periode 0,3 \textbf{--} 5 detik\newline
Pembawa : Kendali motor dan kendali jet\newline
Kecepatan pembawa : 15 dan 30 m/detik\newline
Pantai : Kisi-kisi berupa busur melingkar\newline
\end{description}

\item[\textbf{f.}]  \textbf{\textit{Danish Maritime Institute, Lyngby,
Denmark}}\newline
Ukuran Kolam : 240 m $\times$ 12 m $\times$ 5,5 m\newline
Pembangkit gelombang : Tipe flap-ganda hidrolik yang dikontrol secara numerik%
\newline
Gelombang : Regular dan tak regular; ketinggian maksimum 0,4 m; dan kisaran
periode 0,5 \textbf{--} 7 detik\newline
Kecepatan pembawa : 0 \textbf{--} 11 m/detik (akurat $\pm$ 2 \%)\newline

\item[\textbf{g.}]  \textbf{\textit{Danish Hydraulic Institute, Horsholm,
Denmark}}\newline
Ukuran Kolam : 30 m $\times$ 20 m $\times$ 3 m\newline
Bagian yang dalam : 12 m di tengah\newline
Pembangkit gelombang : 60 flap hidrolik yang dikendalikan pada satu sisi dan
dikontrol oleh suatu komputer-mini\newline
Gelombang : Ketinggian maksimum $\cong$ 0,6 m dan kisaran periode $\cong$
0,5 \textbf{--} 4 detik\newline

\item[\textbf{h.}]  \textbf{\textit{Norwegian Hydrodynamic Laboratory,
Trondheim, Norwegia (MARINTEK)}}\newline
\textbf{The Ocean Basin}\newline
Ukuran Kolam : 80 m $\times$ 50 m $\times$ 10 m\newline
Pembangkit gelombang : Tipe flap-ganda berengsel, 144 dikontrol tersendiri;
Tipe berengsel yang dikendalikan secara hidrolik\newline
Gelombang : Regular dan tak regular; Ketinggian maksimum 0,9 m \newline
Kecepatan gelombang : Kecepatan maksimum 0,2 m/detik\newline
\end{itemize}
